\documentstyle[12pt]{article}

\catcode`\@=11

\ifx\documentclass\undefined
 \def\@sect#1#2#3#4#5#6[#7]#8{\ifnum #2>\c@secnumdepth
     \let\@svsec\@empty\else
     \refstepcounter{#1}\edef\@svsec{\csname prefix#1\endcsname
	\csname the#1\endcsname\hskip 1em}\fi
     \@tempskipa #5\relax
      \ifdim \@tempskipa>\z@
        \begingroup #6\relax
          \@hangfrom{\hskip #3\relax\@svsec}{\interlinepenalty \@M #8\par}%
        \endgroup
       \csname #1mark\endcsname{#7}\addcontentsline
         {toc}{#1}{\ifnum #2>\c@secnumdepth \else
                      \protect\numberline{\csname the#1\endcsname}\fi
                    #7}\else
        \def\@svsechd{#6\hskip #3\relax  
                   \@svsec #8\csname #1mark\endcsname
                      {#7}\addcontentsline
                           {toc}{#1}{\ifnum #2>\c@secnumdepth \else
                             \protect\numberline{\csname the#1\endcsname}\fi
                       #7}}\fi
     \@xsect{#5}}
\else
    \def\@seccntformat#1{\csname prefix#1\endcsname
	\csname the#1\endcsname\quad}
\fi

\@addtoreset{equation}{section}   

\def\thebibliography#1{\section*{References\@mkboth
 {REFERENCES}{REFERENCES}}\list
 {\leftbibmark\arabic{enumi}\rightbibmark}{
 \settowidth\labelwidth{\leftbibmark #1\rightbibmark}\leftmargin\labelwidth
 \advance\leftmargin\labelsep
 \usecounter{enumi}}
 \def\newblock{\hskip .11em plus .33em minus -.07em}
 \sloppy\clubpenalty4000\widowpenalty4000
 \sfcode`\.=1000\relax}

\def\leftbibmark{[}
\def\rightbibmark{]}

\catcode`\@=12

\def\H{{\cal H}}

\def\L{{\cal L}}
\def\M{{\cal M}}

\def\Y{{\cal Y}}

\def\linebreak{\hfill\break}

\def\Then{$\then$ }

\def\Eq#1{Eq.(\ref{#1})}
\def\Eqs#1#2{Eqs.(\ref{#1})-(\ref{#2})}

\def\tend{\rightarrow}
\def\then{\Rightarrow\quad}
\def\equivalent{\quad\Leftrightarrow\quad}
\def\therefore{\mbox{\setbox0=\hbox{X}\hbox{$\ldotp$}\raise0.7\ht0\hbox{$\ldotp$}\hbox{$\ldotp$}} }
\def\because{\mbox{\setbox0=\hbox{X}\raise0.7\ht0\hbox{$\ldotp$}\hbox{$\ldotp$}\raise0.7\ht0\hbox{$\ldotp$}}\kern0pt }
\def\r#1{{\rm #1}}
\def\b#1{{\bf #1}}

\def\bm#1{\mbox{\boldmath $#1$}}

\def\Frac(#1/#2){\left(\frac{#1}{#2}\right)}
\def\Tr{\r{Tr}}
\def\Tp#1{\,{}^t\! #1}

\def\Lie{\hbox{\rlap{$\cal L$}$-$}}

\def\NN{{\bm{N}}}
\def\ZR{{\bm{Z}}}

\def\RF{{\bm{R}}}
\def\CF{{\bm{C}}}

\def\orth{\perp}
\def\bop{\mathchoice{{\scriptstyle\circ}}{{\scriptstyle\circ}}{{\scriptscriptstyle\circ}}{\circ}}

\def\mod{{\rm mod}}

\def\maps{\:\rightarrow\:}
\def\mapsnamed#1{\:\stackrel{#1}{\longrightarrow}\:}

\def\In{\mathrel{\mbox{\setbox0=\hbox{$\cup$}\dimen0=\wd0\divide\dimen0 by 2
\box0\kern -\dimen0\vrule}}}

\def\DiffG{\r{Diff}}
\def\SetDef#1#2{\left\{#1\;\mid \;#2\right\}}

\def\sdp{\tilde\times}
\def\InvG{\r{InvG}}
\def\DiffG{\r{Diff}}
\def\TSUB{\r{TSB}}
\def\HPDG{\r{HPDG}}
\def\Gmax{G_\r{max}{}}
\def\Gmaxo{G^+_\r{max}{}}

\def\Aut{\r{Aut}}
\def\EAut{\r{EAut}}
\def\LieA{\L }
\def\Ad{\r{Ad}}

\def\Isom{\r{Isom}}
\def\Nil{\r{Nil}}
\def\Sol{\r{Sol}}
\def\PSL{\r{PSL}}
\def\SO{\r{SO}}
\def\ISO{\r{ISO}}
\def\IO{\r{IO}}
\def\GL{\r{GL}}
\def\SL{\r{SL}}
\def\SU{\r{SU}}
\def\maprightU#1{\smash{\mathop{\longrightarrow}\limits^{#1}}}
\def\maprightD#1{\smash{\mathop{\longrightarrow}\limits_{#1}}}
\def\mapdownR#1{\Big\downarrow\rlap{$\vcenter{\hbox{$\scriptstyle #1$}}$}}
\def\BlockDiagram#1#2#3#4#5#6#7#8{
\def\normalbaselines{\baselineskip20pt \lineskip3pt \lineskiplimit3pt}
\begin{array}{ccc}#1 & \maprightU{#2} & #3       \\
           \mapdownR{#4} &       & \mapdownR{#5} \\
                  #6 & \maprightD{#7} & #8
\end{array}}
\def\mapseL#1{{\scriptstyle #1}\rlap{$\vcenter{\hbox{$\searrow$}}$}}
\def\mapswR#1{\llap{$\vcenter{\hbox{$\swarrow$}}$}{\scriptstyle #1}}
\def\mapseR#1{\llap{$\vcenter{\hbox{$\searrow$}}$}{\scriptstyle #1}}
\def\mapswL#1{{\scriptstyle #1}\rlap{$\vcenter{\hbox{$\swarrow$}}$}}
\def\DtriangleDiagram#1#2#3#4#5#6{
\def\normalbaselines{\baselineskip20pt \lineskip3pt \lineskiplimit3pt}
\begin{array}{ccc}#1 & \maprightU{#2} & #3    \\
              \mapseL{#4} & &  \mapswR{#5} \\
                     & #6       & 
\end{array}}
\def\UtriangleDiagram#1#2#3#4#5#6{
\def\normalbaselines{\baselineskip20pt \lineskip3pt \lineskiplimit3pt}
\begin{array}{ccc}        & #1  &          \\
               \mapswL{#2} & & \mapseR{#3} \\
                   #4 &\maprightU{#5} & #6
\end{array}}

\newtheorem{theorem}{Theorem}[section]
\newtheorem{proposition}{Proposition}[section]

\renewenvironment{matrix}[1]{\left(\begin{array}{#1}}{\end{array}\right)}

\def\Beq{\begin{equation}}
\def\Eeq{\end{equation}}
\def\Beqr{\begin{eqnarray}}
\def\Eeqr{\end{eqnarray}}
\def\Beqrn{\begin{eqnarray*}}
\def\Eeqrn{\end{eqnarray*}}
\def\Bitm{\begin{itemize}}
\def\Eitm{\end{itemize}}

\font\elevenmib=cmmib10 scaled\magstephalf   \skewchar\elevenmib='177
\def\YUKAWAmark{\hbox{\elevenmib 
 Yukawa\hskip0.05cm Institute\hskip0.05cm Kyoto \hfill}}

\begin{document}

\begin{titlepage}
\hbox to \hsize{\YUKAWAmark \hfill YITP-97-25}
\rightline{May 1997}

\vspace{2cm}

\begin{center}\large\bf
Canonical Structure of Locally Homogeneous Systems on Compact Closed 3-Manifolds
of Types $E^3$, $\Nil$ and $\Sol$
\end{center}

\begin{center}\large
Hideo Kodama
\end{center}

\begin{center}\large\it
Yukawa Institute for Theoretical Physics\\
Kyoto University, Kyoto 606-01, Japan
\end{center}

\vspace{5cm}

\begin{abstract}
In this paper we investigate the canonical structure of diffeomorphism
invariant phase spaces for spatially locally homogeneous spacetimes
with 3-dimensional compact closed spaces. After giving a general
algorithm to express the diffeomorphism-invariant phase space and the
canonical structure of a locally homogeneous system in terms of those
of a homogeneous system on a covering space and a moduli space, we
completely determine the canonical structures and the Hamiltonians of 
locally homogeneous pure gravity systems on orientable compact closed
3-spaces of the Thurston-type $E^3$, $\Nil$ and $\Sol$ for all
possible space topologies and invariance groups. We point out that in
many cases the canonical structure becomes degenerate in the moduli
sectors, which implies that the locally homogeneous systems are not
canonically closed in general in the full diffeomorphism-invariant
phase space of generic spacetimes with compact closed spaces.
\end{abstract}

\end{titlepage}

\section{Introduction}

Locally homogeneous systems are, roughly speaking, extensions of
spatially homogeneous systems on simply connected spaces to spaces
with non-trivial topologies. Recent increase of interest in such
systems are closely related with the investigations of quantum
gravity\cite{Ashtekar.A&Samuel1991,Fujiwara.Y&Kodama&Ishihara1993,%
Koike.T&Tanimoto&Hosoya1994,Tanimoto.M&Koike&Hosoya1997,%
Tanimoto.M&Koike&Hosoya1997a}, although
early work on them was mainly concerned with applications to the
observational cosmology(see Ref.\cite{Lachieze-Rey.M&Luminet1995} for
review).

In quantum gravity spatially homogeneous systems were intensively
studied in the name of minisuperspace models, because they give simple 
gravity systems of finite degrees of freedom. If one applies the quantization
procedure to such systems, however, one encounters the problem that the 
Hamiltonian diverges due to the infinite spatial volume except for
Bianchi IX models. Since the evolution equations for space metric do
not depend on the spatial volume, this difficulty was often avoided 
by considering a virtual finite region of space. Though this
prescription works well in the investigation of classical structure of 
canonical dynamics, it is not satisfactory in the quantum problem
because the virtual volume of the region affects quantum behavior of
the system though the Hamiltonian.

A more natural and reasonable way to resolve this difficulty is to
consider spacetimes with compact spaces instead of simply connected
open spaces. In this approach, however, two new problems
arise. Firstly, most of compact closed 3-manifolds do not allow a
Bianchi-like globally homogeneous metric, i.e., a metric whose
isometry group acts simply transitively on the manifolds. This obliges 
us to replace the requirement of global homogeneity to a local 
one\cite{Ashtekar.A&Samuel1991,Fujiwara.Y&Kodama&Ishihara1993}. 
Secondly, since compact Riemannian manifolds with non-trivial
topologies are not globally isomorphic even if they are locally
isomorphic, the structure of a locally homogeneous space
is not fully described by the components of the  metric with respect
to an invariant basis(even if it is well-defined) as in the Bianchi
models, but depends on the so-called moduli parameters. Thus the
dynamical structure of a locally homogeneous system on a compact
manifold differs from that of a globally homogeneous system on its
simply connected covering manifold.

These problems also occur in $(2+1)$-dimensional systems and are
studied well. There a natural definition of locally homogeneity is
obtained from the requirement that the Ricci scalar curvature is
spatially constant. The locally homogeneous closed two surfaces in
this sense are always covered by a simply connected homogeneous space
and is obtained as a quotient of the latter by its discrete isometry
group. Though the Ricci homogeneity is too weak in the
$(3+1)$-dimensional systems, the latter characterization, i.e., the
requirement that its covering spacetime is globally spatially
homogeneous, can be easily extended to higher dimensions. Furthermore
if we adopt this as the definition of spatial local homogeneity of
$(3+1)$-dimensional systems, we can almost completely determine all
the possible topologies of compact closed spaces and their locally
homogeneous structures including the moduli with the help of
Thurston's theorem.

Recently Koike, Tanimoto and Hosoya\cite{Koike.T&Tanimoto&Hosoya1994}
completely determined the degree of moduli freedom of locally
homogeneous 3-manifolds for all possible topologies by this approach.
They further applied this result to the investigation of dynamics of
spatially locally homogeneous
spacetimes%
\cite{Tanimoto.M&Koike&Hosoya1997,Tanimoto.M&Koike&Hosoya1997a}, 
and gave a systematic
algorithm to determine the dynamics of moduli. Their strategy was as
follows. First, they defined a spatially locally homogeneous spacetime
$(M,g)$ as a spacetime with compact space which is obtained as a 
quotient of a spatially homogeneous spacetime $(\tilde M,\tilde g)$
with a simply connected space by its discrete spacetime isometry group
$K$ contained in the spatial homogeneity subgroup. Here it is
important to distinguish the isometry group $G(t)$ of each constant
time slice of $\tilde M$ and its subgroup $G_\r{e} $ whose
transformations are
extendible to spacetime isometries. Next they classified the covering
spatially homogeneous vacuum solutions and put them into some
standard form $(\tilde M,\tilde g_0)$ to fix the freedom of
extensible homogeneity-preserving diffeomorphisms(EHPDs), which are
defined as transformations of the covering spacetimes $(\tilde
M,\tilde g)$ preserving the spatial homogeneity. By this procedure one
obtains spacetime diffeomorphism classes of $(M,g)$, each of which is
described by the canonical form of the metric $\tilde g_0$ or a
evolutionary family of covering 3-metric $\tilde q_0(t)$, and the
conjugate class of the discrete subgroup $K$ in $G_\r{e}$.  Since
EHPDs are a subset of the transformations of each constant-time slice
which preserve the spatial homogeneity of $\tilde q_0(t)$, one can
further reduce the freedom of $\tilde q_0(t)$ on each slice by the
latter transformations. This time-dependent reduction maps $K$ to a
time-dependent conjugate class $K(t)$ with respect to $G(t)$, which
corresponds the standard moduli/Teichm\"{u}ller freedom. Thus one can
determine the time evolution of the 3-metric and moduli parameters of
a locally homogeneous space.

This algorithm is very useful in determine the time evolution of
moduli parameters in the sense defined above at least in the classical
framework. However, it has some disadvantages in the investigation of
canonical structures of the systems and their quantization. First, the
knowledge on the structure of spacetime solutions is required in
advance in their method. Though this is not an obstacle for the pure
gravity systems, such information is not available for systems coupled
with matter in general. Second, since they start from the spacetime
solutions, the information on the momentum variable conjugate to the
3-metric or on the Hamiltonian is not directly given. It must be
calculated by inserting spacetime solutions in the generic definitions
of the momentums and the Hamiltonian. This procedure often fails
because defining momentums requires the knowledge of the canonical
structure, which is generally not available in the moduli sector in
their approach. Since the information on the canonical structure is
crucial in going to quantum theory, this defect is serious. In fact,
we will show in this paper that the actual canonical structure is
often degenerate in the moduli sector. Such information can not be
obtained in their approach.

On the basis of these considerations, in the present paper, we give a
slightly different scheme to determine the dynamics of locally
homogeneous systems.  The main point is that we directly treat the
diffeomorphism-invariant phase space of locally homogeneous canonical
data including momentums on a compact closed 3-manifold $M$, and
determine its canonical structure and the Hamiltonian directly from
those of the diffeomorphism-invariant phase space of generic canonical
data on $M$.  This enables us to discuss the off-shell behavior of the
systems as well as relations of the canonical structures of locally
homogeneous systems to generic systems, such as the degeneracy of the
canonical structure in the locally homogeneous sector.

The paper is organized as follows. First, in the next section, we
prove a theorem which enables us to express the
diffeomorphism-invariant phase space of locally homogeneous data on a
compact closed 3-manifold $M$ in terms of globally homogeneous data on
its universal covering manifold $\tilde M$ and a moduli freedom of the
embedding of the fundamental group $\pi_1(M)$ into an invariance group
of the covering data. There Thurston's theorem on the classification
and the uniqueness of maximal geometries on compact closed 3-manifolds
play an essential role. Then we explain a general algorithm to
determine the canonical structure and the Hamiltonian of them. We
further prove an important theorem on the dynamics of moduli
parameters. In the subsequent three sections, following this
algorithm, we completely determine the canonical structures and the
Hamiltonians of locally homogeneous pure gravity systems on orientable 
compact closed 3-space of the Thurston-type $E^3$, $\Nil$ and
$\Sol$. As a result we point out that in many of them the canonical
structure becomes degenerate in the moduli sector. Section 6 is
devoted to summary and discussions.

\section{General Theory}

\subsection{Diffeomorphism-Invariant Phase Space}

Let $M$ be a compact closed 3-manifold and $\Phi$ be a set of 
canonical variables on it.
For example, for the pure gravity system, $\Phi$ consists of 
a pair $(q,p)$ of a three metric $q$ on $M$ and its conjugate 
momentum $p$. When it is coupled with matter, canonical pairs of 
matter fields should be included in $\Phi$. 

Naively it is natural to define that canonical data $\Phi$ are 
locally homogeneous when around each point of $M$ there exist 
a set of local fields $\xi_I$ which contain at least three 
linearly independent fields and leave $\Phi$ invariant, 
i.e., $\Lie_{\xi_I}\Phi=0$. However, this local definition is not
convenient for our purpose because we must treat cases in which 
the base space $M$ has non-trivial topology.

If we instead require that the fields $\xi_I$ are globally 
defined on $M$, the analysis of the problems gets much simplified.
However, this requirement is too stringent and excludes many 
interesting cases. For example, consider a compact Riemannian 
manifold $(M,q)$ constructed from the cubic region $0\le x,y,z\le 1$ 
of Euclidean space $E^3$ by identifying the pair of faces $x=0$ 
and $x=1$ by rotating by $\pi$ and the other two pairs of faces 
normally. Clearly $(M,q)$ is locally isometric to $E^3$
and locally homogeneous, but translation Killing vectors parallel 
to the $y-z$ plane cannot be extended to the whole space $M$. 
Another example is given by the class B open Bianchi models 
which cannot be compactified keeping the global transitive 
symmetries\footnote{There had been some misunderstanding on 
this point in the early literature\cite{Fagundes.H1992,Ashtekar.A&Samuel1991,%
Lachieze-Rey.M&Luminet1995}. It was corrected in Ref.\cite{Fujiwara.Y&Kodama&Ishihara1993}.}.

Since the main obstruction against the existence of global symmetries
in these examples is of a topological nature, a better definition
which remedies the defects of the definitions above is obtained by
considering the universal covering data. Let $\tilde M$ be a universal
covering space of $M$, and $j$ be a covering map from $\tilde M$ onto
$M$. Then through the pullback by $j$ we obtain unique data $\tilde
\Phi$ on $\tilde M$ from $\Phi$ on $M$.  Clearly $\tilde \Phi$ is
locally homogeneous in the first definition if $\Phi$ is. Further if
$\tilde \Phi$ is real analytic and $\Phi$ contains the metric data,
the local symmetries of $\tilde \Phi$ can always be extended to global
symmetries\cite{Kobayashi.S&Nomizu1963B}. Hence in this case local
homogeneity of $\Phi$ in the naive definition implies global global
homogeneity of the covering data $\tilde \Phi$.  This global symmetry
defines an invariance group
\Beq
\InvG(\tilde \Phi,\tilde M)
:=\SetDef{f\in\DiffG(\tilde M)}{f_*\tilde\Phi=\tilde \Phi}.
\Eeq
When there occurs no confusion, we often write 
$\InvG(\tilde\Phi,\tilde M)$ as $\InvG(\tilde\Phi)$.

From these observations, in this paper, we define that
canonical data $\Phi$ on $M$ are {\it locally $G$-homogeneous} 
if there exists a universal covering space $j:\tilde M \maps M$ 
such that the invariance group of the pullback $\tilde \Phi=j^*\Phi$,
$\InvG(\tilde \Phi,\tilde M)$, is isomorphic to $G$ and
acts transitively on $\tilde M$. 
The reason why we have introduced the symmetry group $G$ in 
an abstract way is that $\InvG(\tilde \Phi,\tilde M)$ is not
 intrinsic to the original manifold $M$, and manifests itself only 
as local symmetries on $M$. We denote the set of all locally 
$G$-homogeneous data on $M$ by $\Gamma_\r{LH}(M,G)$.

This definition allows us to translate the problem on compact 
manifold with complicated topological structures to that on 
much simpler simply-connected manifolds, and is widely adopted 
in the recent literature on this problem%
\cite{Lachieze-Rey.M&Luminet1995,Fujiwara.Y&Kodama&Ishihara1993,%
Koike.T&Tanimoto&Hosoya1994}.
Now we closely examine the relation between these two problems.

\subsubsection{The freedom of the covering space $\tilde M$}

Our definition gives the description of locally homogeneous
data on $M$ in terms of 
the set of its universal covering space $\tilde M$, a covering
map $j$, and homogeneous covering data $\tilde \Phi$. There
is clearly a redundancy in this description since all the universal 
covering spaces are mutually diffeomorphic. This redundancy is
eliminated with the help of the following well-known fact.

\begin{proposition}\label{prop:LiftOfMap}
Let $j:\tilde M\maps M$ and $j':\tilde M'\maps M'$ be a pair of two 
universal covering spaces. 
\Bitm
\item[1)] For any diffeomorphism $f:M \maps M'$, there exists a 
diffeomorphism $\tilde f:\tilde M\maps \tilde M'$ up to the freedom
of covering transformations such that $f\bop j=j'\bop \tilde f$, i.e.,
the following diagram commutes:
\Beq
\BlockDiagram{\tilde M}{\tilde f}{\tilde M'}{j}{j'}{M}{f}{M'}
\label{diagram:UniversalCover}\Eeq
\item[2)] For any diffeomorphism $\tilde f:\tilde M\maps \tilde M'$ which
preserves fibers, i.e., maps each fiber $j^{-1}(x)\subset \tilde M$ 
onto some fiber $j'{}^{-1}(x')\subset \tilde M'$, there is a unique
diffeomorphism $f:M\maps M'$ such that $f\bop j=j'\bop\tilde f$.
\Eitm
\end{proposition}

The first part of this proposition follows from the uniqueness of 
the lift of curves in the covering space. For example, 
let $x_0$ and $y_0$ be a pair of points such that $y_0=f(x_0)$ 
and pick up a point $\tilde x_0\in j^{-1}(x_0)$ and another 
point $\tilde y_0\in j'{}^{-1}(y_0)$. For a given point 
$\tilde x\in \tilde M$, take a curve $\tilde \gamma$ 
from $\tilde x_0$ to $\tilde x$. Then the curve 
$f\bop j(\tilde \gamma)$ in $M'$ is uniquely lifted to a curve 
$\tilde \gamma'$ in $\tilde M'$ starting from $\tilde y_0$.
Let its endpoint be $\tilde y$. Since any two curves connecting 
$\tilde x_0$ and $\tilde x$ are homotopic due to the 
simple-connectedness of $\tilde M$, $\tilde y$ does not depend 
on the choice of $\tilde \gamma$ and defines a map 
$\tilde y=\tilde f(\tilde x)$. It is a simple task to show that 
$\tilde f$ is a diffeomorphism  and the 
diagram \ref{diagram:UniversalCover} commutes. 
The second part is trivial.
 
Now let $j:\tilde M \maps M$ and $j':\tilde M' \maps M$ be 
two universal covering spaces of $M$ defining the local homogeneity
of data $\Phi$ on $M$. Then the above proposition implies 
that there exists a diffeomorphism 
$\tilde f:\tilde M\maps \tilde M'$
such that the following diagram commutes:

\Beq
\DtriangleDiagram{\tilde M}{\tilde f}{\tilde M'}{j}{j'}{M}
\Eeq
From this it follows that the covering data $\tilde \Phi$ on 
$\tilde M$ and $\tilde \Phi'$ on $\tilde M'$ are related by
$\tilde f_*\tilde\Phi=\tilde f_*j^{-1}_*\Phi
=j'{}^{-1}_*\Phi=\tilde\Phi'$. This implies that 
$\InvG(\tilde\Phi,\tilde M)$ and $\InvG(\tilde\Phi',\tilde M')$ 
are connected by the external automorphism induced by 
$\tilde f$ as $\InvG(\tilde\Phi',\tilde M')=\tilde f
\InvG(\tilde\Phi,\tilde M)\tilde f^{-1}$. In particular, if 
$\InvG(\tilde\Phi,\tilde M)\cong G$ then 
$\InvG(\tilde\Phi',\tilde M')\cong G$, and vice versa. Hence
there is a one-to-one correspondence between the description by
$G$-invariant data on $\tilde M$ and that on $\tilde M'$ 
up to the freedom of the covering transformations. 
Therefore we can work in one fixed universal covering  space 
to construct $\Gamma_\r{LH}(M,G)$.

\subsubsection{Moduli: intrinsic representation of the freedom of 
covering map $j$}

If we apply the argument in the last paragraph to the case 
corresponding to a pair of different covering maps $j$ and $j'$ 
by the same covering space $\tilde M$, we obtain an automorphism 
of the covering data on $\tilde M$ induced by the change of 
the covering map. This implies that we can also fix the 
projection map $j$ to see the correspondence of data on $M$
and those on $\tilde M$. However, this approach is not adequate 
for our purpose because we cannot fix the invariance group of 
$\tilde \Phi$ isomorphic to $G$, which introduces complications 
into the investigation of the structure of allowed covering data.

For example, consider the 3-dimensional Euclidean space $E^3$ as the
covering space and construct the 3-dimensional torus $M=T^3$ as the
quotient manifold by the discrete transformation group $K_0$ generated
by the standard lattice $K_0=\SetDef{(l,m,n)}{l,m,n\in\ZR}$.  The
standard metric on $E^3$ is represented in terms of the natural
Descartes coordinate as $\tilde q_0=dx^2+dy^2+dz^2$ and its invariance
group is given by $\IO(3)$. If we denote the standard projection map
from $E^3$ to $T^3$ by $j_0$, $q_0=j_0{}_* \tilde q_0$ gives locally
IO(3)-homogeneous data on $T^3$. Next let us consider a linear
transformation $\bm{x}'=\tilde f(\bm{x}):=A\bm{x}$. The metric $\tilde
q$ obtained from $\tilde q_0$ by this transformation, $\tilde q=\tilde
f_*\tilde q_0$, gives another data $q$ on $T^3$ through $j_0$. It is
not isomorphic to $q_0$ if $\tilde f$ does not belong to IO(3) because
the lengths of closed geodesics in $T^3$ with respect to $q_0$ and $q$
are different.  Clearly the invariance groups $\InvG(\tilde q_0,E^3)$
and $\InvG(\tilde q,E^3)$ are different, though they are connected by
the external automorphism induced by $f$ and both give locally
IO(3)-homogeneous data on $T^3$.

On the other hand if we consider the projection defined by $j=j_0\bop
\tilde f$, $q=j_0{}_*\tilde f_*\tilde q_0 = j_0{}_*\tilde q$ coincides
with the data induced from the original standard metric $\tilde q_0$
by the new projection map $j$. Hence by changing the projection map,
we can construct non-isomorphic data on $T^3$ from covering data with
the same invariance group.

On the basis of this observation we do not fix the projection map, 
and take the approach to express its freedom in an intrinsic way. 
The basis of this approach is provided by the following fact.

\begin{proposition}\label{prop:LiftOfPi1}
For a given point $\tilde x_0$ in $\tilde M$, each homotopy class of
the loops with the base point $j(\tilde x_0)$ induces a unique
covering transformation of the universal covering space $\tilde
M$. Let us denote the corresponding monomorphic embedding of
$\pi_1(M,j(\tilde x_0))$ into $\DiffG(\tilde M)$ as $j_{\tilde
x_0}^\sharp$. Then the image of this monomorphism, $j_{\tilde
x_0}^\sharp(\pi_1(M,j(\tilde x_0))$, does not depend on the choice of
$\tilde x_0$, and gives a unique discrete subgroup
$j^\sharp(\pi_1(M))$. $\tilde\Phi=j^*\Phi$ is invariant under this
subgroup for any data $\Phi$ on $M$, i.e.,
$j^\sharp(\pi_1(M))\subseteq \InvG(\tilde \Phi,\tilde M)$.
\end{proposition}

The former half of the proposition is a well-known fact and is proved
by considering the special case in Prop. \ref{prop:LiftOfMap} that
$(\tilde M',j',M')=(\tilde M,j,M)$ and $x_0$ and $y_0$ are taken as 
the start point and the end point of the lift of a closed loop. To
show the latter half, take two points $\tilde x_0$ and $\tilde y_0$ 
in $\tilde M$. Let $\tilde \mu$ be a curve from $\tilde x_0$ to
$\tilde y_0$, and $\mu$ be its projection on $M$. Then from the
way of construction of the covering transformation, it is easily 
shown that $j^\sharp_{\tilde y_0}([\mu\alpha\mu^{-1}])
=j^\sharp_{\tilde x_0}([\alpha])$ holds for any closed loop 
$\alpha$ with the base point $j(\tilde x_0)$ where $[\alpha]$ 
represents the homotopy class of $\alpha$. The invariance of 
$\tilde\Phi$ under $j^\sharp(\pi_1(M))$ is 
trivial from the definition of the pullback.

This proposition shows that the discrete subgroup 
$j^\sharp(\pi_1(M))$ yields an intrinsic representation of the
covering map. Though this discrete group does not uniquely 
determine the covering map $j$, this representation is very 
useful in our problem. To see this, suppose that two covering 
maps $j_1$ and $j_2$ map $\pi_1(M)$ to the same discrete group $K$. 
Then since a fiber containing a point $\tilde x$ is given by 
the set $j^\sharp(\pi_1(M))\tilde x$ in general, 
$j_1^{-1}(x)\cap j_2^{-1}(y)\not=\emptyset$ implies $j_1^{-1}(x)=j_2^{-1}(y)$ 
in the present case. This shows that the
identity transformation $\tilde f=$id gives a fiber preserving
diffeomorphism between the two covering space $(\tilde M,j_1,M)$ and
$(\tilde M,j_2,M)$. Hence applying Prop. \ref{prop:LiftOfMap} to the
following diagram,
\Beq
\UtriangleDiagram{\tilde M}{j_1}{j_2}{M}{f}{M}
\Eeq
we obtain a diffeomorphism $f\in\DiffG(M)$ such that this diagram
commutes. Therefore for any covering data $\tilde \Phi$ on $\tilde M$
such that $K\subset \InvG(\tilde\Phi,\tilde M)$, the corresponding
data $\Phi_1=j_1{}^* \tilde \Phi$ and $\Phi_2=j_1{}_*\tilde \Phi$ on
$M$ are related by the diffeomorphism $f$ as $\Phi_2=f_*\Phi_1$. 
Hence the pair $(\tilde \Phi, K)$ uniquely determines a 
diffeomorphism class of locally $G$-homogeneous data $\Phi$ on
$M$. This suggests that the problem of classifying the 
diffeomorphism classes of the data in $\Gamma_\r{LH}(M,G)$ 
can be replaced by that of the pair of data 
$(\tilde \Phi,K)$. In other words, if we denote 
the set of covering data whose invariance group coincides 
with $\tilde G(\subset\DiffG(\tilde M))$ as
\Beq
\Gamma_\r{H}(\tilde M,\tilde
G):=\SetDef{\tilde\Phi}{\InvG(\tilde\Phi,\tilde M)=\tilde G},
\Eeq
and the set of all possible images of $\pi_1(M)$ in $\tilde G$ by
$j^\sharp$ as
\Beq
\M(M,\tilde G):=\SetDef{K: \hbox{a subgroup of}\quad \tilde G}
{K=j^\sharp(\pi_1(M)) \quad\hbox{for some}\quad j:\tilde M\maps M},
\Eeq
the problem is replaced by that of determining the diffeomorphism 
class of 
$\bigcup_{\tilde G\cong G}\Gamma_\r{H}(\tilde M,\tilde G)\times
\M(M,\tilde G)$.

\subsubsection{Eliminating the freedom of $\tilde G$}

Though the above reformulation reduces the problem to a simpler
one, it still has a cumbersome large freedom associated with 
$\tilde G$. In order to eliminate this freedom, we utilize Thurston's
theorem on the geometric structures on 3-dimensional manifolds.

First we give a couple of basic definitions. 
A pair $(\tilde M,\Gmax)$ of a connected and simply connected
manifold $\tilde M$ and its transformation group $\Gmax$ is said
to define a maximal geometry on $\tilde M$ if there exists a metric on 
$\tilde M$ such that its isometry group coincides with $\Gmax$ and 
there exists no metric with a larger isometry group. Two maximal
geometries  $(\tilde M,\Gmax)$ and  $(\tilde M',G'_\r{max})$ are
said to be isomorphic if there exists a diffeomorphism $f:\tilde M
\maps \tilde M'$ such that $G'_\r{max}=f\Gmax f^{-1}$. Further 
a compact closed manifold $M$ is called a compact quotient of a
maximal geometry $(\tilde M,\Gmax )$ if $\Gmax $
has a discrete subgroup $K$ such that the quotient space $\tilde M/K$
is diffeomorphic to $M$. With these definitions Thurston's
theorem is stated as follows\cite{Thurston.W1979B,Thurston.W1982,%
Scott.P1983}:

\begin{table}
\begin{tabular}{lllll}
Space & $\Gmax $ & $G^+_\r{max}$ & $(G_\r{min})_0$ & Bianchi type \\
\\
$E^3$ & $\IO(3)$ & $\ISO(3)$ & $\RF^3$ & I \\
      &         &          & VII$_0^{(\pm)}$ & VII(0) \\
\\
$\Nil$   & $\Isom(\Nil)$ &$\Isom(\Nil)$ & $\Nil$ & II \\
      & $\cong \RF\sdp \IO(2)$ &&&\\
\\
$\Sol$  & $\Isom(\Sol)$ & $\Sol\sdp D_2$ & $\Sol$ & VI(0) \\
     & $\cong \Sol\sdp D_4$ &&&\\
\\
$H^3$ & $\Isom(H^3)$& $\PSL_2\CF$ & $\PSL_2\CF$ & V, VII($A\not=0$) \\  
     & $\cong\PSL_2\CF\sdp \ZR_2$ &&&\\
\\
$\widetilde{\SL_2\RF}$ & $\Isom(\widetilde{\SL_2\RF}$)
& $\Isom(\widetilde{\SL_2\RF}$) & $\RF\sdp \PSL_2\RF$ & III, VIII \\
     & $\cong \RF\sdp \PSL_2\RF\sdp \ZR_2$ &&&\\
\\
$H^2\times E^1$ & $\Isom(H^2\times E^1)$
& $(\PSL_2\RF\times\RF)\sdp\ZR_2$ & $\PSL_2\RF\times\RF$ & III \\
     & $\cong \PSL_2\RF\sdp \ZR_2\times\IO(1)$ &&&\\
\\
$S^3$ & O(4) & $\SO(4)$ & $\SU(2)$ & IX \\
\\
$S^2\times E^1$ & O(3)$\times$ $\IO(1)$ & $\SO(3)\times\RF\sdp\ZR_2$ 
& $\SO(3)\times \RF$ & Kantowski-Sachs \\
&&&& models\\
\\
\end{tabular}
\caption{\label{tbl:ThurstonType}Thurston types}For each Thurston type 
the maximal symmetry group $G_\r{max}$, its orientation-preserving 
component $G^+_\r{max}$, the connected component of minimum transitive 
invariance groups $(G_\r{min})_0$, and the Bianchi types of the simply 
transitive subgroups are shown\cite{Fujiwara.Y&Kodama&Ishihara1993,%
Koike.T&Tanimoto&Hosoya1994}. 
\end{table}

\begin{theorem}[Thurston]
\begin{itemize}
\item[]
\item[1)] [Classification theorem]\\
Any 3-dimensional maximal geometry $(\tilde M, \Gmax )$ is
isomorphic to one of 8 types listed in Table \ref{tbl:ThurstonType},
provided that it allows a compact quotient.
\item[2)] [Uniqueness theorem]\\
For any 3-dimensional compact closed manifold $M$, if $M$ is a compact
quotient of a maximal geometry, the corresponding maximal geometry is
unique up to isomorphism.
\end{itemize}
\end{theorem}

To apply this theorem, first note that from the argument in \S2.1.1
for any covering map $j:\tilde M\maps M$ and $\tilde f\in\DiffG(\tilde
M)$, $j'=\tilde f^{-1}\bop j$ gives a new covering map, and that
conversely for any two covering maps $j$ and $j'$ there exists a
diffeomorphism $\tilde f$ such that $\tilde f\bop j'=j$. Further, in
this case, the invariance groups of covering data $\tilde \Phi$ and
$\tilde \Phi'$ corresponding to the same data $\Phi$ on $M$ are
mutually conjugate by $\tilde f$. On the other hand, from the above
theorem, the base manifold $M$ uniquely determines a diffeomorphism
class of a maximal geometry. Let us pick up one representative pair of
it, $(\tilde M, \Gmax )$, and fix it. Then if the data $\Phi$ contains
metric data $q$, its invariance group $\InvG(\tilde\Phi)$ is contained
in some maximal symmetry group $G'_\r{max}$ since $\InvG(\tilde
\Phi)\subseteq\InvG(\tilde q)$, and again from Thurston's theorem
there exists a diffeomorphism $\tilde f\in\DiffG(\tilde M)$ such that
$G'_\r{max}=\tilde f \Gmax \tilde f^{-1}$. Therefore, for any given
data $\Phi$ on $M$, we can always find a covering map $j$ such that
$\InvG(\tilde\Phi)$ is contained in the fixed transformation group
$\Gmax $.

With the help of this result we can construct a mapping from
$\Gamma_\r{LH}(M,G)$ to a space with a much simpler structure.  First
let us define $\TSUB(\Gmax )$ as a set of subgroups of $\Gmax $ such
that it contains just one representative element for each conjugate
class of the subgroups of $\Gmax $ with respect to $\DiffG(\tilde M)$ .
Then from the above argument, each locally $G$-homogeneous data
$(\Phi,M)$ uniquely determines a subgroup $\tilde G\in\TSUB(\Gmax )$
such that $\InvG(j^*\Phi)=\tilde G$ holds for some covering map
$j$. The corresponding covering map in turn determines data
$(j^*\Phi,j^\sharp(\pi_1(M))$ in $\Gamma_\r{H}(\tilde G,\tilde
M)\times \M(M,\tilde G)$. Though the covering map $j$ here is not
unique, its freedom is quite restricted. In fact, if
$\InvG(j'{}^*\Phi)=\InvG(j{}^*\Phi)=\tilde G$ holds, there should
exists $\tilde f\in \DiffG(\tilde M)$ such that $j'\bop\tilde f=j$,
but this diffeomorphism must satisfy the condition $\tilde f \tilde G
\tilde f^{-1}=\tilde G$. That is, the freedom in the covering map is
represented by homogeneity preserving diffeomorphisms(HPDs for
brevity). Hence, by denoting the set of HPDs with respect to the symmetry
group $\tilde G$ as
\Beq
\HPDG(\tilde M,\tilde G):=\SetDef{\tilde f\in\DiffG(\tilde M)}
{\tilde f \tilde G\tilde f^{-1}=\tilde G},
\Eeq
we naturally obtain the map
\Beq
F: \Gamma_\r{LH}(M,G)\maps \bigcup_{G\cong \tilde G\in
\TSUB(\Gmax )}\left(\Gamma_\r{H}(\tilde M,\tilde G)\times 
\M(M,\tilde G)\right)/\HPDG(\tilde M,\tilde G),
\Eeq
where the action of $\tilde f\in \HPDG(\tilde M,\tilde G)$ is defined
as
\Beq
\tilde f_*(\tilde \Phi,K)=(\tilde f_*\tilde\Phi, \tilde f K \tilde
f^{-1}).
\Eeq

For any data $(\tilde\Phi,K)$ in $\Gamma_\r{H}(\tilde M,\tilde G)\times 
\M(M,\tilde G)$, from the definition of $\M(\pi_1(M),\tilde G)$, there
is a covering map $j:\tilde M\maps M$ such that
$j^\sharp(\pi_1(M))=K$. Since $K\subset \tilde G=\InvG(\tilde\Phi)$,
there exist unique locally $G$-homogeneous data $\Phi$ on $M$ whose 
pullback by $j$ coincide with $\tilde\Phi$. Hence the map $F$ is
surjective.

\subsubsection{An expression of the diffeomorphism-invariant 
phase space}

The map $F$ defined above induces an isomorphism between the
diffeomorphism-invariant phase space of the locally $G$-homogeneous
data and the target space of $F$. 

To see this, we first show that $F$ induces a well-defined map from
$\Gamma_\r{LH}(M,G)/\DiffG(M)$ to 
$\Gamma_\r{H}(\tilde M,\tilde G)\times \M(M,\tilde G)$.  
Suppose that two data $\Phi_1$ and $\Phi_2$ on $M$ are connected by a
diffeomorphism $f$ of $M$: $\Phi_2=f_*\Phi_1$. From the argument
in \S2.1.3, for each data $\Phi_i(i=1,2)$, we can find a covering map 
$j_i$ such that $\InvG(\tilde\Phi_i)\in \TSUB(\Gmax ,\tilde M)$ 
where $\tilde\Phi_i=j_i^*\Phi_i$. On the other hand,
from Prop. \ref{prop:LiftOfMap}, there exists a diffeomorphism 
$\tilde f\in \DiffG(\tilde M)$ such that the diagram
\Beq
\BlockDiagram{(\tilde\Phi_1,\tilde M)}{\tilde f}{(\tilde\Phi_2,\tilde
M)}{j_1}{j_2}{(\Phi_1,M)}{f}{(\Phi_2,M)}
\label{diagram:theorem2}\Eeq
commutes. Clearly the covering data $\tilde \Phi_1$ and $\tilde
\Phi_2$ are related by $\tilde f$ as $\tilde \Phi_2=\tilde f_* \tilde
\Phi_1$, and their invariance groups as $\InvG(\tilde\Phi_2)=\tilde
f\InvG(\tilde\Phi_1)\tilde f^{-1}$. However from the definition of
$\TSUB(\Gmax )$, this implies
$\InvG(\tilde\Phi_1)=\InvG(\tilde\Phi_2)=\tilde G$ for some $\tilde
G\cong G$. Thus $\tilde G=\tilde f\tilde G\tilde f^{-1}$ holds and
$\tilde f$ must be a HPD. Further, since diffeomorphisms preserve the
fundamental group, it follows from Prop. \ref{prop:LiftOfPi1} that
$j_2^\sharp(\pi_1(M))=\tilde f_*j_1^\sharp(\pi_1(M))$. Hence we obtain
$F(\Phi_1)=F(\Phi_2)$, which implies that $F$ induces the map
\Beq
F_*:\Gamma_\r{LH}(M,G)/\DiffG(M) \maps \bigcup_{G\cong \tilde G\in
\TSUB(\Gmax )}\left(\Gamma_\r{H}(\tilde M,\tilde G)\times 
\M(M,\tilde G)\right)/\HPDG(\tilde M,\tilde G).
\Eeq

Next we show that the map $F_*$ is injective. Suppose that 
$F(\Phi_1)=F(\Phi_2)$ holds for $\Phi_1,\Phi_2 \in \Gamma_\r{LH}(M,G)$.
Then there exist covering maps $j_1$ and $j_2$, $\tilde G\in
\TSUB(\Gmax )$, and $\tilde f\in \HPDG(\tilde M,\tilde G)$
such that $\tilde\Phi_2=j_2^*\Phi_2=\tilde f_* j_1^*\Phi_1 =\tilde f_*
\tilde\Phi_1$ and $j_2^\sharp(\pi_1(M))=\tilde f
j_1^\sharp(\pi_1(M))\tilde f^{-1}$. Since $j_2^\sharp(\pi_1(M))
\tilde f(\tilde x)=\tilde f(j_1^\sharp(\pi_1(M)) \tilde x)$
holds for any point $\tilde x\in \tilde M$, $\tilde f$ preserves
fibers. Hence from Prop. \ref{prop:LiftOfMap} there exists a
diffeomorphism $f\in\DiffG(M)$ such that the diagram 
(\ref{diagram:theorem2}) commutes. For this map $f_*\Phi_1=\Phi_2$ 
holds. Hence $\Phi_1$ and $\Phi_2$ belong to the same diffeomorphism
class in $\Gamma_\r{LH}(M,G)$.

Thus we obtain the following theorem:
\begin{theorem}\label{theorem:main}
For a locally $G$-homogeneous system on a compact closed 3-manifold
$M$ of type $(\tilde M, \Gmax )$, if the canonical variables
contain the metric, the diffeomorphism-invariant phase space 
is represented as
\Beq
\Gamma_\r{LH}(M,G)/\DiffG(M)=\bigcup_{G\cong \tilde G\in\TSUB(\Gmax )}
\left(\Gamma_\r{H}(\tilde M,\tilde G)\times 
\M(M,\tilde G)\right)/\HPDG(\tilde M,\tilde G).
\label{LHbyH}\Eeq
\end{theorem}

Here note that this formula holds for $M$ of any dimension 
if we replace $\TSUB(\Gmax )$ by a set of 
representative subgroups for  the general conjugate classes of
transitive transformation groups with respect to $\DiffG(\tilde M)$.
However, this generalization will be of little use in higher 
dimension because possible types of universal covering spaces and 
the structure of the conjugate classes of their transformation groups 
are not known well. On the other hand for two dimension, the theorem
holds in the present form because the uniformization theorem holds.
In this case the maximal geometries are classified into three types:
$(E^2, \IO(2))$, $(S^2, \r{O}(3))$ and $(H^2,\r{PO}(2,1))$.

\subsubsection{Note on orientation}

In physics we usually consider only orientable spaces because the laws
of physics are not invariant under the reversal of space
orientation. When we restrict spaces to orientable ones, it is natural
to restrict covering maps to orientation preserving ones by fixing
orientations of the base manifold and its universal covering
space. Let $\Gamma^+_\r{LH}(M,G)$ be the corresponding phase space of
locally homogeneous data on $M$. Then by inspecting the proof, it is
easy to see that Theorem \ref{theorem:main} holds if we replace
$\DiffG(M)$, $\DiffG(\tilde M)$ and $\InvG(\Phi)$ by the orientation
preserving ones $\DiffG^+(M)$, $\DiffG^+(\tilde M)$ and
$\InvG^+(\Phi)=
\InvG(\Phi)\cap\DiffG^+(\tilde M)$, $\TSUB(\Gmax )$ by
a representative system of the conjugate classes of transitive
subgroups of $\Gmax $ with respect to $\DiffG^+(\tilde M)$, and
$\Gamma_\r{H}(\tilde M,\tilde G)$ and $\HPDG(\tilde M,\tilde G)$ by
$\Gamma^+_\r{H}(\tilde M,\tilde G)=\SetDef{\tilde \Phi}
{\InvG^+(\tilde \Phi)=\tilde G}$ and $\HPDG^+(\tilde M,\tilde
G)=\HPDG(\tilde M,\tilde G)
\cap \DiffG^+(\tilde M)$:
\Beq
\Gamma^+_\r{LH}(M,G)/\DiffG(M)
=\bigcup_{G\cong \tilde G\in\TSUB^+(\Gmax )}
\left(\Gamma^+_\r{H}(\tilde M,\tilde G)\times 
\M(M,\tilde G)\right)/\HPDG^+(\tilde M,\tilde G).
\label{LHbyH+}\Eeq

From now on, in the arguments which apply to the
orientation-preserving case just by these replacements, we indicate it
by putting the suffix $(+)$ on relevant quantities as
$\Gamma^{(+)}_\r{LH}(M,G)$ and $\InvG^{(+)}(\tilde\Phi)$.

\subsection{The algorithm to determine $\Gamma_\r{LH}(M,G)/\DiffG(M)$}

Theorem \ref{theorem:main} gives us a systematic algorithm for the
classification and the determination of the diffeomorphism-invariant
phase space of locally $G$-homogeneous data on a compact closed
manifold of a given Thurston-type $(\tilde M, \Gmax )$. It is divided
into three parts.

\subsubsection{Determination of $\M(M,\Gmax )$}

Each compact closed 3-manifolds modeled on $(\tilde M, \Gmax )$
is diffeomorphic to a quotient space $\tilde M/K$ where $K$ is 
some discrete subgroup of $\Gmax $ acting freely. Two
quotient manifolds $\tilde M/K$ and $\tilde M/K'$ are diffeomorphic 
if and only if $K$ and $K'$ are conjugate with respect to 
$\DiffG(\tilde M)$. In the three-dimensional case this condition
is equivalent to the condition that $K$ and $K'$ are isomorphic as
abstract groups, as in the two-dimensional case, except for the Thurston-type
$\tilde M=S^3$ for the following reason.

First for the Thurston-types other than $S^3$ and $H^3$, $\pi_1(M)$
contain a discrete subgroup isomorphic to
$\ZR$\cite{Scott.P1983}. This implies that $H_1(M)$ is an infinite
group, and $M$ becomes a Haken manifold\cite{Hempel.J1976B}. Hence
from Waldhausen's theorem\cite{Waldhausen.F1968} the diffeomorphism
class of $M$ is completely determined by the isomorphism class of the
fundamental group $\pi_1(M)$. On the other hand for $H^3$ the same result
holds from Mostow's rigidity theorem\cite{Mostow.G1968B}.

Since all the manifolds modeled on $E^3$, $\Nil$,
$\widetilde{\SL_2\RF}$, $H^2\times E^1$ and $S^2\times E^1$ allow the
structure of Seifert fiber space, their abstract fundamental groups
are determined by the indices of base orbifolds and the Seifert index. 
On the other hand compact closed manifolds of type $\Sol$ have the
structure of torus bundle or Klein-bottle bundle over $S^1$, and their
abstract fundamental groups are determined by the gluing map of the
torus or the Klein bottle. Hence from the above fact the moduli space
$\M(M,\Gmax)$ for these types is easily determined by examining all
the possible embedding of the corresponding abstract fundamental group
into $\Gmax$.

All possible fundamental groups of manifolds of type $S^3$ and their
embedding into $O(4)$ are also determined with the help of the Seifert
bundle structure of them\cite{Koike.T&Tanimoto&Hosoya1994}. The
classification of the spaces modeled on $H^3$ is not completed yet.

\subsubsection{Classification of $G$-homogeneous data on $\tilde M$}

This part is divided into the following three steps:
\begin{itemize}
\item[i)] Classification of the conjugate classes of transitive
subgroups of $\Gmax\; \Leftarrow \;\TSUB^{(+)}(\Gmax )$.
\item[ii)] Determination of HPDs for each $\tilde G\in 
\TSUB^{(+)}(\Gmax )$ \Then $\HPDG^{(+)}(\tilde M,\tilde G)$.
\item[iii)] Determination of the structure of $\tilde G$-invariant
$\tilde \Phi$ \Then $\Gamma^{(+)}_\r{H}(\tilde M,\tilde G)$.
\end{itemize}

Some comments are in order. Firstly, in connection with step i), it
should be noted that two subgroups of $\Gmax $ may not be conjugate
with respect to $\Gmax $ even if they are with respect to
$\DiffG^{(+)}(\tilde M)$. However, the following useful proposition
holds.

\begin{proposition}\label{prop:conjugacy}
Let $G_\r{st}$ be a simply transitive subgroup of $\Gmax $.  Then for
$G_\r{st}\subseteq G_1,G_2\subseteq \Gmax $, if $G_1$ and $G_2$ are
conjugate with respect to $\DiffG^{(+)}(\tilde M)$, they are with
respect to $\HPDG^{(+)}(\tilde M,\Gmax)\times
\HPDG^{(+)}(\tilde M,G_\r{st} )$.
\end{proposition}

Let $f\in\DiffG^{(+)}(\tilde M)$ be a transformation such that $G_2=f
G_1 f^{-1}$. Then $f G_\r{st} f^{-1} \subset \Gmax $.  Since explicit
analyses show that all the simply-transitive subgroups of $\Gmax $ are
mutually conjugate with respect to $\HPDG^{(+)}(\tilde M,\Gmax )$(see
\S3, \S4 and \S5 for $\Isom(E^3)$, $\Isom(\Nil)$ and $\Isom(\Sol)$), this
implies that there exists $f'\in \HPDG^{(+)}(\tilde M,\Gmax)$ such
that $f'fG_\r{st}f^{-1}f'{}^{-1} =G_\r{st}$. From this it follows that
$f''=f'f\in\HPDG^{(+)}(\tilde M,G_\r{st})$. Hence $f=f'{}^{-1}f''
\in \HPDG^{(+)}(\tilde M,\Gmax)\times
\HPDG^{(+)}(\tilde M,G_\r{st})$.

Secondly, in step ii), since $f\in \HPDG^{(+)}(\tilde M,\tilde G)$
implies $f_*\in \EAut(\tilde G,\tilde M)\subseteq \Aut(\tilde G)$
where $\EAut(\tilde G,\tilde M)$ is the subset of $\Aut(\tilde G)$
whose elements are induced from transformations of $\tilde M$,
$\HPDG^{(+)}(\tilde M,\tilde G)$ is easily determined from the
automorphism group of the Lie algebra, $\Aut(\LieA(\tilde G))$, as
follows. Let $\phi$ be an element of $\Aut(\LieA(\tilde G))$ which is
expressed in terms of a basis $\xi_I$ of $\LieA(\tilde G)$ as
$\phi(\xi_I)=\xi_J A^J{}_I$.  Then if $\phi$ is induced from
$f\in\DiffG^{(+)}(\tilde M)$, $f$ should satisfy the equation
\Beq
\phi(\xi_I)=f_*\xi_I=\xi_J A^J{}_I .
\Eeq
In a local coordinate system $(x^\mu)$ this gives a set of
differential equations
\Beq
\xi_I^\nu(x)\partial_\nu f^\mu(x)=\xi^\mu_J(f(x))A^J{}_I .
\label{EqForHPD}\Eeq
$\Aut(\LieA(\tilde G))$ is easily determined by simple algebraic 
calculations.

Thirdly, in step iii), even if $\tilde \Phi$ is invariant 
with respect to $\tilde G$ and 
$\InvG^{(+)}(\tilde \Phi)\cap \Gmax =\tilde G$, $\tilde \Phi$
may not belong to $\Gamma_\r{H}(\tilde M,\tilde G)$ because it can be
invariant under some transformation $\tilde f\not\in
\Gmax $. However, in the case $\tilde G$ contains a simply
transitive subgroup $G_\r{st}$, we can check rather easily whether
$\tilde \Phi$ belongs to $\Gamma_\r{H}(\tilde M,\tilde G)$ with help
of the following proposition.

\begin{proposition}\label{prop:InvG}
If $\InvG^{(+)}(\tilde \Phi)$ contains a simply transitive subgroup
$G_\r{st}$, there exists $f\in \HPDG^{(+)}(\tilde M, G_\r{st})$ such
that $\InvG^{(+)}(f_*\tilde \Phi)\subseteq \Gmax $. In particular if
$\InvG^{(+)}(f_*\tilde \Phi)\cap \Gmax =\tilde G$ for any $f\in
\HPDG^{(+)}(\tilde M, G_\r{st})$, then $\InvG^{(+)}(\tilde
\Phi)=\tilde G$.
\end{proposition}

The proof is almost the same as that of Prop. \ref{prop:conjugacy}. 
First from Thurston's theorem there exists
$f'\in \DiffG^{(+)}(\tilde M)$ such that
$f'\InvG^{(+)}(\tilde\Phi)f'{}^{-1}\subseteq \Gmax $. 
From the assumption this implies 
$f' G_\r{st} f'{}^{-1}\subset \Gmax $.
Then, since any subgroup of $\Gmax $ is conjugate to 
$G_\r{st}$ with respect to $\HPDG ^{(+)}(\tilde M, \Gmax )$ 
if it is with respect to $\DiffG^{(+)}(\tilde M)$, 
there exists $f''\in\HPDG^{(+)}(\tilde M, \Gmax )$ 
such that $f''f'G_\r{st}f'{}^{-1}f''{}^{-1}=G_\r{st}$. 
This implies $f=f''f'\in\HPDG^{(+)}(\tilde M, G_\r{st})$, 
and $\InvG^{(+)}(f_*\tilde\Phi) = 
f\InvG^{(+)}(f_*\tilde\Phi)f^{-1}\subseteq f''\Gmax 
f''{}^{-1}=\Gmax $. 

From this proposition it follows that even if $\InvG^{(+)}(\tilde \Phi)\cap
\Gmax =\tilde G$, $\tilde\Phi$ should not be included in
$\Gamma^{(+)}_\r{H}(\tilde M,\tilde G)$  if there exists $f\in \HPDG^{(+)}(\tilde M,
G_\r{st})$ such that $\InvG^{(+)}(f_*\tilde \Phi)\supset \tilde G$ and 
$\not=\tilde G$. For example, as we will show in \S3, for the locally 
homogeneous system $\Phi=(q,p)$ of the type $E^3$, the smallest group
$\tilde G$ for which $\tilde G\supseteq\RF^3$ and
$\Gamma^+_\r{H}(\tilde M,\tilde G)\not=\emptyset$ is given not by
$\RF^3$ but by $\RF^3\sdp D_2$ where $D_2$ is the dihedral group 
of order 4.

In practice we can show by direct analyses that all the transitive
connected subgroups in $G^{(+)}_\r{max}$ are conjugate 
with respect to $\HPDG^{(+)}(\tilde M, G^{(+)}_\r{max})$ 
if they are in $\DiffG^{(+)}(\tilde M)$. Hence the proposition 
still holds even if $G_\r{st}$ is replaced by the connected 
component of $\tilde G$ containing identity.

Finally we comment on the diffeomorphism constraint. When 
$\Gamma^{(+)}_\r{LH}(M,G)/\DiffG^{(+)}(M)$ is regarded as a 
subspace of the diffeomorphism-invariant phase space 
$\Gamma^{(+)}(M)/\DiffG^{(+)}(M)$ of generic data on $M$, 
the canonical structure of the latter induces that of the former.
Let $\Gamma^{(+)}_\r{D}(M)$ be the set of data that satisfy 
the diffeomorphism constraint. Then the canonical structure of 
the generic diffeomorphism-invariant phase space becomes 
non-degenerate only in the subspace 
$\Gamma^{(+)}_\r{D}(M)/\DiffG^{(+)}(M)$. Hence in the study of
the canonical structure of the phase space of locally homogeneous
data we must also restrict consideration to the subspace statisfying the 
diffeormorphism constraint. 

Let us denote this subspace by
\Beq
\Gamma^{(+)}_\r{LH,inv}(M,G):=\left(\Gamma^{(+)}_\r{LH}\cap
\Gamma^{(+)}_\r{D}\right)/\DiffG^{(+)}(M).
\Eeq
Then, since the diffeomorphism constraint on 
$\Phi\in \Gamma^{(+)}_\r{LH}(M,G)$ is represented as that
on the covering data $\tilde\Phi\in \Gamma^{(+)}_\r{H}(\tilde M,\tilde
G)$ and does not depend on the moduli freedom, \Eq{LHbyH} and \Eq{LHbyH+} give
\Beq
\Gamma^{(+)}_\r{LH,inv}(M,G)
=\bigcup_{G\cong \tilde G\in\TSUB^{(+)}(\Gmax )}
\left(\Gamma^{(+)}_\r{H,D}(\tilde M,\tilde G)
\times 
\M(M,\tilde G)\right)/\HPDG^{(+)}(\tilde M,\tilde G),
\label{LHbyHD+}
\Eeq
where 
\Beq
\Gamma^{(+)}_\r{H,D}(\tilde M,\tilde G):=
\Gamma^{(+)}_\r{H}(\tilde M,\tilde G)\cap 
\Gamma^{(+)}_D(\tilde M).
\Eeq
This restriction to the subspace generally selects data with
higher symmetry.  This point should be taken care of in the
determination of $\Gamma^{(+)}_\r{LH,inv}(M,G)$.

\subsubsection{Parametrization of $\Gamma^{(+)}_\r{LH,inv}(M,G)$}

In order to express the canonical structure and the hamiltonian
constraint explicitly in terms of independent canonical variables,
we must parameterize the phase space $\Gamma^{(+)}_\r{LH,inv}(M,G)$
by picking out representative points of the $\HPDG$-orbits
in $\Gamma^{(+)}_\r{H,D}(\tilde M,\tilde G)\times \M(M,\tilde G)$.
In this procedure the following proposition plays a crucial
role.

\begin{proposition}\label{prop:DynamicsOfModuli}
Let $\tilde H(K)$ be the isotropy group at $K$ of the 
action of $\HPDG(\tilde M,\tilde G)$ on $\M(M,\tilde G)$.
If the diffeomorphism constraint  becomes trivial on some
$\tilde G$-homogeneous system on $\tilde M$, i.e., 
$\Gamma^{(+)}_\r{H}(\tilde M,\tilde G)
=\Gamma^{(+)}_\r{H,D}(\tilde M,\tilde G)$,
the maximal connected subgroup of $\tilde H(K)$
is contained in $\tilde G$.
\end{proposition}

To prove this proposition, for a given $K\in \M(M,\tilde G)$, 
take a covering map $j$ such that $K=j^\sharp(\pi_1(M))$ and
fix it. Suppose that there exists a diffeomorphism $\tilde f
\in \HPDG^{(+)}(\tilde M,\tilde G)$ such that 
$\tilde fK\tilde f^{-1}=K$. Then $\tilde f$ is a fiber-preserving 
transformation of the covering space $j:\tilde M\maps M$. Hence 
from Prop. \ref{prop:LiftOfMap} it induces a diffeomorphism 
$f\in \DiffG^{(+)}(M)$ such that $j\bop \tilde f=f\bop j$. 
In particular, if $\tilde H(K)_0$(the connected component 
containing the identity) is non-trivial, there is a vector field
$\xi$ on $M$ such that its pullback $\tilde \xi=j^*\xi$ gives
an infinitesimal HPD on $\tilde M$.

Let $\tilde \Phi=(\tilde Q,\tilde P)$ be a $\tilde G$-homogeneous
canonical system on $\tilde M$ satisfying the assumption, and
$D_K$ be a fundamental region of the action of $K$ on $\tilde M$.
Then, since $K\subset \tilde G$, there exist canonical data
$\Phi=(Q,P)$ on $M$ whose pullback by $j$ coincide with $\tilde \Phi$.
Clearly for these fields and the above vector fields, the following
equality holds:
\Beq
\int_{D_K} \Lie_{\tilde \xi}\tilde Q\cdot\tilde P
=\int_M \Lie_\xi Q\cdot P,
\Eeq
where $\cdot$ denotes natural inner products of $Q$ and $P$
and of $\tilde Q$ and $\tilde P$ as tensor fields. However,
from the definition the right-hand side of this equation
is written in terms of the diffeomorphism constraint
$C_j(Q,P)$ as
\Beq
\int_M \Lie_\xi Q\cdot P=\int_M \xi^i C_j(Q,P)
=\int_{D_K}\tilde \xi^i C_j(\tilde Q, \tilde P)
\Eeq
which vanishes identically for $\tilde\Phi\in 
\Gamma^{(+)}_\r{H}(\tilde M, \tilde G)$ from the assumption.
On the other hand, since HPDs preserve $\tilde G$-invariance
of fields, $\Lie_{\tilde\xi}\tilde Q\cdot\tilde P/\sqrt{|\tilde q_0|}$ becomes constant on $\tilde M$ where $\tilde q_0$ is some
reference $\tilde G$-invariant metric on $\tilde M$. Hence 
$\Lie_{\tilde\xi}\tilde Q\cdot\tilde P$ vanishes identically,
which implies 
\Beq
\Lie_{\tilde\xi}\tilde Q=0
\label{LieDQ}\Eeq
because the inner product $\cdot$ is non-degenerate. By the similar
argument and the identity
\Beq
\int_M \Lie_\xi Q\cdot P=-\int_M Q\cdot\Lie_\xi P,
\Eeq
it follows that 
\Beq
\Lie_{\tilde\xi}\tilde P=0.
\label{LieDP}\Eeq
However, since the isotropy group at $\tilde\Phi$ of the action
$\HPDG^{(+)}(\tilde M,\tilde G)$ on 
$\Gamma^{(+)}_\r{H}(\tilde M, \tilde G)$ coincides with
$\tilde G$, \Eqs{LieDQ}{LieDP} implies that $\tilde \xi$
belongs to the Lie algebra of $\tilde G$. This proves the
proposition.

As will be shown in the following sections, the assumption of the 
proposition is satisfied for the locally homogeneous  pure gravity
systems of types $E^3$, $\Nil$ and $\Sol$. Hence one natural method
of parametrizing $\Gamma^{(+)}_\r{LH,inv}(M,G)$ is obtained by
picking up a subset $\M_0(M,\tilde G)$ of $\M(M,\tilde G)$ 
which is transversal to the HPD-orbits in $\M(M,\tilde G)$. 
If the assumption of the proposition is satisfied, this
leads to 
\Beq
\Gamma^{(+)}_\r{LH,inv}(M,G)\cong \bigcup_{\tilde G}
\left(\Gamma^{(+)}_\r{H}(\tilde M,\tilde G)\times
\M_0(M,\tilde G)\right)/(\r{discrete\ HPDs}).
\label{InvPS:gaugefixed}\Eeq
Hence natural parameterizations of $\tilde G$-homogeneous
data and the reduced moduli space $\M_0$ give canonical
coordinates of the phase space. In the case in which the assumption
is not satisfied, $\Gamma^{(+)}_\r{H,D}(\tilde M,\tilde G)$ 
should be further reduced to a subspace to fix
residual HPD freedoms. As will be illustrated in the following 
sections, this parameterization is quite useful in the study
of the canonical structure.

Of course other methods of parameterization can be adopted depending
on the purpose. For example, if we first pick up a representative
point in each HPD-orbit in $\Gamma^{(+)}_\r{H,D}(\tilde M,\tilde G)$ 
such that the metric data acquire the highest possible symmetry, 
the gauge fixing of the residual HPD freedom reduces 
$\M$ to a larger subspace, which corresponds to the standard
moduli space of locally homogeneous compact Riemann manifolds
\cite{Koike.T&Tanimoto&Hosoya1994}.

\subsection{Canonical Structure and Dynamics}

With the helps of the representation theorem proved in the
previous subsection, we can determine the canonical structure
of locally homogeneous systems rather easily. We next describe 
its procedure. 

\subsubsection{General system}

First we briefly summarize basic features of the canonical 
structure and dynamics of general Einstein-gravity systems.
Let $\Gamma(M)$ be the phase space of general canonical data 
$\Phi=(Q,P)$ on a 3-manifold $M$. Then the canonical 1-form
of $\Gamma(M)$ is given by
\Beq
\Theta=\int_M d^3x \delta Q\cdot P,
\Eeq
where $\delta $ denotes the exterior derivative  in
the phase space as a functional space. Then its exterior
derivative $\omega=\delta \Theta$ gives the symplectic form,
which defines the canonical structure of $\Gamma(M)$ by
the following procedure. First for a functional $F$ on $\Gamma(M)$,
a vector field $X_F$ corresponding to the infinitesimal canonical
transformation is defined by
\Beq
\delta F=-I_{X_F}\omega,
\Eeq
where $I_X$ is the inner product operator on differential forms.
Then the Poisson bracket of two functionals $F$ and $G$ on $\Gamma(M)$
is defined by
\Beq
\{F,G\}=-X_F G.
\Eeq

Since infinitesimal diffeomorphisms are expressed as infinitesimal
canonical transformations generated by the diffeomorphism constraint
functionals $C_\r{D}$, diffeomorphism-invariant functionals are
characterized as those which have vanishing Poisson brackets with
$C_\r{D}$. From this and the above definition, it follows that
diffeomorphism-invariant functionals makes a closed Poisson
algebra. Though this algebra is in general degenerate, it becomes
non-degenerate when restricted to diffeomorphism-invariant functionals
on the subset $\Gamma_\r{D}$ of $\Gamma$ satisfying the diffeomorphism
constraints\cite{Kodama.H1995}. Hence a well-defined canonical
structure is defined on the diffeomorphism-invariant phase space
$\Gamma_\r{inv}(M):=
\Gamma_\r{D}(M)/\DiffG(M)$.

Though the hamiltonian constraint functional
\Beq
C_\r{H}(\Phi;N):=\int_M d^3x N(x)\H_\orth(x;\Phi)
\Eeq
is not diffeomorphism-invariant if $N(x)$ is given as an explicit
function on $M$. However, it becomes diffeomorphism-invariant if we
take as $N(x)$ a functional $\hat N(x;\Phi)$ on the phase space which
takes value in the space of functions on $M$ and transforms
covariantly under diffeomorphism:
\Beq
f_* \hat N(*,\Phi)=\hat N(*,f_*\Phi) \quad \forall f\in \DiffG(M).
\Eeq
Hence $C_\r{H}(\Phi;\hat N)$ gives a hamiltonian constraint 
functional on $\Gamma_\r{inv}(M)$.

Let $\Gamma_\r{h}$ be the subspace defined by
\Beq
\Gamma_\r{h}:=\SetDef{\Phi\in \Gamma_\r{inv}}{C_\r{H}(\Phi;\hat N)=0
\quad \forall \hat N},
\Eeq
where we have denoted a diffeomorphism-invariant class and its
representative data by the same symbol. Let $Y_{\hat N}$ be the
infinitesimal canonical transformation defined by
\Beq
Y_{\hat N}:=X_{C_\r{H}(\hat N)}.
\Eeq
Then, since these vector fields commute on $\Gamma_\r{h}$, it defines
an involutive system there and foliates $\Gamma_\r{h}$ into leaves of
integration subspaces. It is shown that each leaf is in one-to-one
correspondence with a spacetime-diffeomorphism class of the solutions
to the Einstein equations in the sense that two solutions connected by
a spacetime diffeomorphism are mapped into curves in the same leaf,
and any non-degenerate curve in a given leaf corresponds to a solution
in the same spacetime-diffeomorphism class\cite{Kodama.H1995}.

Let $\gamma(t)$ be a non-degenerate curve contained in a leaf.  Then,
since its tangent vector is linear combination of $Y$-vectors, there
exists a one-parameter family of $\hat N$-type functionals, $\hat N_t$
such that $\gamma_*\partial_t=Y_{\hat N_t}$. By operating it on a
functional $F$ on $\Gamma_\r{h}$, we obtain the canonical equation of
motion,
\Beq
\dot F = \{ F, C_\r{H}(\hat N_t)\},
\Eeq
which is equivalent to the Einstein equations. This canonical equation
coincides with the Euler equation for the Lagrangian
\Beq
L=\int_M d^3x \dot Q\cdot P - C_H(Q,P;\hat N_t),
\label{Lagrangian:general}\Eeq
which is obtained from the Einstein-Hilbert action by some
gauge-fixing.

\subsubsection{Reduction to the locally homogeneous system}

Since the diffeomorphism-invariant phase space
$\Gamma_\r{LH,inv}(M,G)$ of locally homogeneous data on $M$ is a
finite-dimensional subspace of $\Gamma_\r{inv}(M)$, $\Theta$ on the
latter uniquely defines a one form on the former. Let us denote it by
the same symbol. Then its exterior derivative $\omega=d\Theta$ defines
a canonical structure of $\Gamma_\r{LH,inv}$. One potential subtle
problem in this procedure is the possibility that the reduced $\omega$
becomes degenerate. As we will show explicitly in the following
sections, this pathology occurs frequently. In such cases, the
canonical structure of $\Gamma_\r{LH,inv}$ is ill-defined, which
implies that the system is not canonically closed in the general phase
space $\Gamma_\r{inv}$. Note that even in such cases the restriction
of $L$ defined by \Eq{Lagrangian:general} to $\Gamma_\r{LH,inv}$ gives
a well-defined canonical Lagrangian for the locally homogeneous system
because $\int_M d^3x \dot Q\cdot P$ is nothing but the value of
$\Theta$ on the tangent vector of the curve, $\gamma_*\partial_t$,
which is well-defined even if the canonical structure is
degenerate. From now on we use this kinetic term to express the
canonical 1-form and denote it by the same symbol $\Theta$:
\Beq
\Theta=\int_M d^3x \dot Q\cdot P.
\Eeq

Now we describe how to obtain explicit expressions for $\Theta$
and the hamiltonian $H$ when a parameterization of 
$\Gamma_\r{LH,inv}(M,G)$ is given. In this subsection 
we omit the symbol $(+)$ on the orientation for simplicity.

Let us denote a representative point of each HPD-orbit in
$\Gamma_\r{D}(M,G)\times \M(M,\tilde G)$ by $(\tilde\Phi, K(\lambda))$
where $\lambda=(\lambda_j)$ is a set of variables parametrizing the
reduced moduli space $\M_0(M,\tilde G)$. We do not introduce an
explicit parameterization for $\tilde \Phi=(\tilde Q,\tilde P)$ here.

First we pick up a point $K_0$ as the base point in $\M(M,\tilde G)$
and identify $M$ with $\tilde M/K_0$. Let us denote the corresponding
natural covering map by $j_0$:
\Beq
j_0: \tilde M \maps M=\tilde M/K_0.
\Eeq
Then it defines an embedding of the phase space $\Gamma_\r{inv}(M)$
into $\Gamma_\r{inv}(D_0)$ through $\tilde\Phi=j_0^*\Phi$ where
$D_0$ is a fundamental region of the action of $K_0$ on $\tilde M$. 
This embedding is isomorphic in the sense that it preserves the
canonical structure:
\Beq
\Theta=\int_M d^3x \dot Q\cdot P 
=\int_{D_0}d^3x \dot{\tilde Q}\cdot\tilde P.
\label{CFonFD}\Eeq

Further, for each $K(\lambda)$ there exists a covering map $j_\lambda$
and a transformation $f_\lambda$ of $\tilde M$ such that 
\Beqr
&& j_\lambda=j_0\bop f_\lambda^{-1},\\
&& j_\lambda^\sharp(\pi_1(M))=K(\lambda),\\
&& K(\lambda)=f_\lambda K_0 f_\lambda^{-1}\equivalent
K(\lambda)=(f_\lambda)_* K_0.
\Eeqr
From the second condition $\tilde \Phi$ defines canonical data
$\Phi$ on $M$ such that $\tilde\Phi=j_\lambda^*\Phi$. 
Its embedding by $j_0$ into $\Gamma_\r{inv}(D_0)$ is 
represented by the data $j_0^*\Phi=f_\lambda^*\tilde\Phi$ from
the first condition. Hence from \Eq{CFonFD} the canonical
1-form of $\Gamma_\r{LH,inv}(M,G)$ is expressed in terms
of the coordinates $(\tilde Q,\tilde P, \lambda)$ as
\Beq
\Theta=\int_{D_0}d^3x f_\lambda^*(\dot{\tilde Q}\cdot\tilde P
+ \dot\lambda^j\Lie_{\zeta_j(\lambda)}\tilde Q\cdot\tilde P),
\label{CFonFD1}\Eeq
where $\zeta_j(\lambda)$ is the vector field defined by
\Beq
\left(\zeta_j(\lambda)\right)_x:=(\partial_{\lambda_j}f_\lambda)
(f_\lambda^{-1}(x)).
\Eeq
From now on we call the transformation $f_\lambda$ {\it a deformation
map} from the base point $K_0$ to $K(\lambda)$. 

In terms of the components with respect to an invariant basis
this expression can be rewritten in a more useful form.
Let $G_\r{st}$ be a simply transitive subgroup of $\tilde G$,
and $\tau^a$ and $\tau_b$ be bases of $G_\r{st}$-invariant
tensors of types $\tilde Q$ and $\tilde P$ which are normalized 
as
\Beq
\tau^a\cdot\tau_b=\delta^a_b.
\label{Normalization:InvTensor}\Eeq
Then, since $\tilde P$ is a tensor density, $\tilde Q$ and $\tilde P$
are expressed in terms of them as
\Beq
\tilde Q=Q_a\tau^a,\quad \tilde P=\sqrt{|\tilde q|}P^b\tau_b,
\Eeq
where $|\tilde q|$ is the determinant of the metric data 
in $\tilde Q$, and $Q_a$ and $P^b$ are spatially constant 
components.

Let us write the pullback of $\tau^a$ by $f_\lambda$ as
\Beq
f_\lambda^* \tau^a=F^a{}_b \tau^b.
\Eeq
Then from \Eq{Normalization:InvTensor} the pullback of $\tau_a$ 
is written as
\Beq
f_\lambda^*\tau_a=\tau_b(F^{-1})^b{}_a.
\Eeq
Hence the components $Q_a$ and $P^b$ transform as
\Beqr
&& (f_\lambda^* \tilde Q)_a=Q_b F^b{}_a,\\
&& (f_\lambda^* \tilde P)^a= (F^{-1})^a{}_b P^b.
\Eeqr

Inserting these expressions into \Eq{CFonFD}, we obtain
\Beqr
&\Theta &= \int_{D_0}d^3x (f_\lambda^*\sqrt{|\tilde q|})
(\dot Q_aP^a + (\dot F F^{-1})^a{}_b Q_a P^b) \nonumber\\
&&= V(Q,\lambda)\dot Q_a P^a + \dot\lambda^j C^a_{jb}Q_aP^b,
\Eeqr
where
\Beqr
&& V(Q,\lambda):=\int _{D_0}d^3x f_\lambda^*\sqrt{|\tilde q|},\\
&& C^a_{jb}=\int_{D_0}d^3x (f_\lambda^*\sqrt{|\tilde q|})
(\partial_{\lambda_j}F F^{-1})^a{}_b.
\Eeqr
Here note that the matrix $F$ depends on the position as well as
on $\lambda$ in general because $f_\lambda$ maps $\tau_a$ 
to invariant tensors if and only if $f_\lambda$ is in 
$\HPDG(\tilde M,G_\r{st})$.

Similarly the hamiltonian gets a simple expression in terms of 
the components $Q_a$ and $P^b$. Since $\H_\orth(Q,P)$ behaves as
a scalar density, its pullback to $\tilde M$ by $j_\lambda$ 
is written in terms of the metric and a function $H_\orth(Q_a,P^b)$
as
\Beq
j_\lambda^*\H_\orth(Q,P)=\sqrt{|\tilde q|}H_\orth(Q_a,P^b).
\Eeq
Further the pullback of the lapse function $\hat N$ should be
represented by a spatially constant function $N$ on $\tilde M$
in the locally homogeneous sector. Hence the Hamiltonian
is written as
\Beqr
&H &= \int_M d^3x \hat N \H_\orth(Q,P)\nonumber\\
&& = \int_{D_0} d^3x N (f_\lambda^*\sqrt{|\tilde q|})
H_\orth(Q_a,P^b)\nonumber\\
&&= N V(Q,\lambda) H_\orth(Q_a,P^b).
\Eeqr

In particular for the pure gravity system in which $\Phi=(q,p)$,
$\tilde q$ and $\tilde p$ are expressed in terms of the basis of
invariant vectors, $X_I$, and its dual basis $\chi^I$ as
\Beq
\tilde q=Q_{IJ}\chi^I\otimes\chi^J,\quad
\tilde p=\sqrt{|\tilde q|}P^{IJ}X_I\otimes X_J,
\Eeq
where $|\tilde q|$ is now written as 
\Beq
|\tilde q|=|Q||\chi|^2.
\Eeq
For the pullback of the basis 
\Beq
f_\lambda^*\chi^I=F^I{}_J \chi^J,
\label{def:F-matrix}\Eeq
the transformation of the components is written  in the matrix form as
\Beqr
&&f_\lambda^*Q = \Tp F Q F,\\
&&f_\lambda^*P = F^{-1} Q \Tp F^{-1},\\
&& f_\lambda^*\sqrt{|\tilde q|}=|F||\chi|\sqrt{|Q|}.
\Eeqr
Hence the canonical 1-form and the Hamiltonian are expressed as
\Beqr
&& \Theta=\Omega(\lambda)\sqrt{Q}\left(\Tr(\dot Q P)
+ 2\dot\lambda^j \Tr(C_j(\lambda) QP)\right),
\label{Theta:LH:general}\\
&& H=2\kappa^2 N\Omega(\lambda)\sqrt{|Q|}
\left[\Tr(QPQP)-{1\over2}\left(\Tr(QP)\right)^2
-{1\over 4\kappa^2}R(Q)\right],
\label{H:LH:general}
\Eeqr
where 
\Beqr
&& \Omega(\lambda):=\int_{D_0}d^3x |\chi||F|,\\
&& C_j(\lambda):={1\over\Omega(\lambda)}
\int_{D_0}d^3x |\chi||F|\partial_{\lambda_j}F F^{-1},
\Eeqr
and $R(Q)$ is the Ricci scalar curvature. 

Finally note that $(Q,P)$ must satisfy the diffeomorphism 
constraint. This constraint is expressed in terms of these
components with respect to the invariant basis as
\Beq
H_I:=2(c_K Q_{IJ}P^{KJ} + c^K_{JI}Q_{KL}P^{LJ})=0,
\label{DC:H}\Eeq
where $c^I_{JK}$ is the structure constant of $G_\r{st}$,
\Beq
d\chi^I=-{1\over2}C^I_{JK}\chi^J\wedge \chi^K,
\Eeq
and $c_I:=c^J_{IJ}$.

\subsubsection{Dynamics of the moduli parameters}

As touched upon in the previous subsection, the canonical 
structure of $\Gamma_\r{LH,inv}(M,G)$ often gets degenerate.
In such cases the time evolution of moduli parameters is not 
fully determined by the canonical equation of motion. However,
it does not mean that their time evolution is really uncertain, but
it just means that the locally homogeneous system is not canonically
closed in the generic phase space as the following theorem shows.

\begin{theorem}\label{theorem:ModuliDynamics}
For a locally $G$-homogeneous system $(\Phi,M)$ of type 
$(\tilde M,\Gmax )$, let $(\tilde\Phi(t),K(t))$ be 
a representative trajectory in $\Gamma^{(+)}_\r{H}(\tilde M,\tilde G)
\times \M(M,\tilde G)$ corresponding to a locally 
$G$-homogeneous solution $\Phi(t)$ on $M$ to
the Einstein equations. Then $K(t)$ is expressed in terms of
a family of transformations $f_t\in \HPDG^{(+)}(\tilde M,\tilde G)$
as
\Beq
K(t)=f_t K(t_0) f_t^{-1}.
\Eeq
\end{theorem}

As in the previous subsection, let us identify $M$ with
$\tilde M/K(t_0)$ and let the corresponding covering map be
$j_0:\tilde M\maps M$. Then there exists a family of transformations
$f_t\in\DiffG^{(+)}(\tilde M)$ such that 
\Beqr
&& K(t)=f_t K(t_0) f_t^{-1} (\subset \tilde G),\\
&& \tilde \Phi(t)=(f_t)_*j_0^*\Phi(t).
\Eeqr
Hence the Einstein equation
\Beq
\dot \Phi=\{\Phi,C_H(\Phi;N)\}
\Eeq
is written in terms of $\tilde\Phi$ and $f_t$ as
\Beq
\partial _t(f_t^*\tilde \Phi)
=\{f_t^*\tilde\Phi,C_H(f_t^*\tilde \Phi;N)\}.
\Eeq
Since the Poisson bracket operation commutes with diffeomorphism
and $N$ is spatially constant, this equation is rewritten as
\Beq
\partial_t\tilde\Phi+\Lie_{\eta(t)}\tilde\Phi
=\{\tilde\Phi,C_H(\tilde\Phi,N)\},
\Eeq
where $\eta(t)$ is the vector field defined by
\Beq
\eta(t)_x =(\partial_t f_t)(f_t^{-1}(x)).
\Eeq

Since $\partial_t\tilde \Phi$ and $\{\tilde\Phi,C_H(\tilde\Phi,N)\}$
are $\tilde G$-homogeneous, it follows from this equation that
$\Lie_\eta\tilde\Phi$ is also $\tilde G$-homogeneous. Hence 
for any $g\in \tilde G$, $\Lie_\eta\tilde\Phi=g_*\Lie_\eta\tilde\Phi
=\Lie_{g_*\eta}\tilde \Phi$. Since $\InvG^{(+)}(\tilde \Phi)=\tilde G$, this is equivalent to the condition
\Beq
g_*\eta(t)- \eta(t)\in \LieA(\tilde G).
\label{Dcondforf}\Eeq
From this it immediately follows that
\Beq
[\xi,\eta(t)]\in \LieA(\tilde G)\quad \forall \xi\in\LieA(\tilde G).
\label{DDcondforf}\Eeq
This equation implies that $\eta(t)$ induces an automorphism
of $\LieA(\tilde G)=\LieA(\tilde G_0)$ where $\tilde G_0$ is the connected 
component of $\tilde G$ containing the identity transformation.

We first show that $f_t$ belongs to $\HPDG^{(+)}(\tilde M,
\tilde G_0)$.  For $\xi\in\LieA(\tilde G)$, let $\xi(t)$ be a vector field
defined by
\Beq
\xi(t):=\Ad(f_t)\xi=(f_t)_*\xi.
\Eeq
By differentiating this equation with respect to $t$, we obtain
\Beq
{d\over dt}\xi(t)=-[\eta(t),\xi(t)].
\Eeq
Let us introduce a basis $\xi_I,\xi'_\alpha$ of $\LieA(\DiffG(\tilde
M))$ such that $\xi_I$ gives a basis of $\LieA(\tilde G_0)$, and
expand $\xi(t)$ in terms of them as
\Beq
\xi(t)=c^I(t)\xi_I + d^\alpha(t)\xi'_\alpha.
\Eeq
Since $\xi(t)$ belongs to a finite dimensional subspace of
$\LieA(\DiffG(\tilde M))$ due to \Eq{DDcondforf}, the right-hand side
of this equation has well-defined meaning.  Then since
$[\eta(t),\xi_I]\in \LieA(\tilde G_0)$, $d^\alpha(t)$ satisfies a
closed evolution equation of the form
\Beq
\dot d^\alpha=f^\alpha_\beta d^\beta.
\Eeq
However, since $d^\alpha(0)=0$, we obtain $d^\alpha(t)=0$.  Hence
$\Ad(f_t)\xi\in\LieA(\tilde G_0)$ and $f_t\in \HPDG^+(\tilde M,\tilde
G_0)$.

Next, in order to show that $f_t$ preserves $\tilde G$, let us 
define $\Psi(t)$ by
\Beq
\Psi(t):=(f_t)_* \Psi_0
\Eeq
where $\Psi_0$ is arbitrary $\tilde G$-invariant data. 
Then for any $g\in \tilde G$, we obtain
\Beq
\partial_t(g_*\Psi-\Psi)=\Lie_{\eta}(g_*\Psi-\Psi)
+\Lie_{g_*\eta-\eta}g_*\Psi.
\Eeq
Since $f_t$ maps $\tilde G_0$-invariant data to data with the same
invariance and $\tilde G_0$ is a normal subgroup of $\tilde G$,
$g_*\Psi$ is $\tilde G_0$-invariant. Hence from \Eq{Dcondforf} the
second term on the right-hand side of this equation vanishes, and we
get a closed evolution equation for $g_*\Psi-\Psi$. If we expand this
quantity in terms of a basis $\sigma_a$ of the $\tilde G_0$-invariant
fields as $g_*\Psi-\Psi=\psi^a\sigma_a$, then this equation is written
as
\Beq
\partial_t \psi^a \sigma_a=\psi^a\Lie_\eta \sigma_a.
\Eeq
Since $\Lie_\eta\sigma_a$ is again a $\tilde G_0$-invariant field,
and written as a linear combination of $\sigma_a$, it gives a
closed linear evolution equation for $\psi^a$. Hence
from $g_*\Psi(0)-\Psi(0)=0$, we obtain $g_*\Psi(t)-\Psi(t)=0$,
which implies that $f_t$ maps $\tilde G$-invariant fields
to themselves. Therefore $f_t$ must be HPDs with respect to
$\tilde G$.

This theorem states that the dynamics of the moduli parameters
is essentially frozen as pointed out by Tanimoto, Koike 
and Hosoya\cite{Tanimoto.M&Koike&Hosoya1997}. In particular,
in the parameterization in which  the continuous HPD-freedom with 
respect to $\tilde G$ is completely removed, the moduli
parameters become constants of motion. We will confirm this
in the following sections explicitly.

\section{LHS of type $E^3$}

In this section we completely determine the canonical structure
of locally homogeneous pure gravity systems 
on compact closed orientable manifolds of type $E^3$.

\subsection{Basic properties}

\subsubsection{G$^{(+)}_\r{max}$ and $\HPDG^+(G^+_\r{max})$}

The maximal symmetry group of $E^3$ is given by
$\Gmax=\Isom(E^3)=\IO(3)$ and its orientation-preserving 
subgroup by $\Gmaxo=\Isom^+(E^3)=\ISO(3)$. 
We denote each element of $\ISO(3)$,
\Beq
f(\bm{x})=A\bm{x}+\bm{a};\quad A\in\SO(3),\bm{a}\in\RF^3
\Eeq
by the pair $(\bm{a},A)$. In this notation the product of two
elements are written as
\Beq
(\bm{a},A)(\bm{b},B)=(\bm{a}+A\bm{b},AB).
\Eeq

The structure of the Lie algebra $\LieA(\ISO(3))$ of $\ISO(3)$ is
expressed in terms of the basis
\Beq
T_I=\partial_I,\quad J_I=\epsilon_{IJK}x^J\partial_K
\Eeq
as
\Beq
[T_I,T_J]=0,\quad [J_I,T_J]=-\epsilon_{IJK}T_K,
\quad [J_I,J_J]=-\epsilon_{IJK}J_K.
\Eeq
In terms of this basis a generic element $\phi$ of 
the automorphism group $\Aut(\LieA(\ISO(3)))$ is written 
in the matrix form as
\Beq
\phi\begin{matrix}{cccccc}T_1& T_2 & T_3& J_1 & J_2 & J_3\end{matrix}
=\begin{matrix}{cccccc}T_1& T_2 & T_3 & J_1 & J_2 & J_3\end{matrix}
\begin{matrix}{cc} kR & RV\\ 0 & R\end{matrix},
\Eeq
where $k$ is a non-zero constant, $R\in\SO(3)$, and $V$ is a 
2-dimensional anti-symmetric matrix. From \Eq{EqForHPD}
this automorphism belongs to
$\EAut(\LieA(\ISO(3)))$ if and only if $V$ is written
in terms of a vector $\bm{c}=(c^1,c^2,c^3)$ as
\Beq
V_{IJ}=-\epsilon_{IJK}(R^{-1})^K{}_L c^L,
\Eeq
and if $k>0$, it is induced from $f\in \HPDG^+(E^3,\ISO(3))$
given by
\Beq
f(\bm{x})=kR\bm{x}+ \bm{c}.
\label{HPD:ISO3}\Eeq

\subsubsection{Transitive subgroups of $G^+_\r{max}$}

Conjugate classes of connected transitive subgroups of $G^+_\r{max}$
are determined as follows. For a connected transitive subgroup $\tilde
G$, let $\xi_a=A_a^I T_I + B_a^I J_I$ be a basis of its Lie
algebra. Since $\tilde G$ is transitive, the rank of the matrix
$(A^I_a)$ must be equal to 3. Further from the exact sequence
\Beq
0 \maps \RF^3 \mapsnamed{j} \ISO(3) \mapsnamed{p} \SO(3)
\maps 1,
\label{ExactSequence:E3}\Eeq
$p(\tilde G)$, the isotropy group at $\bm{x}=0$,
must be a connected Lie subgroup of $\SO(3)$, 
which is either 1 or $\SO(2)$ or $\SO(3)$. 

First, in the case $p(\tilde G)=1$, $\tilde G$ coincides with the
normal subgroup $\RF^3$, which is simply transitive and
corresponds to the Bianchi-type I group. Its invariant basis
is given by
\Beq
\tilde G=\RF^3: \quad 
\chi^1=dx^1, \quad \chi^2=dx^2,\quad \chi^3=dx^3.
\label{InvariantBasis:I}\Eeq

Second, in the case $p(\tilde G)=\SO(2)$, its generator
can be put to $J_3$ by conjugate transformation in
$\ISO(3)$. Then $\tilde G\cap \RF^3$ must be an invariant
subspace of $\RF^3$ for rotations around $x^3$-axis, 
and its dimension is larger than one. 

If the dimension 
is two, $\tilde G\cap \RF^3$ should coincides with 
the subgroup generated by $T_1$ and $T_2$. Since
$\tilde G$ becomes a simply transitive group,  by
a linear transformation if necessary, $\xi_I$ can be put
in the form $\xi_1=T_1, \xi_1=T_2,\xi_3=c T_3 -J_3$. Further
the constant $c$ can be transformed to $\pm1$ by  a scaling of 
the $x^3$-coordinate by a positive constant which is 
in $\HPDG^+(E^3,\ISO(3))$. Thus in this case $\tilde G$
is conjugate to a group generated by
\Beq
\xi_1=T_1,\quad \xi_2=T_2,\quad \xi_3=-\epsilon T_3-J_3,
\Eeq
where $\epsilon=\pm1$. Its algebra is given by
\Beq
[\xi_1,\xi_2]=0,\quad [\xi_3,\xi_1]=\xi_2,\quad
[\xi_3,\xi_2]=-\xi_1,
\Eeq
and a generic element of this group is written as 
$(\bm{a},R_3(\epsilon a^3))$. Since this group is isomorphic
to the Bianchi-type VII(0) group, we denote it as VII$^\epsilon$(0).
The corresponding invariant basis is given by
\Beq
G=\r{VII}^\epsilon(0): \quad 
\begin{matrix}{c}\chi^1\\ \chi^2 \end{matrix}
=R(-\epsilon x^3)\begin{matrix}{c}dx^1\\ dx^2 \end{matrix},\quad
\chi^3=dx^3.
\label{InvariantBasis:VII}\Eeq
Note that if we allow orientation-reversing transformations, 
VII$^+(0)$ and VII$^-(0)$ become conjugate with each other.

On the other hand if the dimension of $\tilde G\cap\RF^3$ is three,
we obtain a four-dimensional transitive subgroup isomorphic 
to $\RF^3\sdp \SO(2)$. In terms of the basis 
\Beq
\xi_1=T_1,\quad \xi_2=T_2,\quad \xi_3=T_3,\quad
\xi_4=J_3,
\Eeq
the structure of its Lie algebra is given by
\Beqr
&&[\xi_I,\xi_J]=0\quad(I,J=1,2,3),\nonumber\\
&&[\xi_4,\xi_1]=-\xi_2,\quad [\xi_4,\xi_2]=\xi_1,\quad
[\xi_4,\xi_3]=0.
\Eeqr
Note that this subgroup contains both of the simply transitive
groups $\RF^3$ and VII$^\pm$(0).

Finally in the case $p(\tilde G)=\SO(3)$, $\tilde G$ obviously
coincides with $\ISO(3)$.

Next we determine invariance groups with disconnected components.
The maximal connected subgroup of each transitive invariance 
group  must be isomorphic to one of the 
connected subgroups determined above. 
Hence after an appropriate
conjugate transformation, $\tilde G$ is decomposed as
\Beq
0 \maps \tilde G_0 \mapsnamed{j} \tilde G \mapsnamed{p} H
\maps 1 \quad(\r{exact}),
\Eeq
where $H$ is a discrete subgroup of $\SO(3)$. Each element of $H$
is expressed as a rotation around some vector $\bm{n}$, 
$R_n(\theta)$. It is easily shown that if covering data 
$(\tilde q,\tilde p)$ on $E^3$ is 
invariant by the rotation $\theta\not=\pi(\mod 2\pi)$, 
it is invariant under $R_n(\theta)$ with arbitrary value of
$\theta$. Hence $\theta$ should be equal to $\pi$, which
implies that $H$ is isomorphic to a subgroup of the
dihedral group  $D_2=\{1,R_1(\pi),R_2(\pi),R_3(\pi)\}$, where
$R_j(\pi)$($j=1,2,3$) is the rotation matrix of angle $\pi$ around
the $x^j$-axis.

In the case $\tilde G_0=\RF^3$, it follows from this that
$H$ is transformed to either  $1$, $\{1,R_3(\pi)\}$
or $D_2$ by a $\SO(3)$-transformation, which leaves $\RF^3$
invariant. On the other hand, in the case 
$\tilde G_0=$VII$^\pm$(0), $R_n(\pi)$ leaves $\tilde G_0$
invariant only when $\bm{n}$ is parallel to or orthogonal to the 
$x^3$-axis. Hence by a rotation around the $x^3$-axis, which 
leaves $\tilde G_0$ invariant, $H$ is transformed to 
either $1$, $\{1,R_1(\pi)\}$, $\{1,R_3(\pi)\}$ or $D_2$.
Similarly in the case $\tilde G_0=\RF^3\sdp\SO(2)$, $\bm{n}$
must be parallel to the $x^3$-axis, and $H$
is transformed to either $1$ or $\{1,R_1(\pi)\}$ by
a rotation around the $x^3$-axis. Thus every transitive invariance 
group contained in $\ISO(3)$ is conjugate to one of the
following groups with respect to $\HPDG^+(E^3,\ISO(3))$:
$\RF^3$, $\RF^3\sdp \{1,R_1(\pi)\}$, $\RF^3\sdp D_2$, 
VII$^\epsilon (0)$, VII$^\epsilon (0)\sdp\{1,R_1(\pi)\}$,
VII$^\epsilon (0)\sdp\{1,R_3(\pi)\}$,
VII$^\epsilon (0)\sdp D_2$, $\RF^3\sdp \SO(2)$,
$\RF^3\sdp\SO(2)\times\{1,R_1(\pi)\}$, and $\ISO(3)$.

\begin{table}
\begin{tabular}{ll}
\bf Space & \bf Fundamental group and  representation \\
\\
$T^3$ 
& $<\alpha,\beta,\gamma| [\alpha,\beta]=1,[\beta,\gamma]=1,
[\gamma,\alpha]=1>$ \\
\\
& $\alpha=(\bm{a},1),\beta=(\bm{b},1),\gamma=(\bm{c},1)$;\\
& $(\bm{a},\bm{b},\bm{c})\in\r{GL}(3,\RF).$\\
\\
$T^3/\ZR_2$ 
& $<\alpha,\beta,\gamma| [\alpha,\beta]=1,
\gamma\alpha\gamma^{-1}\alpha=1,\gamma\beta\gamma^{-1}\beta=1>$\\
\\
& $\alpha=(\bm{a},1),\beta=(\bm{b},1),\gamma=(\bm{c},R_\gamma)$;
$R_\gamma=R_{a\times b}(\pi)$,\\
& $(\bm{a},\bm{b},\bm{c})\in\r{GL}(3,\RF).$\\
\\
$T^3/\ZR_2\times\ZR_2$ 
& $<\alpha,\beta,\gamma|\alpha\beta\gamma=1,
\alpha\beta^2\alpha^{-1}\beta^2=1,\beta\alpha^2\beta^{-1}\alpha^2=1>$\\
\\
& $\alpha=(\bm{a},R_\alpha),\beta=(\bm{b},R_\beta),
\gamma=(\bm{c},R_\gamma)$;\\
& $(\bm{a}+R_\alpha\bm{a},\bm{b}+R_\beta\bm{b},\bm{c}+R_\gamma\bm{c})
\in\r{GL}(3,\RF)$\\
&$R_\alpha,R_\beta$ and $R_\gamma$ are rotations of the angle $\pi$
around mutually \\
&orthogonal axes. 
$R_\beta\bm{a}+R_\gamma\bm{b}+R_\alpha\bm{c}=0.$\\
\\
$T^3/\ZR_3$ 
& $<\alpha,\beta,\gamma| [\alpha,\beta]=1,
\gamma\alpha\gamma^{-1}=\beta,
\gamma\beta\gamma^{-1}=\alpha^{-1}\beta^{-1}>$\\
\\
& $\alpha=(\bm{a},1),\beta=(\bm{b},1),\gamma=(\bm{c},R_\gamma)$;
$R_\gamma=R_{a\times b}({2\pi\over3})$,\\
& $ \bm{b}=R_\gamma\bm{a}$,
$(\bm{a},\bm{b},\bm{c})\in\r{GL}(3,\RF).$\\
\\
$T^3/\ZR_4$ 
& $<\alpha,\beta,\gamma| [\alpha,\beta]=1,
\gamma\alpha\gamma^{-1}=\beta^{-1},
\gamma\beta\gamma^{-1}=\alpha>$\\
\\
& $\alpha=(\bm{a},1),\beta=(\bm{b},1),\gamma=(\bm{c},R_\gamma)$;
$R_\gamma=R_{a\times b}({\pi\over2})$, \\
& $\bm{b}=R_\gamma\bm{a}$,
$(\bm{a},\bm{b},\bm{c})\in\r{GL}(3,\RF).$\\
\\
$T^3/\ZR_6$ 
& $<\alpha,\beta,\gamma| [\alpha,\beta]=1,
\gamma\alpha\gamma^{-1}=\beta,
\gamma\beta\gamma^{-1}=\alpha^{-1}\beta>$\\
\\
& $\alpha=(\bm{a},1),\beta=(\bm{b},1),\gamma=(\bm{c},R_\gamma)$;
$R_\gamma=R_{a\times b}({\pi\over3})$, \\
& $\bm{b}=R_\gamma\bm{a}$,
$(\bm{a},\bm{b},\bm{c})\in\r{GL}(3,\RF).$\\
\\
\end{tabular}
\caption{\label{tbl:E3:Pi1}Fundamental groups and their 
representation in $\Isom(E^3)$ of compact closed orientable 
3-manifolds of type $E^3$}
\end{table}

\subsubsection{Topology of orientable compact quotients}

All the compact closed manifolds modeled on $(E^3,IO(3))$ admit
Seifert fibration with $\chi=0$ and $e=0$, where $\chi$ is the Euler
number of the base orbifold and $e$ is the Euler number of the Seifert
bundle\cite{Scott.P1983}. From this it is shown that they are all
covered by three torus $T^3$.  In particular the orientable ones
are diffeomorphic to either $T^3$, $T^3/\ZR_2$, $T^3/\ZR_2\times
\ZR_2$, $T^3/\ZR_3$, $T^3/\ZR_4$, and $T^3/\ZR_6$, whose fundamental
groups and their representation in $IO(3)$ are listed in
Table\ref{tbl:E3:Pi1}.

We need the volume $\Omega_K$ of the fundamental region 
$D_K$ for the action of each $K=j^\sharp(\pi_1(M))$ in writing
down the expression for the canonical 1-form and the 
Hamiltonian(see \Eqs{Theta:LH:general}{H:LH:general}).
It is determined as follows.

First for $M=T^3$, $D_K$ is given by $\bm{x}=u\bm{a}
+v\bm{b}+w\bm{c}$ with $0\le u,v,w\le 1$. Its volume is given by
$\Omega_K=|(\bm{a}\times\bm{b})\cdot\bm{c}|$. 

Second for $M=T^3/\ZR_k$($k=2,3,4,6$), since $\alpha$, $\beta$
and $\gamma^k$ generate the fundamental group of
the covering $T^3$, $D_K$ is given by 
$\bm{x}=u\bm{a}+v\bm{b}+w\tilde{\bm{c}}$($0\le u,v,w\le 1$)
where $\tilde{\bm{c}}=(1+R+\cdots+R^{k-1})\bm{c}/k$. 
Hence its volume is again given by 
$\Omega_K=|(\bm{a}\times\bm{b})\cdot\bm{c}|$.

Finally for $M=T^3/\ZR_2\times\ZR_2$, $\alpha^2$, $\beta^2$
and $\gamma^2$ generate $\pi_1(T^3)$, and the quotient group
$\pi_1(M)/\pi_1(T^3)$ is isomorphic to the Klein group
$\ZR_2\times\ZR_2$ generated by the cosets $[\alpha]$ and $[\beta]$.
Hence the volume of $D_K$ is given by the fourth of
that of the covering torus. In terms of 
$\tilde{\bm{a}}=(1+R_\alpha)\bm{a}/2$,
$\tilde{\bm{b}}=(1+R_\beta)\bm{b}/2$ and
$\tilde{\bm{c}}=(1+R_\gamma)\bm{c}/2$, it is expressed as
$\Omega_K=2|\tilde{\bm{a}}||\tilde{\bm{b}}||\tilde{\bm{c}}|$.

\subsection{$\tilde G_0=\RF^3$}

The automorphism group of $\RF^3$ coincides with GL$(3,\RF)$, and
its action on the Lie algebra is expressed as
\Beq
\phi(\xi_I)=\xi_J A^J{}_I \quad A\in\r{GL}(3,\RF).
\label{Aut(R3)}\Eeq
For $\det A>0$ it is induced from $f\in \HPDG^+(E^3,\RF^3)$ given by
\Beq 
f(\bm{x})=A\bm{x}+\bm{a}\quad \bm{a}\in\RF^3.
\label{HPDG(R3)}\Eeq
Since the invariant basis (\ref{InvariantBasis:I}) transforms by $f$
as
\Beq
f^* \chi^I=A^I{}_J\chi^J,
\Eeq
the components of $\RF^3$-invariant data $\tilde\Phi=(\tilde q,\tilde p)$ with
respect to the invariant basis, $Q=(Q_{IJ})$ and $P=(P^{IJ})$,
transforms as
\Beq
(f^*Q)=\Tp{A}QA,\quad (f^*P)=A^{-1}P\Tp{A^{-1}}.
\label{Trf:QP:R3}\Eeq

With the helps of these formula we can show that  $\RF^3$-invariant
data always have higher symmetry. First from \Eq{Trf:QP:R3}, $Q$ can 
be always transformed to the unit matrix $I_3$ by a HPD in
$\HPDG^+(E^3,\RF^3)$. This leaves residual HPDs given by 
$A\in \SO(3)$,
which can be used to diagonalize $P$ as $[P^1,P^2,P^3]$. If all of
these eigenvalues of $P$ are distinct with each other, 
$\InvG^+(\tilde\Phi)\cap \Gmax =\RF^3\times D_2$. Since it is clear
that there does not exist $f\in \HPDG^+(E^3,\RF^3)$ such that 
$\InvG^+(f_*\tilde\Phi)\cap \Gmax $ increases, from
Prop.\ref{prop:InvG} it follows that $\InvG^+(\tilde \Phi)=\RF^3\times
D_2$. On the other hand, if two of the eigenvalues coincide, 
$\InvG^+(\tilde\Phi)$ is obviously equal to or larger than  a group
isomorphic to $\RF^3\sdp\SO(2)$. Thus the smallest invariance group
containing $\RF^3$ is conjugate to $\RF^3\times D_2$.

The HPDs for $\RF^3\times D_2$ is easily determined from
$\HPDG^+(E^3,\RF^3)$ by finding $f$ such that $fD_2f^{-1}\subset
\RF^3\times D_2$. The result is 
\Beq
f\in \HPDG^+(E^3,\RF^3\times D_2) 
 \equivalent A=\begin{matrix}{ccc}p&0&0\\0&q&0\\0&0&r\end{matrix}
 B
\label{HPDG:R3xD2}\Eeq
where $p$, $q$, and $r$ are positive constants and $B$ is an element
of the octahedral group. That is, $A$ must be a regular matrix with
positive determinant whose components vanishes except for three
entries. 

If $Q$ and $P$ are invariant under $\RF^3\times D_2$, they must be
given by the diagonal matrices, $Q=[Q_1,Q_2,Q_3]$ and
$P=[P^1,P^2,P^3]$. By a transformation in $\HPDG^+(E^3,\RF^3\times
D_2)$ they can be put in the form $Q=I_3$ and
$P=[Q_1P^1,Q_2P^2,Q_3P^3]$. Hence by the same argument as above, we
obtain
\Beq
\Gamma^+_\r{H}(E^3,\RF^3\times D_2)=
\SetDef{Q=[Q_1,Q_2,Q_3],P=[P^1,P^2,P^3]}{Q_IP^I\not=Q_JP^J(I\not=J)}.
\label{HPS:R3}\Eeq

From the general formula (\ref{H:LH:general}) we can easily 
write down the Hamiltonian for this system. Let us define the variable
$\alpha$ by 
\Beq
e^{3\alpha}:=|F|(Q_1Q_2Q_3)^{1/2},
\Eeq
where $F$ is the matrix defined by \Eq{def:F-matrix}, which turns out
to be always spatially constant in the present case. Then, since the
metric is flat($R(Q)=0$) and $|\chi|=1$, the Hamiltonian is written as
\Beq
H={\kappa^2\over 12\Omega}Ne^{-3\alpha}(-p_\alpha^2 + p_+^2 + p_-^2),
\Eeq
where $\Omega$ is the coordinate volume of a standard fundamental
region to be defined later, and $p_\alpha$ and $p_\pm$ are the
momentum variables defined by
\Beqr
&&p_\alpha:=2\Omega e^{3\alpha}(Q_1P^1+Q_2P^2+Q_3P^3),
\label{def:palpha}\\
&& p_-:=2\sqrt{3}\Omega e^{3\alpha}(Q_1P^1-Q_2P^2),
\label{def:p-}\\
&& p_+:=2\Omega e^{3\alpha}(Q_1P^1+Q_2P^2-2Q_3P^3).
\label{def:p+}
\Eeqr
The difference of topology affects the Hamiltonian only through the
value of $\Omega$. 

On the other hand the structure of the canonical 1-form is sensitive
to topology. Next we determine it for each topology of
$M$. Note that $K=j^\sharp\pi_1(M)$ is contained in 
$\tilde G=\RF^3\times D_2$
only when $M$ is either $T^3$, $T^3/\ZR_2$ or $T^3/\ZR_2\times\ZR_2$
because $R_n(2\pi/k)$($k=3,4,6$) is not contained in it(see
Table \ref{tbl:E3:Pi1}).

\subsubsection{$T^3$}

Following the procedure explained in \S2.2.3, we fix the freedom of
continuous HPDs by reducing the moduli space in order to obtain 
the expression
for $\Gamma^+_\r{LH,inv}(T^3,\RF^3\times D_2)$ of the form
(\ref{InvPS:gaugefixed}). From Table \ref{tbl:E3:Pi1} the moduli space
$\M(T^3,\RF^3\times D_2)$ is parameterized by a matrix
$\begin{matrix}{ccc}\bm{a}&\bm{b}&\bm{c}\end{matrix}\in$GL$(3,\RF)$ 
apart from the discrete $\GL(3,\ZR)$ modular transformations. 
By $\HPDG^+(E^3,\RF^3\times D_2)$ it can be always put in the form
\Beq
F=\begin{matrix}{ccc}1 & X & Y \\ 0 & 1 & Z\\0 & 0 & 1\end{matrix}
R(\phi,\theta,\psi),
\Eeq
where $X$, $Y$, and $Z$ are arbitrary real constants, and
$R(\phi,\theta,\psi)=R_3(\psi)R_1(\theta)R_3(\phi)$ is a $\SO(3)$ matrix
parameterized by the Euler angles. By this gauge fixing the HPDs
reduced to a discrete subgroup. Note that it does not fix the
discrete modular transformations.

Let us take $K_0\cong \SetDef{(l,m,n)}{l,m,n\in\ZR}$ corresponding 
to the parameters $X=Y=Z=0$ and $ R(\theta,\phi,\psi)=1$ as 
the base point in the moduli space, and identify $T^3$ with 
$E^3/K_0$. Then the deformation map is given by 
$f_\lambda(\bm{x})=F\bm{x}$, and the volume $\Omega$ 
of the fundamental region $D_0$ is unity. 
Hence from \Eq{Theta:LH:general} the canonical 1-form is
expressed as
\Beq
\Theta=e^{3\alpha}[\dot Q_1P^1+\dot Q_2P^2+\dot Q_3P^3
+ 2\Tr(\dot F F^{-1}QP)].
\label{Theta:LH:diagonal}\Eeq
From the expression for $F$ given above, after a short calculation, 
we find that it is put in the canonical form
\Beq
\Theta=\dot\alpha p_\alpha + \dot \beta_+p_+ + \dot \beta_-p_-
+\dot\theta p_\theta + \dot\phi p_\phi + \dot\psi p_\psi,
\Eeq
if we change the variables $Q_1$, $Q_2$ and $Q_3$ to $\alpha$ and
$\beta_\pm$ defined by
\Beqr
&& Q_1=e^{2(\alpha+\beta_++\sqrt{3}\beta_-)},
\label{def:alphabeta:1}\\
&& Q_2=e^{2(\alpha+\beta_+-\sqrt{3}\beta_-)},
\label{def:alphabeta:2}\\
&& Q_3=e^{2(\alpha-2\beta_+)},
\label{def:alphabeta:3}
\Eeqr
and introduce the momentum variables conjugate to $\phi,\theta$ and
$\psi$ by
\Beqr
&\begin{matrix}{c}p_\phi \\ p_\theta \\ p_\psi\end{matrix}
=&\begin{matrix}{ccc}1 & 0 & 0\\ 0 & \sin\phi & \cos\phi\\
\cos\theta & -\sin\theta\cos\phi & \sin\theta\sin\phi \end{matrix}
\nonumber\\
&& \quad \times 
\begin{matrix}{cc} {1\over\sqrt{3}}p_- X \\
-{1\over 2\sqrt{3}}p_-(Y+XZ)-{1\over 2}p_+(Y-XZ)\\
\left(-{1\over2\sqrt{3}}p_- + {1\over2}p_+\right)Z\end{matrix}.
\Eeqr

\subsubsection{$T^3/\ZR_2$}

For $T^3/\ZR_2$, as given in Table \ref{tbl:E3:Pi1}, 
the generator $\gamma$ is represented by a
glide rotation. In order for this to be contained in 
$\RF^3\sdp D_2$, the corresponding rotation matrix $R_\gamma$ 
coincides
with either $R_1(\pi)$, $R_2(\pi)$ or $R_3(\pi)$. These three
cases are connected with each other by HPDs because 
$\HPDG^+(E^3,\RF^3\sdp D_2)$ contains rotations exchanging
two of $x^1$, $x^2$ and $x^3$. Hence we can choose the gauge
such that $\alpha=(\bm{a},1)$, $\beta=(\bm{b},1)$ and 
$\gamma=(\bm{c},R_3(\pi))$ where $a^3=b^3=0$. Further, since
$\gamma$ changes by a translation $(\bm{d},1)$ as
\Beq
(\bm{d},1)(\bm{c},R_3(\pi))(\bm{d},1)^{-1}
=(\bm{c}+\bm{d}-R_3(\pi)\bm{d},R_3(\pi))
\Eeq
we can put $c^1$ and $c^2$ to zero. Finally
by automorphisms(i.e., modular transformation) 
$\alpha\leftrightarrow \beta$ and $\gamma\maps\gamma^{-1}$
if necessary, 
$\det\begin{matrix}{ccc}\bm{a}&\bm{b}&\bm{c}\end{matrix}$ 
and $c^3$ can be made positive. 

Thus the matrix $\begin{matrix}{ccc}\bm{a}&\bm{b}&\bm{c}\end{matrix}$
parametrizing $\M(T^3/\ZR_2,\RF^3\sdp D_2)$ is reduced by
HPDs to $\M_0(T^3/\ZR_2,\RF^3\sdp D_2)$ parameterized by
\Beq
F=\begin{matrix}{ccc}1& X& 0\\ 0 & 1 & 0\\ 0& 0 & 1
\end{matrix}R_3(\phi),
\Eeq
where $X$ and $\phi$ are arbitrary real numbers. The residual
HPDs form a discrete group. This gauge fixing reduces the
freedom of modular transformations to $\SL(2,\ZR)$.

If we take $K_0$ with $X=0$ and $\phi=0$ as the base point of the 
reduced moduli space, $f_\lambda(\bm{x})=F\bm{x}$ gives the
deformation map from  $K_0$ to $K$ corresponding to the above 
moduli matrix $F$. Further from the argument in \S3.1.3, 
the fundamental region $D_0$ and its volume coincide
with those in the case $M=T^3$. Hence the canonical 1-form
is obtained from that in the previous case by putting $Y=Z=0$
and $\theta=\psi=0$:
\Beq
\Theta=\dot\alpha p_\alpha + \dot \beta_+p_+ + \dot \beta_-p_-
+ \dot\phi p_\phi,
\Eeq
where 
\Beq
p_\phi={1\over\sqrt{3}}p_- X.
\Eeq

\subsubsection{$T^3/\ZR_2\times\ZR_2$}

In this case, from Table \ref{tbl:E3:Pi1}, all the generators
are represented by glide rotations, whose axes are mutually
orthogonal. Since the corresponding rotation matrices must 
be taken from $D_2$, by the HPDs producing permutations in
$D_2$, the generators can be put in the form
\Beq
\alpha=(\bm{a},R_1(\pi)),\quad
\beta=(\bm{b},R_2(\pi)),\quad
\gamma=(\bm{c},R_3(\pi)).
\Eeq
Then the constraint $R_\beta\bm{a}+R_\gamma\bm{b}+R_\alpha\bm{c}=0$
gives
\Beq
c^1=a^1+b^1,\quad
a^2=b^2+c^2,\quad
b^3=a^3+c^3.
\Eeq

Hence, if we make $b^1$, $c^2$ and $a^3$ vanish by the 
transformation corresponding to a translation $(\bm{d},1)$,
\Beqr
&& (a^1,a^2,a^3) \maps (a^1,a^2 +2d^2,a^3+2d^3),\\
&& (b^1,b^2,b^3) \maps (b^1+2d^1,b^2,b^3+2d^3),\\
&& (c^1,c^2,c^3) \maps (c^1+2d^1,c^2+2d^2,c^3),
\Eeqr
we obtain $a^1=c^1$, $a^2=b^2$ and $b^3=c^3$.  
Since we can make $b^2c^3$ positive by the modular transformation 
$\beta^{-1}\maps\beta$ and $\beta^2\gamma\maps \gamma$, 
these constants can be put to $1$ by rescalings
of $x^1$, $x^2$ and $x^3$ in $\HPDG^+(E^3,\RF^3\sdp D_2)$.
Thus the moduli space is reduced to a single point corresponding to
\Beq
\begin{matrix}{ccc}\bm{a}&\bm{b}&\bm{c}\end{matrix}=
\begin{matrix}{ccc}1 & 0 & 1\\1&1&0\\0&1&1\end{matrix}.
\Eeq
There exists no residual HPD or modular transformation.

From the argument in \S3.1.3, the volume of the fundamental region
is given by $\Omega=2$. Hence by putting $F=I_3$ in 
\Eq{Theta:LH:diagonal}, we obtain
\Beq
\Theta=\dot\alpha p_\alpha + \dot \beta_+p_+ + \dot \beta_-p_-,
\Eeq
where $\alpha$ and $\beta_\pm$ are defined by the same relations
\Eqs{def:alphabeta:1}{def:alphabeta:3} as in the previous case.

\subsection{$\tilde G_0=$VII$^{\pm}(0)$}

The automorphism group of VII$^{\epsilon}(0)$ is 
isomorphic to $\RF_+\sdp$IO(2) and its element $\phi$ is
represented as
\Beq
\phi(\xi_I)=\xi_J T^J{}_I,\quad
T=\{1,R_1(\pi)\}\times
\begin{matrix}{cc} kR(\theta) & \begin{array}{c}-c^2\\c^1\end{array}\\
\begin{array}{cc}0 & 0 \end{array} & 1\end{matrix},
\Eeq
where $k>0$, and $\theta$, $c^1$ and $c^2$ are arbitrary real numbers.
It is induced from $f\in \HPDG^+(E^3,\r{VII}^{(\epsilon)}(0))$ 
given by
\Beq
f(\bm{x})=\{1,R_1(\pi)\}\left[
\begin{matrix}{cc} kR(\theta) & \begin{array}{c}0\\0\end{array}\\
\begin{array}{cc}0 & 0 \end{array} & 1\end{matrix}\bm{x}
+\begin{matrix}{c}c^1\\c^2\\c^3\end{matrix}
+R_3(\epsilon x^3)\begin{matrix}{c}d^1\\d^2\\0\end{matrix}
\right].
\Eeq
By this HPD the invariant basis (\ref{InvariantBasis:VII})
transforms as
\Beq
f^*\chi^I=A^I{}_J\chi^J,
\Eeq
where
\Beq
A=\{1,R_1(\pi)\}\times
R_3(-\epsilon c^3)
\begin{matrix}{cc} kR(\theta) & 
\begin{array}{c}-\epsilon v^2\\\epsilon v^1\end{array}\\
\begin{array}{cc}0 & 0 \end{array} & 1\end{matrix}.
\Eeq

As in the case $\tilde G_0=\RF^3$, we can show with the helps of
these formula that VII(0)-homogeneous data always have higher 
symmetries. To see this, first note that we can transform 
the matrix $Q$ representing the components of the metric data 
$\tilde q$ with respect to the invariant basis to the diagonal 
matrix $[Q_1,1/Q_1,Q_3]$ by the above HPDs. After this transformation,
the diffeomorphism constraint (\ref{DC:H}) is expressed in terms
of the coefficient matrix $P$ of the momentum data $\tilde p$ as
\Beq
P^{13}=P^{23}=0, \quad (Q_1-Q_1^{-1})P^{12}=0.
\Eeq
If $Q_1\not=0$, this implies that $P$ is diagonal. Hence
$\InvG(\tilde\Phi)$ contains VII$^{(\epsilon)}(0)\sdp D_2$
at least. On the other hand, if $Q_1=1$, by the residual
HPDs of the form $A=R_3(-\epsilon c^3)$, we can put $P$
diagonal. Hence the same conclusion holds also in this case.

As in the case of $\tilde G_0=\RF^3$, 
$\HPDG^+(E^3,\r{VII}^{(\epsilon)}\sdp D_2)$ is obtained from
$f\in\HPDG^+(E^3,\r{VII}^{(\epsilon)})$ 
such that $fD_2 f^{-1}\subset $VII$^{(\epsilon)}(0)\sdp D_2$.
Its explicit form is given by
\Beq
f(\bm{x})=\{1,R_1(\pi)\}\left[
\begin{matrix}{cc} kR(\epsilon c^3+{l\pi\over2}) 
& \begin{array}{c}0\\0\end{array}\\
\begin{array}{cc}0 & 0 \end{array} & 1\end{matrix}\bm{x}
+\begin{matrix}{c}c^1\\c^2\\c^3\end{matrix}
\right],
\label{HPDG:VIIxD2}\Eeq
where $l$ is an arbitrary integer. This induces 
the following transformation of the invariant basis:
\Beq
f^*\chi^I=A^I{}_J\chi^J;\quad
A=\{1,R_1(\pi)\}
\begin{matrix}{cc} kR(l\pi/2) & 
\begin{array}{c}0\\0\end{array}\\
\begin{array}{cc}0 & 0 \end{array} & 1\end{matrix}.
\Eeq

Covering data invariant under VII$^{(\epsilon)}(0)\sdp D_2$
are represented by diagonal matrices $Q=[Q_1,Q_2,Q_3]$ 
and $P=[P^1,P^2,P^3]$. From Prop.\ref{prop:InvG} it is easily
shown that the data has higher symmetries if and only if $Q_1=Q_2$
and $P^1=P^2$. Hence $\Gamma^+_\r{H,D}$ for 
$\tilde G=$VII$^{(\epsilon)}(0)\sdp D_2$ is given  by
\Beqr
&\Gamma^+_\r{H,D}(E^3,\r{VII}^{(\epsilon)}(0)\sdp D_2)
=&\left\{Q=[Q_1,Q_2,Q_3],P=[P^1,P^2,P^3]\mid \right.\nonumber\\
&&\quad \left. Q_1\not=Q_2 \quad\r{or}\quad P^1\not=P^2\right\}.
\Eeqr

Now we determine explicit expressions for the canonical 1-form
and the Hamiltonian of 
$\Gamma^+_\r{LH,inv}(M,\r{VII}^{(\epsilon)}(0)\sdp D_2)$.

\subsubsection{$T^3$}

All elements of $\pi_1(T^3)$ are represented by translations. However,
since each element of $\tilde G$=VII$^{(\epsilon)}(0)\sdp D_2$ is 
expressed in the form $(\bm{a},R_3(\epsilon a^3))$ or its 
product with $R_I(\pi)$, a translation $(\bm{a},1)$ is 
contained in $\tilde G$ if and only if $a^3$ is an integer multiple of
$\pi$. Hence the matrix 
$\begin{matrix}{ccc}\bm{a}&\bm{b}&\bm{c}\end{matrix}$ parametrizing
the moduli space must take the form
\Beq
\begin{matrix}{ccc}\bm{a}&\bm{b}&\bm{c}\end{matrix}
=\begin{matrix}{ccc}
a^1& b^1 & c^1 \\ a^2 & b^2 & c^2\\ l\pi & m\pi & n\pi
\end{matrix},
\Eeq
where $l,m,n\in\ZR$. However, since for any integer vector
$(l,m,n)$ there exists $C\in$GL$(3,\ZR)$ such that
$(l,m,n)C=(0,0,k)$ where $k>0$ is the greatest common divisor
of $l$, $m$ and $n$, we can always put $l=m=0$ and $n>0$ by modular 
transformations. Further by the modular transformation 
$\alpha\leftrightarrow \beta$ if necessary, we can make 
$\det\begin{matrix}{ccc}\bm{a}&\bm{b}&\bm{c}\end{matrix}$ positive. 
Hence by
applying HPDs of the form (\ref{HPDG:VIIxD2}), we can finally 
put $\begin{matrix}{ccc}\bm{a}&\bm{b}&\bm{c}\end{matrix}$ in the form
\Beq
B=\begin{matrix}{ccc}X& Y& Z\\ 0 & X^{-1} & W\\ 0& 0 & n\pi
\end{matrix},
\Eeq
where $X>0$, $n$ is a positive integer, and $Y$, $Z$ and $W$ are
arbitrary real numbers. This completely fixes the HPD freedom of the
moduli parameters except in the subspace $Y=0$, which is invariant 
by the residual HPDs $R_3(\pm\pi/2)$. These residual HPD 
transformations produce  a singularity in the invariant phase space
with the structure of corned lens space. On the other hand
the freedom of modular transformations represented by upper 
triangle matrix in $\SL(3,\ZR)$ still remains. Since
these residual modular transformations do not change the value of
$n$, the invariant phase space $\Gamma^+_\r{LH,inv}(T^3,\r{VII}(0)\sdp
D_2)$ consists of two families(VII$^{(\pm)}(0)$) each of which has
countably infinite number of connected components.

To find the expression for $\Theta$, let us take $K_0$ corresponding
to $X=1$ and $Y=Z=W=0$ as the base point of the moduli space. Then
$f_\lambda$ defined by
\Beq
f_\lambda(\bm{x})=A\bm{x};\quad 
A=\begin{matrix}{ccc}X& Y& Z\\ 0 & X^{-1} & W\\ 0& 0 & 1
\label{CanonicalMap:VII:T3}\end{matrix}
\Eeq
gives the deformation map from  $K_0$ to $K$ corresponding to 
the matrix $B$. The fundamental region $D_0$ is given by 
$0\le x^1,x^2\le1$
and $0\le x^3\le n\pi$, and its volume by $\Omega=n\pi$.

Since $f_\lambda$ transforms the invariant basis as
\Beq
f_\lambda^*\chi^I=F^I{}_J\chi^J;\quad
F=R_3(-\epsilon x^3)A R_3(\epsilon x^3),
\Eeq
the last term in \Eq{Theta:LH:general} is expressed as
\Beq
2\Tr(\dot F F^{-1}PQ)
=\left[2{\dot X\over X}\cos(2\epsilon x^3)+(X\dot Y-Y\dot X)
\sin(2\epsilon x^3)\right](Q_1P^1-Q_2P^2).
\Eeq
Though it is not non-zero, its integration over $D_0$
vanishes. Hence, noting that $|\chi|=1$ and $|F|=1$, we obtain
from \Eq{Theta:LH:general} the following expression for $\Theta$:
\Beq
\Theta=\Omega\sqrt{Q_1Q_2Q_3}(\dot Q_1 P^1+\dot Q_2P^2+\dot Q_3P^3),
\Eeq
which is diagonalized by the change of variables 
(\ref{def:palpha})-(\ref{def:p-}) and 
(\ref{def:alphabeta:1})-(\ref{def:alphabeta:3}) as
\Beq
\Theta=\dot\alpha p_\alpha + \dot\beta_+ p_+ + \dot\beta_-p_-.
\label{Theta:VII:T3}\Eeq
Thus the canonical structure is completely degenerate in the
4-dimensional moduli sector.

For the present system the Ricci scalar curvature of the metric 
$\tilde q$ is given by
\Beq
R=-{1\over 2Q_3}\left({Q_1\over Q_2}+ {Q_2\over Q_1}-2\right)
=-2e^{-2\alpha+4\beta_+}\sinh^2(2\sqrt{3}\beta_-).
\Eeq
Hence from \Eq{H:LH:general} the Hamiltonian is expressed as
\Beq
H={\kappa^2\over 12\Omega}Ne^{-3\alpha}
\left[-p_\alpha^2+p_+^2+p_-^2+{12\Omega^2\over\kappa^4}
e^{4(\alpha+\beta_+)}\sinh^2(2\sqrt{3}\beta_-)\right]
\label{H:VII:T3}\Eeq

\subsubsection{$T^3/\ZR_2$}

Since each transformation in VII$^{\epsilon}(0)$ is a glide rotation
around the $x^3$-axis, the rotation matrix  $R_\gamma$ associated 
with the generator $\gamma$ in VII$^{(\epsilon)}(0)\sdp D_2$ must
be of the form $R_3(\theta)R_I(\pi)$($I=1,2$) or $R_3(\pi)$.
The former can be always transformed to $R_2(\pi)$ by some 
transformation in VII$^{\epsilon}(0)$ because they can be written as
$R_3(\theta/2)R_I(\pi)R_3(-\theta/2)$, and 
$R_2(\pi)=R_3(\pi/2)R_1(\pi)R_3(-\pi/2)$. On the other hand, 
there exists no transformation in 
$\HPDG^+(E^3,\r{VII}^{\epsilon}(0)\sdp D_2)$ which transforms
$R_2(\pi)$ to $R_3(\pi)$. Hence in this case the moduli space 
$\M(T^3/\ZR_2,\r{VII}^{(\epsilon)}(0)\sdp D_2)$ is divided into two
disconnected families.

\paragraph{a) $R_\gamma=R_3(\pi)$:} In this case $a^3=b^3=0$, 
and $c^3$ must be written as $n\pi$ with non-vanishing integer 
$n$ as in the previous case. On the other hand $c^1$ and $c^2$ 
can be put to zero by translations along $x^1-x^2$ plane. 
Further by modular transformations, $\alpha\leftrightarrow \beta$ 
and $\gamma\maps \gamma^{-1}$, 
$\det\begin{matrix}{ccc}\bm{a}&\bm{b}&\bm{c}\end{matrix}$ and 
$n$ can be made positive. Hence by HPDs the moduli matrix 
$\begin{matrix}{ccc}\bm{a}&\bm{b}&\bm{c}\end{matrix}$ can be 
put into the canonical form
\Beq
B=\begin{matrix}{ccc}X& Y& 0\\ 0 & X^{-1} & 0\\ 0& 0 & n\pi
\end{matrix},
\Eeq
where $X>0$, $n$ is a positive integer and $Y$ is an arbitrary real
number. There remains no residual freedom of HPDs except in
the subspace $Y=0$, while there remains residual modular 
transformations corresponding $\alpha^p\beta\maps\beta$ with
$p\in\ZR$. 

This moduli matrix is the special case $Z=W=0$ of that in 
the previous case. Hence the canonical structure is degenerate in
the 2-dimensional moduli sector, and the the value of $\Omega$ and 
the forms of $\Theta$
and $H$ are exactly the same as in the previous case.

\paragraph{b) $R_\gamma=R_2(\pi)$:} In this case the moduli matrix
$\begin{matrix}{ccc}\bm{a}&\bm{b}&\bm{c}\end{matrix}$ takes the form
\Beq
\begin{matrix}{ccc}\bm{a}&\bm{b}&\bm{c}\end{matrix}
=\begin{matrix}{ccc}a^1& b^1 & c^1\\
0&0&c^2\\ l\pi & n\pi & m\pi \end{matrix},
\Eeq
where $l,m$ and $n$ are integers. $l$ can be put to zero by unimodular
transformations among $\alpha$ and $\beta$, and $m$ to 0 or 1
by a transformation in VII$^{\epsilon}(0)\sdp D_2$;
\Beq
(\bm{d},1)\left((c^1,c^2,m\pi),R_2(\pi)\right)(\bm{d},1)^{-1}
=\left((c^1,c^2,m\pi+k\pi),R_2(\pi)\right),
\Eeq
where $\bm{d}=(0,0,k\pi)$. Further by
modular transformations $\alpha\maps \alpha^{-1}$,
$\beta\maps\beta^{-1}$ and $\gamma\maps\gamma^{-1}$ if necessary,
$a^1$ and $c^2$ and $n$ can be made positive. Finally by translation
along the $x^1$-axis and HPDs, 
$\begin{matrix}{ccc}\bm{a}&\bm{b}&\bm{c}\end{matrix}$ is transformed 
to
\Beq
B=\begin{matrix}{ccc}X& Y& {m\over n}Y\\ 0 & 0 &X^{-1} \\ 0& n\pi & m\pi
\end{matrix},
\Eeq
where $X>$, $n$ is a positive integer, $m=0,1$ and $Y$ is an 
arbitrary real number. As in the case a), this gauge fixing 
leaves only the residual modular transformations isomorphic to
$\ZR$.

Let us take $K_0$ with $X=1$ and $Y=0$ as the base point
for each of the connected component of the reduced moduli space
$\M_0(T^3/\ZR_2,\r{VII}^{(\epsilon)}(0)\sdp D_2)$. Then $f_\lambda$
defined by
\Beq
f(\bm{x})=A\bm{x};\quad
A=\begin{matrix}{ccc} X& 0 & Y/n\pi\\ 0& 1/X &0\\0&0&1\end{matrix}
\Eeq
maps $K_0$ to $K$ corresponding to the generic matrix $B$. This is
again the special case obtained from that for $M=T^3$ by putting
$Y\tend 0$, $Z\tend Y/n\pi$ and $W\tend0$. Hence the canonical 
structure is completely degenerate in the 2-dimensional moduli 
sector. $\Omega$, $\Theta$
and $H$ are the same as those in the previous case.

\subsubsection{$T^3/\ZR_2\times\ZR_2$}

As in the case of $\tilde G=\RF^3\sdp D_2$, three generators can be
put into the form
\Beq
\alpha=(\bm{a},R_1(\pi)),\quad
\beta=(\bm{b},R_2(\pi)),\quad
\gamma=(\bm{c},R_3(\pi)),
\label{K:T3xZ2xZ2}\Eeq
where $\begin{matrix}{ccc}\bm{a}&\bm{b}&\bm{c}\end{matrix}$ is 
expressed in the present case as
\Beq
\begin{matrix}{ccc}\bm{a}&\bm{b}&\bm{c}\end{matrix}
=\begin{matrix}{ccc}a^1& b^1 & c^1\\
a^2&b^2&c^2\\ l\pi & m\pi & n\pi \end{matrix}
\Eeq
with integers $l$, $m$ and $n$. As in \S3.2.3, $b^1$ and $c^2$ can be
put to zero  by translations parallel to the $x^1-x^2$ plane, which
entails $a^1=c^1$ and $a^2=b^2$. Further $l$ can be put to  
0 or 1 by translation along the $x^3$-axis. 
Hence taking account of the constraint $m=l+n$, 
$\begin{matrix}{ccc}\bm{a}&\bm{b}&\bm{c}\end{matrix}$ is finally 
put by HPDs into the form
\Beq
B=\begin{matrix}{ccc}X& 0& X\\ X^{-1} & X^{-1} & 0\\ l\pi& (l+n)\pi 
& n\pi\end{matrix},
\Eeq
where $X>0$, $l=0,1$ and $n$ is a non-zero integer. 
Thus the invariant phase space consists of two families($l=0,1$) 
of countably infinite number of connected components.

Let us take $K_0$ with $X=0$ as the base point of each connected
component. Then $f_\lambda$ defined by
\Beq
f_\lambda(\bm{x})=A\bm{x};\quad
A=\begin{matrix}{ccc}X&0&0\\0&1/X&0\\0&0&1\end{matrix}
\Eeq
yields the deformation map to the generic point corresponding to $B$. 
This is a special case of \Eq{CanonicalMap:VII:T3} with $Y=Z=W=0$.
Hence the canonical structure is degenerate in the 1-dimensional
moduli sector, and $\Theta$ and $H$ are again given by 
\Eq{Theta:VII:T3} and
\Eq{H:VII:T3}. The only difference is the value of $\Omega$, which is
given by $2|n|\pi$ in the present case. 

\subsubsection{$T^3/\ZR_k$($k=3,4,6$)}

In order for $K\subset $VII$^{(\epsilon)}(0)\sdp D_2$, from
Table \ref{tbl:E3:Pi1}, its generator
should be of the form
\Beq
\alpha=(\bm{a},1),\quad
\beta=(\bm{b},1),\quad
\gamma=(\bm{c},R_3(\pm 2\pi/k)),
\Eeq
where the translation part 
$\begin{matrix}{ccc}\bm{a}&\bm{b}&\bm{c}\end{matrix}$ is given by
\Beq
\begin{matrix}{ccc}\bm{a}&\bm{b}&\bm{c}\end{matrix}=\begin{matrix}{ccc}a^1& b^1 & c^1\\
a^2&b^2&c^2\\ 0 & 0 & c^3 \end{matrix},\quad
\begin{matrix}{c}b^1\\b^2\end{matrix}
=R(\pm{2\pi\over k})\begin{matrix}{c}a^1\\a^2\end{matrix},
\Eeq
and $c^3$ is a discrete number of the form $\pm
2\pi/k+n\pi$($n\in\ZR$). By the rotation $R_1(\pi)\in
\HPDG^+(E^3,\r{VII}^{(\epsilon)}\sdp D_2)$, $R_\gamma$ can be put to
$R_3(2\pi/k)$ and $a^1b^2-a^2b^1$ can be made positive. Further by
translations along the $x^1-x^2$ plane we can put $c^1$ and $c^2$ to
zero, and $a^2$ to zero by a translation along $x^3$ in 
VII$^\epsilon(0)\sdp D_2$. Hence by HPDs the moduli matrix is 
finally transformed to 
\Beq
\begin{matrix}{ccc}\bm{a}&\bm{b}&\bm{c}\end{matrix}=\begin{matrix}{ccc}1 &\cos(2\pi/k)& 0 \\
0 & \sin(2\pi/k) & 0\\ 0&0& {2\pi\epsilon/ k}+n\pi\end{matrix}.
\Eeq

Thus the reduced moduli space $\M_0$ consists of countably
infinite number of
discrete points. The canonical 1-form and the Hamiltonian are given by
the same expressions as for $T^3$ except that 
$\Omega=\sin(2\pi/k)|{2\pi\epsilon/ k}+n\pi|$ in the present case.

\begin{table}
\noindent
\begin{tabular}{llllc}
\bf Symmetry &\bf Topology & \bf Moduli & $\Omega$ & $\Theta$ \\
$\RF^3\sdp D_2$ 
& $T^3$ 
& $\begin{matrix}{ccc}1& X& Y\\ 0 & 1 & Z\\ 0& 0 & 1
\end{matrix}R(\theta,\phi,\psi)$ & 1 & 
$\begin{array}{l}
\dot\alpha p_\alpha +\dot\beta_+p_+ + \dot\beta_- p_-\\
+\dot\theta p_\theta+\dot\phi p_\phi + \dot\psi p_\psi\end{array}$\\
\\
& $T^3/\ZR_2$ 
& $\begin{matrix}{ccc}1& X& 0\\ 0 & 1 & 0\\ 0& 0 & 1
\end{matrix}R_3(\phi)$ & 1 & 
$\begin{array}{l}
\dot\alpha p_\alpha +\dot\beta_+p_+ + \dot\beta_- p_- \\
+\dot\phi p_\phi\end{array}$\\
&& ($R_\gamma=R_3(\pi)$) &&\\
& $T^3/\ZR_2\times\ZR_2$ & 
$\begin{matrix}{ccc}1& 0& 1\\ 1 & 1 & 0\\ 0& 1 & 1
\end{matrix}$ & 2 & 
$\dot\alpha p_\alpha +\dot\beta_+p_+ + \dot\beta_- p_- $\\
&& ($R_\alpha=R_1(\pi),R_\beta=R_2(\pi),$ &&\\
&& $R_\gamma=R_3(\pi)$) && \\
\\
VII$^{(\epsilon)}\sdp D_2$
& $T^3$ 
& $\begin{matrix}{ccc}X& Y& Z\\ 0 & X^{-1} & W\\ 0& 0 & n\pi
\end{matrix}$& $n\pi$ & 
$\dot\alpha p_\alpha +\dot\beta_+p_+ + \dot\beta_- p_-$\\
&&($X>0,n\in\NN$) && \\
& $T^3/\ZR_2$ 
& $\begin{matrix}{ccc}X& Y& 0\\ 0 & X^{-1} & 0\\ 0& 0 &
n\pi \end{matrix}$ & $n\pi$ & ${}''$\\
&& $(X>0,R_\gamma=R_3(\pi),n\in\NN)$&&\\
&  & $\begin{matrix}{ccc}X& Y& {m\over n}Y\\ 0 & 0 & X^{-1}\\ 0& n\pi
& m\pi \end{matrix}$ & $n\pi$ & ${}''$\\
&& $(X>0,R_\gamma=R_2(\pi),$&&\\
&& $ n\in\NN,m=0,1)$ &&\\
& $T^3/\ZR_2\times\ZR_2$ & 
$\begin{matrix}{ccc}X& 0& X\\ X^{-1} & X^{-1} & 0\\ l\pi& (l+n)\pi 
& n\pi\end{matrix}$ & $2|n|\pi$ & ${}''$\\
&& ($X>0,n\not=0\in\ZR,l=0,1$, &&\\
&& $R_\alpha=R_1(\pi),R_\beta=R_2(\pi),$&&\\
&& $R_\gamma=R_3(\pi)$) &&\\
& $T^3/\ZR_k$ & $\begin{matrix}{ccc}1 &\cos{2\pi\over k}& 0 \\
0 & \sin{2\pi\over k} & 0\\ 0&0& 
{2\pi\epsilon\over k}+n\pi\end{matrix}$
& $\begin{array}{l}|{2\pi\epsilon\over k}+n|\\ \times\sin{2\pi\over k}\end{array}$ & ${}''$\\
&$(k=3,4,6)$&($n\not=0\in\ZR,R_\gamma=R_3({2\pi/ k})$) &&\\
\end{tabular}
\caption{\label{tbl:CanonicalStr:E3}Canonical Structure of compact
orientable closed 3-manifold of type $E^3$.}
\end{table}

\noindent
\begin{table}
\begin{tabular}{llllc}
\bf Symmetry & \bf Topology & \bf Moduli & $\Omega$ & $\Theta$ \\
$\RF^3\sdp$O(2)
& $T^3$ 
& $\begin{matrix}{ccc}X& Y& Z\\ 0 & X^{-1} & W\\ 0& 0 & 1
\end{matrix}R(\theta,\phi)$& 1 & 
$\begin{array}{l}\dot\alpha p_\alpha +\dot\beta_+p_+ \\
+ \dot\theta p_\theta + \dot\phi p_\phi\end{array}$\\
&&($X>0$) && \\
& $T^3/\ZR_2$ 
& $\begin{matrix}{ccc}0& 0& X^{-1}\\ X & Y & 0\\ 0& 1 &
0 \end{matrix}R_3(\phi)$ & 1 & 
$\begin{array}{l}\dot\alpha p_\alpha +\dot\beta_+p_+ \\
+ \dot\phi p_\phi\end{array} $\\
&& $(X>0,R_\gamma=R_2(\pi))$&&\\
&  & $\begin{matrix}{ccc}X& Y& 0\\ 0 & X^{-1} & 0\\ 0& 0 & 1
\end{matrix}$ & 1  &$\dot\alpha p_\alpha +\dot\beta_+p_+ $ \\
&& $(X>0,R_\gamma=R_3(\pi))$&&\\
& $T^3/\ZR_2\times\ZR_2$ & 
$\begin{matrix}{ccc}X& 0& X\\ X^{-1} & X^{-1} & 0\\ 0 & 1
& 1 \end{matrix}$ & $2$ & ${}''$\\
&& ($X>0,R_\alpha=R_1(\pi),$&&\\
&& $R_\beta=R_2(\pi), R_\gamma=R_3(\pi)$) && \\
& $T^3/\ZR_k$ & $\begin{matrix}{ccc}1 &\cos{2\pi\over k}& 0 \\
0 & \sin{2\pi\over k} & 0\\ 0&0& \pm1 \end{matrix}$
& $\sin{2\pi\over k}$ & ${}''$ \\
&$(k=3,4,6)$&($R=R_3({2\pi/k})$) &&\\
\\
$\ISO(3)$ 
& $T^3$ 
& $\begin{matrix}{ccc}X& Z& W\\ 0 & Y & U\\ 0& 0 & 1/XY
\end{matrix}$& 1 & 
$\dot\alpha p_\alpha $\\
&&($X,Y>0$) && \\
& $T^3/\ZR_2$ 
& $\begin{matrix}{ccc}X& Z& 0\\ 0 & Y & 0\\ 0& 0 & 1/XY
\end{matrix}$ & 1  & ${}''$\\
&& $(X,Y>0,R_\gamma=R_3(\pi))$&&\\
& $T^3/\ZR_2\times\ZR_2$ & 
$\begin{matrix}{ccc}X& 0& X\\ Y & Y & 0\\ 0 & 1/XY & 1/XY 
\end{matrix}$ & $2$ & ${}''$\\
&& ($X,Y>0,R_\alpha=R_1(\pi)$,&&\\
&& $R_\beta=R_2(\pi),R_\gamma=R_3(\pi)$) && \\
& $T^3/\ZR_k$ & $\begin{matrix}{ccc}X &X\cos{2\pi\over k}& 0 \\
0 & X\sin{2\pi\over k} & 0\\ 0&0& \pm 1/X^2 \end{matrix}$
& $\sin{2\pi\over k}$& ${}''$ \\
&$(k=3,4,6)$&($R_\gamma=R_3({2\pi/k})$) &&\\
\end{tabular}
\begin{center}
Table \ref{tbl:CanonicalStr:E3}(continued).
\end{center}
\end{table}

\subsection{$\tilde G_0=\RF^3\sdp\SO(2)$}

An automorphism of $\RF^3\sdp\SO(2)$ is expressed as
\Beq
\phi(\xi_a)=\xi_b T^b{}_a;\quad
T=\begin{matrix}{cccc}1&0&0&0\\0&\pm1&0&0\\0&0&\pm1&0\\0&0&0&\pm1\end{matrix}
\begin{matrix}
{cc}k_1R(\theta)&\begin{array}{cc}0&c^2\\0&-c^1\end{array}\\
\begin{array}{cc}0&0\\0&0\end{array} &\begin{array}{cc}k_2&d\\0&1\end{array}
\end{matrix},
\Eeq
where $k_1>0$. When $d=0$ and $k_2>0$, it is induced from 
 $f\in \HPDG^+(E^3,\RF^3\sdp\SO(2))$ given by
\Beq
f(\bm{x})=\{1,R_1(\pi)\}\left[
\begin{matrix}{cc} k_1R(\theta) & \begin{array}{c}0\\0\end{array}\\
\begin{array}{cc}0 & 0 \end{array} & k_2\end{matrix}\bm{x}
+\begin{matrix}{c}c^1\\c^2\\c^3\end{matrix}
\right],
\Eeq
where $c^3$ is an integration constant. 
By this HPD the invariant basis \Eq{InvariantBasis:I} transforms as
\Beq
f^*\chi_I=A^I{}_J\chi^J;\quad
A=\{1,R_1(\pi)\}\begin{matrix}{cc} k_1R(\theta) & \begin{array}{c}0\\0\end{array}\\
\begin{array}{cc}0 & 0 \end{array} & k_2\end{matrix}.
\Eeq

From the symmetry the components $Q$ and $P$ of covering data 
$\tilde\Phi=(\tilde q,\tilde p)$ with respect to the invariant 
basis \Eq{InvariantBasis:I} must be represented by the
diagonal matrices with constant entries, $[Q_1,Q_1,Q_3]$ and
$[P^1,P^1,P^3]$. In particular the covering geometry is flat, and the
diffeomorphism constraint becomes trivial. Further the data are always
invariant under $R_1(\pi)$. Hence the invariance group contains
$\RF^3\sdp \SO(2)\sdp\{1,R_1(\pi)\}\cong \RF^3\sdp $O(2). Since 
$\HPDG^+(E^3,\RF^3\sdp $O$(2))=\HPDG^+(E^3,\RF^3\sdp\SO(2))$, 
$Q$ and $P$ are transformed by HPDs to $[1,1,1]$ and
$[Q_1P^1,Q_1P^1,Q_2P^2]$. Hence $\InvG^+(\tilde \Phi)$ coincides 
with $\RF^3\sdp $O(2) if and only if $Q_1P^1\not=Q_3P^3$. Thus we
obtain
\Beq
\Gamma^+_\r{H,D}(E^3,\RF^3\sdp \r{O}(2))=
\SetDef{Q=[Q_1,Q_1,Q_3],P=[P^1,P^1,P^3]}{Q_1P^1\not=Q_3P^3}.
\Eeq

\subsubsection{$T^3$}

The translation vectors generating the moduli $K$ can be put into the
form
\Beq
\begin{matrix}{ccc}\bm{a}&\bm{b}&\bm{c}\end{matrix}=
F=\begin{matrix}{ccc}X& Y& Z\\ 0 & X^{-1} & W\\ 0& 0 & 1
\end{matrix}R(\theta,\phi)
\Eeq
by modular transformations $\alpha\leftrightarrow \alpha^{-1}$, 
$\beta\leftrightarrow\beta^{-1}$ and
$\gamma\leftrightarrow\gamma^{-1}$, and HPDs, where
$X>0$ and $R(\theta,\phi):=R_3(\phi-\pi/2)R_1(\theta)R_3(\pi/2-\phi)$
is a matrix which rotates a vector with the angular direction
$(\theta,\phi)$ to a vector parallel to the $x^3$-axis. 
This fixes the gauge freedom up to residual discrete HPDs and modular
transformations. The invariant phase space is connected.

If we take $K_0$ with $X=1$ and $Y=Z=W=\theta=\phi=0$ as the base
point of the reduced moduli space, $f_\lambda(\bm{x})=F\bm{x}$ maps
$K_0$ to a generic $K$ corresponding to $F$. Hence from
\Eq{Theta:LH:general} and \Eq{H:LH:general}, after a short
calculation, we obtain
\Beqr
&&\Theta=\dot\alpha p_\alpha +\dot\beta_+p_+ +\dot\theta p_\theta
+\dot\phi p_\phi,\\
&& H={\kappa^2\over 12\Omega}Ne^{-3\alpha}(-p_\alpha^2+p_+^2).
\label{H:R3xO2}\Eeqr
Here $\alpha$ and $\beta_+$ are related to $Q_1$ and $Q_2$ by 
the equations obtained from \Eq{def:alphabeta:1} and \Eq{def:alphabeta:3}
by putting $\beta_-=0$, $p_\alpha$ and $p_+$ are defined by 
\Eq{def:palpha} and \Eq{def:p+} with $Q_1P^1=Q_2P^2$, and $p_\phi$ and
$p_\theta$ are given by
\Beqr
&& p_\phi:={1\over2}p_+\left[\left(YW-{Z\over X}\right)\sin\phi
+XW\cos\phi\right]\sin\theta,\\
&& p_\theta:={1\over2}p_+\left[-\left(YW-{Z\over X}\right)\cos\phi
+XW\sin\phi\right].
\Eeqr
Of course $\Omega=1$ is understood in the present case.
The canonical structure is partially degenerate in the moduli 
sector.

\subsubsection{$T^3/\ZR_2$}

As in the corresponding case with $\tilde G=$VII$^{(\pm)}(0)\sdp D_2$, 
the rotation matrix $R$ associated with the generator $\gamma$ can be
put to $R_3(\pi)$ or $R_1(\pi)$. Accordingly the invariant phase 
space consists of two connected components.

\paragraph{a) $R_\gamma=R_3(\pi)$:} By a procedure similar to that in
\S3.3.2, the moduli matrix 
$\begin{matrix}{ccc}\bm{a}&\bm{b}&\bm{c}\end{matrix}$ can be put
in the form
\Beq
F=\begin{matrix}{ccc}X& Y& 0\\ 0 & X^{-1} & 0\\ 0& 0 & 1
\end{matrix},
\Eeq
where $X>0$. This gauge fixing eliminates the freedom of HPDs 
except in the subspace $Y=0$, while it leaves modular transformations
isomorphic to $\ZR$. By taking $K_0$ with $X=1$ and $Y=0$ as 
the base point, the deformation map $f_\lambda$ is given by
$f_\lambda(\bm{x})=F\bm{x}$. This is obtained from the
corresponding map for $T^3$ by putting $Z=W=0$ and
$R(\theta,\phi)=1$. Hence $\Theta$ is given by
\Beq
\Theta=\dot\alpha p_\alpha + \dot \beta_+p_+,
\label{Theta:T3/Z2}\Eeq
and $H$ by \Eq{H:R3xO2} with $\Omega=1$. Thus 
the canonical structure is completely degenerate in the
2-dimensional moduli sector.

\paragraph{b) $R_\gamma=R_1(\pi)$: } In this case the moduli matrix 
has the form
\Beq
\begin{matrix}{ccc}\bm{a}&\bm{b}&\bm{c}\end{matrix}
=\begin{matrix}{ccc}0& 0 & c^1\\
a^2&b^2&c^2\\ a^3 & b^3 & c^3 \end{matrix}.
\Eeq
$c^2$ and $c^3$ can  be put to zero by translations along the
$x^2-x^3$ plane. Hence by modular transformations and HPDs it can be
put in the form
\Beq
B=\begin{matrix}{ccc}0& 0& X^{-1}\\ X & Y & 0\\ 0& 1 &
0 \end{matrix}R_3(\phi),
\Eeq
where $X>0$. There remains residual discrete HPDs as well as
modular transformations isomorphic to $\SL(2,\ZR)$. 

If we take $K_0$ with $X=1$ and $Y=0$ and $\phi=0$ as the base point, 
$f_\lambda$ given by
\Beq
f_\lambda(\bm{x})=F\bm{x};\quad
F=\begin{matrix}{ccc}1/X&0 &0\\ 0&X&Y\\0&0&1\end{matrix}R_1(\phi)
\Eeq
maps $K_0$ to $K$ corresponding to $B$. Inserting this into
\Eq{Theta:LH:general} with $\Omega=1$, we obtain
\Beq
\Theta=\dot\alpha p_\alpha+\dot\beta_+p_++\dot\phi p_\phi,
\Eeq
where
\Beq
p_\phi:={Y\over 2X}p_+.
\Eeq
The canonical structure is again completely degenerate
in the moduli sector.

\subsubsection{$T^3/\ZR_2\times\ZR_2$}

As in \S3.3.3, the generator of $K$ can be transformed to
the same form as \Eq{K:T3xZ2xZ2}.
Further by HPDs and modular transformations, the moduli matrix can 
be put into the form
\Beq
\begin{matrix}{ccc}\bm{a}&\bm{b}&\bm{c}\end{matrix}
=\begin{matrix}{ccc}X& 0& X\\ 1/X & 1/X & 0\\ 0 & 1 & 1 \end{matrix},
\label{MM:T3xZ2xZ2}\Eeq
where $X>0$. No freedom of HPD and modular transformation remains. 
Hence the phase space $\Gamma^+_\r{LH,inv}(T^3/\ZR_2\times\ZR_2,\RF^3\sdp \r{O}(2))$ is connected.

For the base point $K_0$ with $X=1$ and  $\Omega=2$, the deformation
map $f_\lambda$ is given by 
\Beq
f_\lambda(\bm{x})=F\bm{x};\quad
F=\begin{matrix}{ccc}X&0&0\\0&1/X&0\\0&0&1\end{matrix}.
\Eeq
This is the special case of $f_\lambda$ in \S3.4.2-a) with $Y=0$. 
Hence $\Theta$ is given by the same equation as \Eq{Theta:T3/Z2},
and the canonical structure is degenerate in the 1-dimensional
moduli sector.

\subsubsection{$T^3/\ZR_k$($k=3,4,6$)}

By an argument similar to that in \S3.3.4, the generators of $K$ can
be put in the form
\Beq
\alpha=(\bm{a},1),\quad
\beta=(\bm{b},1),\quad
\gamma=(\bm{c},R_3(2\pi/k)),
\label{K:R3xO2:T3/Zk}\Eeq
with 
\Beq
\begin{matrix}{ccc}\bm{a}&\bm{b}&\bm{c}\end{matrix}=\begin{matrix}{ccc}1 &\cos(2\pi/k)& 0 \\
0 & \sin(2\pi/k) & 0\\ 0&0& \pm1\end{matrix}.
\label{MM:R3xO2:T3/Zk}\Eeq
Thus the reduced moduli space consists of two points which are
connected by orientation-reversing HPDs. The volume of the 
fundamental region is given by $\Omega=\sin(2\pi/k)$, 
and $\Theta$ and $H$ 
by \Eq{Theta:T3/Z2} and \Eq{H:R3xO2}, respectively.

\begin{table}
\begin{tabular}{llccccccccc}
\bf Symmetry & \bf Space & $Q$ & $P$ & $N_\r{m}$ & $N$ & $N_\r{c}$ &
$N_\r{d}$ & $N_\r{cc}$ & \small HPD & \small Modular\\ 
&&&&&&&&\\
$\RF^3\sdp D_2$ & $T^3$ & 3 & 3 & 6 & 12& 12& 0 & 1 &$\bigcirc$ &$\bigcirc$ \\ 
& $T^3/\ZR_2$ & 3 & 3 & 2 & 8 & 8 & 0 & 1 &$\bigcirc$ &$\bigcirc$\\ 
&$T^3/\ZR_2\times\ZR_2$& 3 & 3 & 0 & 6 & 6 & 0 & 1 &$\times$ &$\times$\\ 
&&&&&&&&\\
VII(0)$\sdp D_2$ & $T^3$ & 3 & 3 & 4 & 10& 6 & 4 & $\infty$ &
$\triangle$ & $\bigcirc$\\ 
& $T^3/\ZR_2$ & 3 & 3 & 2 & 8 & 6 & 2 & $\infty$ &$\triangle$ & $\bigcirc$ \\ 
& $T^3/\ZR_2\times\ZR_2$& 3 & 3 & 1 & 7 & 6 & 1 & $\infty$ &$\times$ &$\times$\\ 
& $T^3/\ZR_k(k=3,4,6)$ & 3 & 3 & 0 & 6 & 6 & 0 & $\infty$ &$\times$ &$\times$\\ 
&&&&&&&&\\
$\RF^3\sdp$O(2) & $T^3$ & 2 & 2 & 6 & 10& 8 & 2 & 1 &$\bigcirc$ &$\bigcirc$\\ 
& $T^3/\ZR_2$ &2 & 2 & 2 & 6 & 4 & 2 & 1 &$\triangle$ & $\bigcirc$\\ 
& & 2 & 2 & 3 & 7 & 6 & 1 & 1 &$\bigcirc$ &$\bigcirc$\\ 
& $T^3/\ZR_2\times\ZR_2$& 2 & 2 & 1 & 5 & 4 & 1 & 1 &$\times$ &$\times$\\ 
& $T^3/\ZR_k(k=3,4,6)$ & 2 & 2 & 0 & 4 & 4 & 0 & 2 &$\times$ &$\times$\\ 
&&&&&&&&\\
$\ISO(3)$ & $T^3$ & 1 & 1 & 5 & 7 & 2 & 5 & 1 &$\triangle$ & $\bigcirc$\\ 
& $T^3/\ZR_2$ & 1 & 1& 3 & 5 & 2 & 3 & 1 &$\triangle$ & $\bigcirc$\\ 
& $T^3/\ZR_2\times\ZR_2$& 1 & 1 & 2 & 4 & 2 & 2& 1 &$\times$ &$\times$\\ 
& $T^3/\ZR_k(k=3,4,6)$ & 1 & 1 & 1 & 3 & 2 & 1 & 2 &$\times$ &$\times$\\
&&&&&&&&\\
\end{tabular}
\caption{\label{tbl:freedom:E3}Canonical and degenerate degrees of
freedom for LHS of type $E^3$} The last two columns show whether there 
exist residual discrete HPDs and modular transformations. $\bigcirc$
indicates the existence, $\times$ the non-existence, and $\triangle$
the exsistence at lower-dimensional subspaces.
\end{table}

\subsection{$\tilde G=\ISO(3)$}

For $\tilde G=\ISO(3)$, the phase space of homogeneous covering 
data is obviously given by
\Beq
\Gamma^+_\r{H,D}(E^3,\ISO(3))=
\left\{Q=[Q_1,Q_1,Q_1],P=[P^1,P^1,P^1]\right\}.
\Eeq

The arguments to determine $\Theta$ and $H$ for each spaces are 
almost the same as those in the previous case. So I just give 
outlines of derivations and results.

\subsubsection{$T^3$}

By HPDs (\ref{HPD:ISO3}) and modular transformations the modular
matrix can be put in the form
\Beq
\begin{matrix}{ccc}\bm{a}&\bm{b}&\bm{c}\end{matrix}=
F=\begin{matrix}{ccc}X& Z& W\\ 0 & Y & U\\ 0& 0 & 1/XY
\end{matrix},
\Eeq
where $X,Y>0$. There exists no residual HPD except in
the subspace $ZUX=0$, but the freedom of modular transformations
isomorphic to $\SL(2,\ZR)$ remains. 

The deformation map $f_\lambda$ is given by the
linear transformation by $F$. Inserting it into
\Eqs{Theta:LH:general}{H:LH:general}, we find that the moduli
parameters do not appear in the canonical 1-form, which is simply
written as
\Beq
\Theta=\dot\alpha p_\alpha.
\label{Theta:ISO3:T3}\Eeq
The Hamiltonian is given by
\Beq
H=-{\kappa^2\over 12\Omega}Ne^{-3\alpha}p_\alpha^2,
\Eeq
\label{H:ISO3:T3}
where $\Omega=1$.

\subsubsection{$T^3/\ZR_2$}

By $\SO(3)$ transformation $\gamma$ can be put in the form
$\gamma=(\bm{c},R_3(\pi))$. The freedom of HPDs is fixed 
by putting the moduli matrix to the form
\Beq
\begin{matrix}{ccc}\bm{a}&\bm{b}&\bm{c}\end{matrix}=
F=\begin{matrix}{ccc}X& Z& 0\\ 0 & Y & 0\\ 0& 0 & 1/XY
\end{matrix},
\Eeq
except at the subspace $Z=0$ where $X,Y>0$. The residual freedom
of modular transformations is isomorphic to $\ZR$. 
The phase space is connected and the canonical
structure is completely degenerate in the moduli sector as in the
previous case. $\Omega$, $\Theta$ and $H$ are the same as those for
$T^3$.

\subsubsection{$T^3/\ZR_2\times\ZR_2$}

The canonical form of the moduli matrix is given by
\Beq
\begin{matrix}{ccc}\bm{a}&\bm{b}&\bm{c}\end{matrix}
=\begin{matrix}{ccc}X&0&X\\Y&Y&0\\0&1/XY&1/XY\end{matrix},
\Eeq
where $X,Y>0$. For the base point $K_0$ with $X=Y=1$, the
deformation map is given by
\Beq
f_\lambda(\bm{x})=F\bm{x};\quad
F=\begin{matrix}{ccc}X&0&0\\0&Y&0\\0&0&1/XY\end{matrix},
\Eeq
and the volume of the fundamental region by $\Omega=2$. 
The phase space is connected, and its canonical structure 
becomes completely degenerate in the 2-dimensional moduli sector.
Hence $\Theta$ and $H$
are given by \Eqs{Theta:ISO3:T3}{H:ISO3:T3} with $\Omega=2$.

\subsubsection{$T^3/\ZR_k$($k=3,4,6$)}

The moduli matrix for the reduced moduli space is given by
\Beq
\begin{matrix}{ccc}\bm{a}&\bm{b}&\bm{c}\end{matrix}
=F=\begin{matrix}{ccc}X&X\cos(2\pi/k)&0\\0&X\sin(2\pi/k)&0\\
0&0&\pm 1/X^2\end{matrix},
\Eeq
and the deformation map by $f_\lambda(\bm{x})=F\bm{x}$. 
There remains no freedom of residual HPDs and modular transformations.
Thus the phase space has two connected components, and 
$\Theta$ and $H$ are given by
 \Eqs{Theta:ISO3:T3}{H:ISO3:T3} with $\Omega=\sin(2\pi/k)$

\subsection{Degeneracy of the canonical structure}

The reduced form of the moduli matrix, the corresponding canonical
1-form $\Theta$ and the volume $\Omega$ of the standard fundamental
region obtained so far are summarized in Table
\ref{tbl:CanonicalStr:E3}.

As we have seen, the canonical 1-form often becomes degenerate in the
sector of moduli parameters. The number of degenerate degrees of
freedom, $N_\r{d}$, for each space and symmetry are listed in Table
\ref{tbl:freedom:E3} with  the
independent degrees of freedom of $Q$ and $P$, the dimension 
$N_\r{m}$ of
moduli space, the total dimension $N$ of the invariant phase space,
the number $N_\r{c}$ of the non-degenerate canonical degrees of
freedom and the number of connected components $N_\r{cc}$.

From this table we see that the degrees of degeneracy increases as the
covering data have higher symmetries in general. However, it is
difficult to regard this to be simply brought about by the decrease of
the freedom of homogeneous data due to symmetry because invariant
phase spaces corresponding to different invariance groups have no
simple relations with each other. For example the moduli sector of the
invariance phase spaces for $G\cong \RF^3\sdp D_2$ are all compact due
to the modular transformations, while their counter parts for
$G\cong\RF^3\sdp $O$(2)$ are all non-compact.

Furthermore the degeneracy for the systems with $G=$VII$(0)$ cannot be
a result of the existence of isotropy groups because the invariance
group is simply transitive. The calculation of the
canonical 1-form in \S3.3.1 rather shows that the degeneracy
arises partly because the couplings among variables vanishes by
integration over the fundamental region. This implies that canonical
variables describing the locally homogeneous sector has non-trivial
couplings with locally inhomogeneous degrees of freedom.

In any case, when the canonical structure is degenerate, the 
canonical equations of motion alone do not determine the time
evolution of locally homogeneous systems completely. As touched upon
in \S2.3.3, however, this does not implies that the degenerate moduli
freedom is non-dynamical, because, from Theorem
\ref{theorem:ModuliDynamics}, the full Einstein equations completely
determine the time evolution of the parameters describing the reduced
moduli space obtained by eliminating the HPD freedom. They are
actually constants of motion. For the locally homogeneous systems of
type $E^3$ (as well as other types considered in the present paper), it
is directly confirmed that the moduli parameters in the non-degenerate 
sectors are conserved  from the structures of the canonical 1-form and
the Hamiltonian. Thus the degeneracy in the canonical structure of
locally homogeneous systems means that they are not canonically closed
in the full diffeomorphism-invariant phase space.

\section{LHS of type $\Nil$}

In this section we investigate the canonical structure of locally
homogeneous pure gravity systems of type $\Nil$.

\subsection{Basic properties}

\subsubsection{$\Gmax $ and $\HPDG^+(\Gmax )$}

The space $\Nil$ has the structure of the Heisenberg group,
and classified as type II in the Bianchi scheme. As a subgroup
of $\GL(3,\RF)$, $\Nil$ is expressed as
\Beq
\Nil\cong\SetDef{\begin{matrix}{ccc}1&x&z+xy/2\\0&1&y\\0&0&1\end{matrix}}
{x,y,z\in\RF}.
\Eeq
Here we have adopted a parameterization which is different from the
standard one. This parameterization is more convenient for our purpose
because it enables us to express all the HPDs by linear 
transformations. In this
section we denote each element of $\Nil$ by the coordinate $(x,y,z)$.
In this notation the product of two elements is expressed as
\Beq
(a,b,c)(x,y,z)=(a+x, b+y, c+z+{ay-bx\over2}).
\Eeq

In terms of the generators
\Beq
\xi_1=\partial_x + {1\over2}y\partial_z,\quad
\xi_2=\partial_y - {1\over2}x\partial_z,\quad
\xi_3=\partial_z,
\Eeq
the structure of the Lie algebra $\LieA(\Nil)$ is expressed as
\Beq
[\xi_1,\xi_2]=-\xi_3,\quad
[\xi_3,\xi_1]=0,\quad
[\xi_3,\xi_2]=0.
\label{LieA:Nil}\Eeq
The corresponding invariant basis is given by
\Beqr
&& \chi^1=dx, \quad \chi^2=dy,\quad \chi^3=dz+{1\over2}(ydx-xdy),
\label{InvariantBasis:Nil}\\
&& d\chi^1=0,\quad d\chi^2=0,\quad d\chi^3=-\chi^1\wedge\chi^2.
\Eeqr
As in the case of $E^3$, $|\chi|=1$ holds for this basis.

The maximally symmetric metric on $\Nil$ is given by
\Beqr
&ds^2&=Q_1\left((\chi^1)^2+(\chi^2)^2\right)+Q_3(\chi^3)^2\nonumber\\
&&=Q_1(dx^2+dy^2)+Q_3[dz+{1\over2}(ydx-xdy)]^2,
\Eeqr
where $Q_1$ and $Q_3$ are positive constants. From this we find that
the maximal symmetry group of $\Nil$ is 4-dimensional and has
following decomposition:
\Beq
0\maps \Nil \maps \Isom(\Nil) \maps \r{O}(2) \maps 1 \quad
(\r{exact}),
\label{ExactSequence:Isom(Nil)}\Eeq
where O(2) is the group generated by rotations  around the $z$-axis,
$R_3(\theta)$, and the rotation of angle $\pi$ around the $x$-axis,
$R_1(\pi)$. Hence a generic transformation $f$ in $\Isom(\Nil)$ is
expressed as
\Beq
f(\bm{x})=\{1,R_1(\pi)\}\times R_3(\theta)\left[
\begin{matrix}{ccc}1&0&0 \\0&1&0\\-{1\over2}d^2&
{1\over2}d^1&0\end{matrix}\bm{x}
+\begin{matrix}{c}d^1\\d^2\\d^3\end{matrix}\right].
\Eeq
The above invariant basis transforms by $f$ as
\Beq
f^*\chi^I=A^I{}_J \chi^J;\quad
A=\{1,R_1(\pi)\}\times R_3(\theta).
\Eeq
$\Isom(\Nil)$ has two connected components, $\Isom_0(\Nil)$ and
$R_1(\pi)\times\Isom_0(\Nil)$, both of which preserve
orientation. 

The generators of $\LieA(\Isom(\Nil))$ are given by $\xi_I$($I=1,2,3$)
and 
\Beq
\xi_4=-y\partial_x + x\partial_y,
\Eeq
whose commutation relations with $\xi_I$ are 
\Beq
[\xi_4,\xi_1]=-\xi_2,\quad [\xi_4,\xi_2]=\xi_1,\quad [\xi_4,\xi_3]=0.
\label{LieA:Isom(Nil)}\Eeq
From these commutation relations an automorphism of $\LieA(\Isom(\Nil))$
is represented as
\Beq
\phi(\xi_1,\xi_2,\xi_3,\xi_4)=(\xi_1,\xi_2,\xi_3,\xi_4)
\begin{matrix}{ccc}\begin{matrix}{cc}k & 0\\ 0 &\pm k\end{matrix}
R(\theta) & \begin{array}{cc}0&0\\0&0\end{array}\\
\begin{array}{cc}a^3 & b^3\\0&0\end{array}&
\begin{array}{cc}\pm k^2 & d \\ 0&\pm 1\end{array}\end{matrix},
\Eeq
where $k>0$ and $R(\theta)\in\SO(2)$.
This automorphism is induced from $f$ in
$\HPDG^+(\Isom_0(\Nil))$ if $a^3,b^3$ and $d$ are written as
\Beq
(a^3,b^3)=\pm k(-d^2,d^1),\quad 
d=\pm{1\over2}\left((d^1)^2+(d^2)^2\right).
\Eeq
The explicit form of $f$ are
\Beq
f(\bm{x})=\{1,R_1(\pi)\}\times R_3(\theta)\left[
\begin{matrix}{ccc} k &0 &0\\0 & k & 0\\
-{k\over 2}d^2 & {k\over 2}d^1 & k^2\end{matrix}\bm{x}
+\begin{matrix}{c}d^1\\d^2\\d^3\end{matrix}\right],
\label{HPD:Isom(Nil)}\Eeq
where $d^3$ is an integration constant. This is a product of
a scaling represented by a diagonal matrix $[k,k, k^2]$ 
and an element of $\Isom(\Nil)$. 

\subsubsection{Transitive subgroups of $\Isom(\Nil)$}

Let $\tilde G$ be a connected transitive subgroup of $\Isom(\Nil)$.
Then from \Eq{ExactSequence:Isom(Nil)} its projection on O(2)
is either 1 or $\SO(2)$. In the former case $\tilde G$ should be 
equal to $\Nil$. In the latter case, $\tilde G\cap\Nil$ must have a
dimension greater than one. If the dimension is two, 
from \Eq{LieA:Nil}, it should be generated by $\xi_3$ and 
a linear combination of $\xi_1$ and $\xi_2$, $a\xi_1+b\xi_2$. 
As the third generator of $\tilde G$ we can take one with the
form $\xi'_4=\xi_4+c^I\xi_I$. However, since $[\xi'_4,a\xi_1+b\xi_2]$
is written as $-b\xi_1+a\xi_2+c\xi_3$ with some constant $c$, these
three generators do not form a closed algebra. Hence $\tilde G$
must coincide with $\Isom_0(\Nil)$. Thus the connected transitive
subgroups of $\Isom(\Nil)$ are given by $\Nil$ and $\Isom_0(\Nil)$.

Since every invariance subgroup of  $\Isom(\Nil)$ contains $\Nil$, 
homogeneous covering data $\tilde \Phi=(\tilde q,\tilde p)$
is expressed by their coefficient matrices $Q$ and $P$
with respect to the invariant basis (\ref{InvariantBasis:Nil}).
From the exact sequence (\ref{ExactSequence:Isom(Nil)}), if
the invariance group $\tilde G$ is not equal to $\Nil$, it should
contain an element of O(2), i.e., $R_1(\pi)$ or $R_3(\theta)$
or their product.
Hence by the same argument in \S3.1.2, if $\tilde G_0=\Nil$,
$\tilde G$ should be a product of $\Nil$ and a subgroup
of $D_2=\{1,R_1(\pi),R_2(\pi),R_3(\pi)\}$. On the other hand,
if $\tilde G_0=\Isom_0(\Nil)$, $\tilde G$ is either $\Isom_0(\Nil)$
or $\Isom(\Nil)$.

\subsubsection{Topology of orientable compact quotients}

Compact closed 3-manifolds modeled on $(\Nil,\Isom(\Nil))$
allow Seifert fibering with $\chi=0$ and $e\not=0$. Hence
we can determine all possible fundamental groups from
the fundamental groups of base orbifolds with $\chi=0$ and 
Seifert index.
They are classified into 7 types as listed in Table 
\ref{tbl:Nil:Pi1}, where their general embeddings into 
$\Isom(\Nil)$  are also given. 

The fundamental groups have similar
structure to those of type $E^3$ except that each has 
a parameter $n$ of positive integer. Six of them have 
counter parts in those of type $E^3$, and five of them
are covered by the simplest one denoted as $T^3(n)$. 
Their covering transformation groups coincide with the
corresponding ones for $E^3$ as will be explicitly shown 
later.

\begin{table}
\noindent
\begin{tabular}{ll}
\bf Space & \bf Fundamental group and  representation \\
\\
$T^3(n)$ 
& $<\alpha,\beta,\gamma| [\alpha,\gamma]=1,[\beta,\gamma]=1,
[\alpha,\beta]=\gamma^n>  \quad (n\in\NN)$ \\
& $\alpha=\bm{a},\beta=\bm{b},\gamma=(0,0,\Delta(a,b)/n)$;
 $\Delta(a,b)\not=0$\\
\\
$K^3(n)$ 
& $<\alpha,\beta,\gamma| [\alpha,\gamma]=1,
\beta\gamma\beta^{-1}\gamma=1,
\alpha\beta\alpha\beta^{-1}=\gamma^n>\quad(n\in\NN)$\\
& $\alpha=\bm{a},\beta=R_1(\pi)R_3(\theta)\bm{b},
\gamma=(0,0,\Delta(a,b)/n)$;\\
& $R(\theta)\begin{matrix}{c}a^1\\ a^2\end{matrix}=\begin{matrix}{c}-a^1\\
a^2\end{matrix},
\Delta(a,b)\not=0$\\
\\
$T^3(n)/\ZR_2$ 
& $<\alpha,\beta,\gamma| \gamma\alpha\gamma^{-1}\alpha=1,
\gamma\beta\gamma^{-1}\beta=1, [\alpha,\beta]=\gamma^{2n}>$\\
& $\alpha=(a^1,a^2,\Delta(a,c)/2),
\beta=(b^1,b^2,\Delta(b,c)/2),$\\
& $\gamma=R_3(\pi)(c^1,c^2,\Delta(a,b)/2n); \Delta(a,b)\not=0$\\
\\
$T^3(2n)/\ZR_2\times\ZR_2$ 
& $<\alpha,\beta,\gamma|
\alpha\gamma^2\alpha^{-1}\gamma^2=1,\gamma\alpha^2\gamma^{-1}\alpha^2=\gamma^{-2n}>
\quad (n\in \NN)$\\
& $\alpha=R_1(\pi)R_3(\theta)\bm{a},\gamma=R_3(\pi) \bm{c}$;\\
& $2n c^3=-(a^2,a^1)R(\theta)\begin{matrix}{c}a^1-c^1\\
a^2-c^2\end{matrix}+\Delta(a,c)$\\
\\
$T^3(n)/\ZR_3$ 
& $<\alpha,\beta,\gamma| \gamma\alpha\gamma^{-1}=\beta,
\gamma\beta\gamma^{-1}=\alpha^{-1}\beta^{-1},[\alpha,\beta]=\gamma^{3n}> 
\quad(n\in\NN)$\\
& $\alpha=\bm{a},\beta=\bm{b},\gamma=R_3(\pm{2\pi\over3}) \bm{c}$;
\\
&$\begin{matrix}{c}b^1\\ b^2\end{matrix}=R\begin{matrix}{c}a^1\\a^2\end{matrix},
\begin{matrix}{c}c^1\\c^2\end{matrix}
=\left({a^3-b^3\over\Delta(a,b)}R+{1\over2} 
-{a^3+2b^3\over\Delta(a,b)}\right)
\begin{matrix}{c}a^1\\a^2\end{matrix}$,\\
&$c^3={\Delta(a,b)\over 3n}+{1\over 6}(c^1,c^2)R
\begin{matrix}{c}-c^2\\c^1\end{matrix}$
with $R=R(\pm{2\pi\over3})$ \\
\\
$T^3(n)/\ZR_4$ 
& $<\alpha,\beta,\gamma| \gamma\alpha\gamma^{-1}=\beta^{-1},
\gamma\beta\gamma^{-1}=\alpha,[\alpha,\beta]=\gamma^{4n}> 
\quad(n\in\NN)$\\
& $\alpha=(a^1,a^2,0),\beta=(b^1,b^2,0),
\gamma=R_3(\pm{\pi\over2})(0,0,\Delta(a,b)/4n)$;\\
& $\begin{matrix}{c}b^1\\ b^2\end{matrix}=R(\pm{\pi\over2})
\begin{matrix}{c}a^1\\a^2\end{matrix}$\\
\\
$T^3(n)/\ZR_6$ 
& $<\alpha,\beta,\gamma| \gamma\alpha\gamma^{-1}=\beta,
\gamma\beta\gamma^{-1}=\alpha^{-1}\beta,[\alpha,\beta]=\gamma^{6n}> 
\quad(n\in\NN)$\\
& $\alpha=\bm{a},\beta=\bm{b},\gamma=R_3(\pm{\pi\over3})\bm{c}$;
\\
&$\begin{matrix}{c}b^1\\ b^2\end{matrix}=R\begin{matrix}{c}a^1\\a^2\end{matrix},
\begin{matrix}{c}c^1\\c^2\end{matrix}=\left(-{a^3\over\Delta(a,b)}
-{1\over2} +{a^3-b^3\over\Delta(a,b)}R\right)
\begin{matrix}{c}a^1\\a^2\end{matrix}$, \\
&$c^3={\Delta(a,b)\over 6n}+{1\over 2}(c^1,c^2)R
\begin{matrix}{c}-c^2\\c^1\end{matrix}$
with $R=R(\pm{\pi\over3})$\\
\\
\end{tabular}
\caption{\label{tbl:Nil:Pi1}Fundamental groups and their
representation in $\Isom^+(\Nil)$ of compact closed orientable
3-manifolds of type $\Nil$}
In this table $\Delta(a,b)=a^1b^2-a^2b^1$.
\end{table}

\subsection{$\tilde G_0=\Nil$}

The automorphism group of $\Nil$ consists of elements whose 
action on $\LieA(\Nil)$ is expressed as
\Beq
\phi(\xi_1,\xi_2,\xi_3)=(\xi_1,\xi_2,\xi_3)
\begin{matrix}{cc} \hat A &\begin{array}{c}0\\0\end{array}\\
\begin{array}{cc}a^3&b^3\end{array}& \Delta
\end{matrix};\quad
\Delta:=\det \hat A.
\Eeq
Every automorphism $\phi$ is induced from  
$f\in\HPDG^+(\Nil,\Nil)$ of the form
\Beq
f(\bm{x})=
\begin{matrix}{cc} \hat A &\begin{array}{c}0\\0\end{array}\\
(a^3,b^3)+{1\over2}(d^2,-d^1)\hat A & \Delta
\end{matrix}\bm{x}
+\begin{matrix}{c}d^1\\ d^2\\ d^3\end{matrix}.
\label{HPD:Nil}\Eeq
The invariant basis (\ref{InvariantBasis:Nil}) transforms
by $f$ as
\Beq
f^*\chi^I=A^I{}_J\chi^J;\quad
A=
\begin{matrix}{cc} \hat A &\begin{array}{c}0\\0\end{array}\\
(a^3,b^3)+(d^2,-d^1)\hat A & \Delta
\end{matrix}.
\Eeq

By these HPDs the metric matrix $Q$ can be put in the
diagonal form $[1,1,Q_3]$. For this metric the diffeomorphism
constraint \Eq{DC:H} is expressed as
\Beq
P^{13}=0, \quad P^{23}=0.
\Eeq
Hence the momentum matrix can be diagonalized as $P=[P^1,P^2,P^3]$ 
by residual HPDs keeping the form of $Q$. Hence the invariance
group of covering data $\tilde \Phi$ always contains
$\Nil\sdp D_2$.

For $f$ given by \Eq{HPD:Nil}, the condition $fD_2 f^{-1}\subset
\Nil\sdp D_2$ holds
if and only if the matrix $\hat A$ is of the form
\Beq
\hat A=\begin{matrix}{cc}p&0 \\ 0 & q\end{matrix},\quad
\begin{matrix}{cc}0&p\\ q&0\end{matrix},
\Eeq
and $(a^3,b^3)=(-d^2,d^1)\hat A$. Hence $f\in\HPDG^+(\Nil,\Nil\sdp D_2)$ is expressed as
\Beq
f(\bm{x})=
\begin{matrix}{cc} \hat A &\begin{array}{c}0\\0\end{array}\\
{1\over2}(-d^2,d^1)\hat A & \Delta
\end{matrix}\bm{x}
+\begin{matrix}{c}d^1\\ d^2\\ d^3\end{matrix},
\label{HPD:NilxD2}\Eeq
by which the invariant basis transforms as
\Beq
f^*\chi^I=A^I{}_J\chi^J;\quad
A=
\begin{matrix}{cc} \hat A &\begin{array}{c}0\\0\end{array}\\
\begin{array}{cc}0 &0\end{array}& \Delta
\end{matrix}.
\Eeq

Covering data invariant under $\Nil\sdp D_2$ are represented by
diagonal matrices $Q=[Q_1,Q_2,Q_3]$ and $P=[P^1,P^2,P^3]$. 
If we transform $Q$ to the $\Isom(\Nil)$-invariant form, 
$[1,1,Q_3/Q_1Q_2]$, by HPDs of $\Nil\sdp D_2$, $P$
transforms to $[Q_1P^1,Q_2P^2,Q_1Q_2P^3]$, which becomes
$\Isom_0(\Nil)$-invariant only when $Q_1P^1=Q_2P^2$. Hence
from Prop.\ref{prop:InvG}, we obtain
\Beq
\Gamma^+_\r{H,D}(\Nil,\Nil\sdp D_2)
=\SetDef{Q=[Q_1,Q_2,Q_3],P=[P^1,P^2,P^3]}
{Q_1P^1\not=Q_2P^2}.
\Eeq

For the covering metric $\tilde q$ in this phase space, 
the Ricci scalar curvature is given by
\Beq
R=-{Q_3\over 2Q_1Q_2}.
\Eeq
Hence in terms of $\alpha$ and $\beta_\pm$ defined by
\Eqs{def:alphabeta:1}{def:alphabeta:2}, and $p_\alpha$ 
and $p_\pm$ defined by \Eqs{def:palpha}{def:p+}, 
the Hamiltonian is written as
\Beq
H={\kappa^2\over 12\Omega}Ne^{-3\alpha}
\left[-p_\alpha^2+p_+^2+p_-^2
+{3\Omega^2\over\kappa^2}e^{4(\alpha-2\beta_+)}\right],
\label{H:NilxD2}\Eeq
where $\Omega$ is the volume of the canonical fundamental 
region to be determined below.

Next we determine the canonical 1-form for each topology.
Since the rotation matrices of angle $\pm 2\pi/k$ do not
belong to $\Nil\sdp D_2$ for $k=3,4,6$, $T^3(n)/\ZR_k$
($k=3,4,6$) do not allow locally $\Nil\sdp D_2$-invariant
data as in the case of $E^3$.

\subsubsection{$T^3(n)$}

For this space the fundamental group is embedded in $\Nil$,
as is shown in Table \ref{tbl:Nil:Pi1}, and the moduli 
is determined by the two generators $\alpha=(a^1,a^2,a^3)$
and $\beta=(b^1,b^2,b^3)$. Since $(p,q,r)\in \Nil$ 
is transformed by HPDs (\ref{HPD:NilxD2}) as
\Beq
f(p,q,r)f^{-1}=\left((p,q)\Tp{\hat A},\Delta r + 
\begin{matrix}{cc}p &q\end{matrix}\Tp{\hat A}
\begin{matrix}{c}-d^2\\d^1\end{matrix}\right),
\Eeq
$a^3$ and $b^3$ can be put to zero by HPDs with $\hat A=1$.
Further by using the freedom of $\hat A$, we can transform
the moduli matrix 
$\begin{matrix}{ccc}\bm{a}&\bm{b}&\bm{c}\end{matrix}$ to
\Beq
B=\begin{matrix}{ccc}1&X & 0\\0&1&0\\0&0&1/n\end{matrix}
R_3(\phi).
\label{ModuliMatrix:T3(n)}\Eeq
Then the HPD freedom is eliminated except for residual 
discrete ones. There remains the freedom of modular
transformations among $\alpha$ and $\beta$ represented
by $\SL(2,\ZR)$ matrix as well.

Taking $K_0$ with $X=0$ and $\phi=0$ as the base point,
a deformation map $f_\lambda$ from $K_0$ to a generic
moduli $K$ is given by 
\Beq
f_\lambda(\bm{x})=F\bm{x};\quad
F=\begin{matrix}{ccc}1&X & 0\\0&1&0\\0&0&1\end{matrix}
R_3(\phi).
\label{DeformationMap:T3(n)}\Eeq
For the moduli $K_0$ its generic element
is written as
\Beq
\alpha^p\beta^q\gamma^r=(p,q,{r\over n}),
\Eeq
where $p,q,r\in\ZR$. Hence the fundamental region $D_0$ of $K_0$
is given by $0\le x,y\le1$ and $0\le z\le 1/n$, and its volume
by $\Omega=1/n$.  

Inserting the above $F$ into \Eq{Theta:LH:general}, we obtain
\Beq
\Theta=\Omega\sqrt{Q_1Q_2Q_3}\left[
\Tr(\dot QP) + 2\dot\phi X(Q_1P^1-Q_2P^2)\right],
\Eeq
which is written in terms of $\alpha$, $\beta_\pm$, $p_\alpha$,
$p_\pm$ and $p_\phi$ defined by
\Beq
p_\phi:={1\over\sqrt{3}} p_- X
\Eeq
as
\Beq
\Theta=\dot\alpha p_\alpha +\dot\beta_+p_+ +\dot\beta_- p_-
+ \dot\phi p_\phi.
\Eeq
Thus no degeneracy occurs in the canonical structure.

\subsubsection{$K^3(n)$}

From Table \ref{tbl:Nil:Pi1}, the generator $\beta$ is a product
of a transformation $\bm{b}$ in Nil and $R_\beta=R_1(\pi)R_3(\theta)$,
which is a rotation of angle $\pi$ around an axis in the $x-y$ plane.
Hence, in order for $\beta$ to be in $\Nil\sdp D_2$, $R_\beta$ should
coincides with $R_1(\pi)$ or $R_2(\pi)$. Since these two cases are
connected by the HPD with 
$\hat A=\begin{matrix}{cc}0&-1\\1&0\end{matrix}$, we can put
$R_\beta=R_2(\pi)$, i.e. $\theta=\pi$, 
which implies $a^2=0$ from the definition of
$R(\theta)$ in Table \ref{tbl:Nil:Pi1}. By the HPD given by
\Eq{HPD:NilxD2} with
\Beqr
&& \hat A=\begin{matrix}{cc}1/a^1&0\\ 0 & 1/b^2\end{matrix},\\
&& (d^1,d^2,d^3)=\left({b^1\over 2a^1},{a^3\over a^1b^2},
{b^3\over 2a^1b^2}-{a^3b^1\over 4 (a^1)^2 b^2}\right)
\Eeqr
the corresponding moduli is transformed into one with
the moduli matrix
\Beq
\begin{matrix}{ccc}\bm{a}&\bm{b}&\bm{c}\end{matrix}
=\begin{matrix}{ccc}1& 0& 0\\ 0 & 1 & 0\\ 0& 0 & 1/n \end{matrix}.
\Eeq
Hence the reduced moduli space consists of one point.

Transformations in the discrete group $K_0$ are expressed in
one of the following two forms:
\Beqr
&& \alpha^p\beta^{2q}\gamma^r=(p,2q,{r\over n}+pq),\\
&& \alpha^p\beta^{2p+1}\gamma^r=(p,2q+1,-{r\over n}+{1\over2}
p(2q+1))R_2(\pi),
\Eeqr
where $p,q,r\in\ZR$. By these transformation, any point in $\Nil$
can be moved into the region $0\le x,y\le1$. All the transformations
in $K_0$ that leave this region invariant are written in the form
\Beq
\gamma^r (x,y,z)=(x,y,z+{r\over n}).
\Eeq
Hence the fundamental region $D_0$ of $K_0$ is given by
$0\le x,y\le1$ and $0\le z\le 1/n$, and its volume
by $\Omega=1/n$. Since there exists no moduli freedom,
the canonical 1-form is given by
\Beq
\Theta=\dot\alpha p_\alpha + \dot\beta_+ p_+ 
+\dot\beta_-p_-.
\label{Theta:K3(n)}\Eeq

\subsubsection{$T^3(n)/\ZR_2$}

In this case all the generators of the moduli are always contained
in $\Nil\sdp D_2$. It is easy to see that by a similar argument 
in the case $T^3(n)$, HPDs transform a generic moduli matrix
to the form (\ref{ModuliMatrix:T3(n)}) with $n$ replaced by
$2n$. Hence the deformation map is given by \Eq{DeformationMap:T3(n)}
with the same replacement.

Transformations in $K_0$ are written as
\Beq
\alpha^p\beta^q\gamma^r=(p,q,{r\over 2n}+{1\over 2}pq)R_3(\pi r).
\Eeq
Hence by the transformations of the forms $\alpha^p\beta^q$,
we can move any point in $\Nil$ to the region $|x|,|y|\le 1/2$.
This region is left invariant only by the transformations 
of the form $\gamma^r$. It is easy to see that we can move any
point to $0\le z\le 1/2n$ by these transformations. Hence 
the fundamental region $D_0$ of $K_0$ is given by 
$|x|,|y|\le1/2$ and $0\le z\le 1/2n$, and its volume by
$\Omega=1/2n$. The canonical 1-form and the Hamiltonian 
are given by the same expressions for $T^3(n)$. 

Finally note that $\alpha$, $\beta$ and $\gamma^2$ satisfy
the same relations as those among $\alpha$, $\beta$ and $\gamma$
for $T^3(n)$. Hence we obtain the exact sequence for the 
fundamental group of the present manifold $M$, 
\Beq
1\maps \pi_1(T^3(n)) \maps \pi_1(M) \maps \ZR_2 \maps 1,
\Eeq
which implies that $M\approx T^3(n)/\ZR_2$.

\subsubsection{$T^3(2n)/\ZR_2\times\ZR_2$}

In this case, in order for $\alpha$ to be in $\Nil\sdp D_2$,
$R_\alpha=R_1(\pi)R_3(\theta)$ should be $R_1(\pi)$ or $R_2(\pi)$.
Hence, as in the previous case, we can put $R_\alpha=R_1(\pi)$
by a HPD. Further by the freedom $d^1,d^2$ and $d^3$ of HPDs,
we can put $a^3,c^1$ and $c^2$ to zero. Then the HPD
with $\hat A=\begin{matrix}{cc}1/a^1&0\\0& -1/a^2\end{matrix}$ brings
the generators to canonical forms
\Beq
\alpha=R_1(\pi)(1,-1,0),\quad
\gamma=R_3(\pi)(0,0,1/n).
\Eeq
Hence the moduli space is reduced to a single point. 

To determine the fundamental region of this moduli $K_0$, 
let us introduce the transformation $\beta$ defined by
\Beq
\beta:=\alpha^{-1}\gamma=R_2(\pi)(1,1,1/n).
\Eeq
Then we obtain
\Beq
[\alpha^2,\gamma^2]=1,\quad
[\beta^2,\gamma^2]=1,\quad
[\alpha^2,\beta^2]=(\gamma^2)^{2n}.
\Eeq
From these relations we easily see that any transformation $g$
in $K_0$ is expressed in the form 
\Beq
g=\alpha^{2p}\beta^{2q}\gamma^{2r}\times\{1,\alpha,\beta,\gamma\}.
\Eeq
In particular, since $\{1,\alpha,\beta,\gamma\}$ forms the Klein 
group $\mod(\alpha^2,\beta^2,\gamma^2)$, we obtain the
following exact sequence for the fundamental group of the
present manifold $M$:
\Beq
1\maps \pi_1(T^3(2n))\maps \pi_1(M)\maps \ZR_2\times\ZR_2\maps 1,
\Eeq
which justifies our notation for $M$. 

The fundamental region $D_0$ is determined as follows. First
by transformations of the form $\alpha^{2p}\beta^{2q}
\times\{1,\alpha\}$, we can move any point to the region
$0\le x\le 1$, and by those of $\{1,\alpha^2\}\times \beta^{2q}$
to $0\le y\le1$. Only transformations which leave the
region $0\le x,y\le 1$ invariant are $\gamma^{2r}=(0,0,2r/n)$. Hence
$D_0$ is given by $0\le x,y\le1$ and $0\le z\le 2/n$, 
and its volume by $\Omega=2/n$. Since there exists
no moduli freedom, the canonical 1-form is given by
the same expression as \Eq{Theta:K3(n)} for $K^3(n)$.

\begin{table}
\noindent
\begin{tabular}{llllc}
\bf Symmetry & \bf Topology &\bf Moduli & $\Omega$ & $\Theta$ \\
\\
Nil$\sdp$D$_2$ 
& $T^3(n)$ 
& $\begin{matrix}{ccc}1& X& 0\\ 0 & 1 & 0\\ 0& 0 & 1/n
\end{matrix}R_3(\phi)$
& ${1\over n}$ 
& $\begin{array}{l}\dot\alpha p_\alpha +\dot\beta_+p_+ 
+\dot\beta_- p_-\\ + \dot\phi p_\phi\end{array}$ \\
& $K^3(n)$ 
& $\begin{matrix}{ccc}1& 0& 0\\ 0 & 1 & 0\\ 0& 0 & 1/n \end{matrix} $
& ${1\over n}$ 
&$\dot\alpha p_\alpha +\dot\beta_+p_++\dot\beta_- p_-$\\
&& $(R_\beta=R_2(\pi))$ &&\\
& $T^3(n)/\ZR_2$ 
& $\begin{matrix}{ccc}1& X& 0\\ 0 & 1 & 0\\ 0& 0 &
1/n\end{matrix}R_3(\phi)$
& ${1\over 2n}$ 
& $\begin{array}{l}\dot\alpha p_\alpha +\dot\beta_+p_+ +\dot\beta_-
p_-\\ + \dot\phi p_\phi\end{array}$ \\
&& $(R_\gamma=R_3(\pi))$&&\\
& $T^3(2n)/\ZR_2\times\ZR_2$ 
& $\begin{matrix}{ccc}1& 1& 0\\ -1 & 1 & 0\\ 0 &1/n&1/n \end{matrix}$ 
& $2\over n$ 
&$\dot\alpha p_\alpha +\dot\beta_+p_++\dot\beta_- p_-$\\
&& ($R_\alpha=R_1(\pi),R_\beta=R_2(\pi),$ &&\\
&& $R_\gamma=R_3(\pi)$) && \\
\\
\end{tabular}
\caption{\label{tbl:CanonicalStr:Nil}Canonical structure of compact
orientable closed 3-manifold of type Nil}
\end{table}

\begin{table}
\noindent
\begin{tabular}{llllc}
\bf Symmetry & \bf Topology &\bf Moduli & $\Omega$ & $\Theta$ \\
\\
Isom(Nil) 
& $T^3(n)$ 
& $\begin{matrix}{ccc}X& Y& 0\\ 0 & X^{-1} & 0\\ 0& 0 &1/n\end{matrix}$
& $1\over n$ 
& $\dot\alpha p_\alpha +\dot\beta_+ p_+$\\
&&($X>0$) 
&& \\
& $K^3(n)$ 
& $\begin{matrix}{ccc}X& 0& 0\\ 0 & X^{-1} & 0\\ 0& 0 &1/n\end{matrix}$
& $1\over n$ & ${}''$\\
&&($X>0, R_\beta=R_3(\pi)$) 
&& \\
& $T^3(n)/\ZR_2$ 
& $\begin{matrix}{ccc}X& Y& 0\\ 0 & X^{-1} & 0\\ 0& 0 & 1/2n
\end{matrix}$ 
& $1\over 2n$  & ${}''$\\
&& $(X>0,R_\gamma=R_3(\pi))$
&&\\
& $T^3(2n)/\ZR_2\times\ZR_2$ 
& $\begin{matrix}{ccc}X& X& 0\\ -X^{-1} & X^{-1} & 0\\ 0 & 1/n
& 1/n \end{matrix}$ 
& $2\over n$ & ${}''$\\
&& ($X>0,R_\alpha=R_1(\pi)$,&&\\
&& $R_\beta=R_2(\pi),R_\gamma=R_3(\pi)$) 
&& \\
& $T^3(n)/\ZR_3$ 
& $\begin{matrix}{ccc}1 & -1/2& 1/2 \\
0 & \sqrt{3}/2 & 0\\ 0&0& {\sqrt{3}\over 6n}-{\sqrt{3}\over48}\end{matrix}$
& $1\over 4n$ & ${}''$\\
&&($R_\gamma=R_3({2\pi/3})$) 
&&\\
& $T^3(n)/\ZR_4$ 
& $\begin{matrix}{ccc}1 & 0& 0 \\ 0 & 1 & 0\\ 0&0& 1/4n\end{matrix}$
& $1\over 4n$ & ${}''$\\
&&($R_\gamma=R_3({\pi/2})$) 
&&\\
& $T^3(n)/\ZR_6$ 
& $\begin{matrix}{ccc}1 & 1/2& -1/2 \\
0 & \sqrt{3}/2 & 0\\ 0&0& {\sqrt{3}\over 12n}-{\sqrt{3}\over16}\end{matrix}$
& $1\over 8n$ & ${}''$\\
&&($R_\gamma=R_3({\pi/3})$) 
&&\\
\end{tabular}
\par
\begin{center}
Table \ref{tbl:CanonicalStr:Nil}(continued)
\end{center}
\end{table}

\subsection{$\tilde G_0=\Isom_0(\Nil)$}

From the arguments in \S4.1, 
$\Isom_0(\Nil)$-invariant data are always invariant by $\Isom(\Nil)$,
and the corresponding phase space is given by
\Beq
\Gamma^+_\r{H,D}(\Nil,\Isom(\Nil))=
\left\{Q=[Q_1,Q_1,Q_3],P=[P^1,P^2,P^3]\right\}.
\Eeq
HPDs (\ref{HPD:Isom(Nil)}) for $\Isom(\Nil)$ transform
the invariant basis (\ref{InvariantBasis:Nil}) as
\Beq
f^*\chi^I=A^I{}_J\chi^J;\quad
A=\{1,R_1(\pi)\}
\begin{matrix}{cc} kR & \begin{array}{c} 0\\0\end{array}\\
\begin{array}{cc}0&0\end{array} & k^2\end{matrix}.
\Eeq

As in the case of $E^3$ with $\tilde G=\RF^3\sdp\SO(2)$,
the Hamiltonian for the present case is obtained 
from that for $\tilde G=\Nil\sdp D_2$,
\Eq{H:NilxD2},  simply by
putting $Q_1P^1=Q_2P^2$, or $\beta_-=0$ and $p_-=0$:
\Beq
H={\kappa^2\over 12\Omega}Ne^{-3\alpha}
\left[-p_\alpha^2+p_+^2
+{3\Omega^2\over\kappa^2}e^{4(\alpha-2\beta_+)}\right].
\label{H:Isom(Nil)}
\Eeq

The arguments to determine the reduced moduli space and
the canonical 1-form are almost the same as those for 
$\tilde G=\Nil\sdp D_2$ except for $T^3(n)/\ZR_k$($k=3,4,6$).
Hence we just outline the derivation for $T^3(n)$,
$K^3(n)$, $T^3(n)/\ZR_2$ and $T^3(n)/\ZR_2\times\ZR_2$.

\subsubsection{$T^3(n)$}

The moduli matrix is reduced by HPDs to the canonical form
\Beq
\begin{matrix}{ccc}\bm{a}&\bm{b}&\bm{c}\end{matrix}=\begin{matrix}{ccc}X& Y& 0\\ 0 & X^{-1} & 0\\ 0& 0 &1/n\end{matrix}.
\Eeq
There exists no freedom of residual HPDs except in the subspace
$Y=0$ for which the transformation $X\maps X^{-1}$ gives
the residual HPD. There remain modular transformations
isomorphic to $\ZR$.
For the base point $K_0$ with $X=1$ and $Y=0$, the deformation
map is given by
\Beq
f_\lambda(\bm{x})=F\bm{x};\quad
F=\begin{matrix}{ccc}X& Y& 0\\ 0 & X^{-1} & 0\\ 0& 0 &1\end{matrix}.
\Eeq
For this transformation $\Tr\dot FF^{-1}QP$ vanishes. Hence
the canonical structure is degenerate in the moduli sector, and
the canonical 1-form is given by
\Beq
\Theta=\dot\alpha p_\alpha + \dot\beta_+ p_+.
\label{Theta:T3(n):Isom(Nil)}\Eeq
The volume of the fundamental region is $\Omega=1/n$.

\subsubsection{$K^3(n)$}

By the rotation $R_3(-\pi/2+\theta/2)$ in $\HPDG^+(\Nil,\Isom(\Nil))$,
$\alpha$ and $\beta$ transform to generators of the forms 
$(a^1,0,a^3)$ and $R_2(\pi)(b^1,b^2,b^3)$, respectively. 
Further by a transformation $\bm{d}\in\Nil$, they can be
put into the form $(a^1,0,0)$ and $R_2(\pi)(0,b^2,0)$, and 
by rotations $R_1(\pi)$ and $R_2(\pi)$ if necessary, $a^1$ and
$b^2$ can be made positive.  Hence the reduced moduli matrix is
given by
\Beq
\begin{matrix}{ccc}\bm{a}&\bm{b}&\bm{c}\end{matrix}=\begin{matrix}{ccc}X& 0& 0\\ 0 & X^{-1} & 0\\ 0& 0 &1/n\end{matrix}.
\Eeq
There exists no residual freedom of modular transformations, but
the discrete HPD $X\maps X^{-1}$ remains.

For the base point $K_0$ with $X=1$, the deformation map is
given by
\Beq
f_\lambda(\bm{x})=F\bm{x};\quad
F=\begin{matrix}{ccc}X& 0& 0\\ 0 & X^{-1} & 0\\ 0& 0 &1\end{matrix},
\Eeq
which is a special form of that in the previous case. Hence
the canonical structure becomes degenerate in the moduli sector
again, and the canonical 1-form is given by 
\Eq{Theta:T3(n):Isom(Nil)}. $\Omega$ is equal to $1/n$.

\subsubsection{$T^3(n)/\ZR_2$}

By the freedom of $\Nil$-translation $\bm{d}$ of HPDs, 
$c^1$ and $c^2$ can be put to zero. Hence by rotations 
around the $z$-axis and scalings in $\HPDG^+(\Nil,\Isom(\Nil))$,
the moduli matrix is put in the form
\Beq
\begin{matrix}{ccc}\bm{a}&\bm{b}&\bm{c}\end{matrix}=\begin{matrix}{ccc}X& Y& 0\\ 0 & X^{-1} & 0\\ 0& 0 &1/2n\end{matrix}.
\Eeq
This is obtained from that for $T^3(n)$ by replacing $n$ by $2n$,
and the same deformation map can be used.
Hence the results for $T^3(n)$ apply to the present case 
by the same replacement.

\subsubsection{$T^3(2n)/\ZR_2\times\ZR_2$}

By a translation parallel to the $x-y$ plane and a rotation
around the $z$-axis, $\alpha$ and $\gamma$ is put into
the forms $R_1(\pi)(a^1,a^2,0)$ and $R_3(\pi)(0,0,-a^1a^2/n)$,
respectively. By the rotation $R_1(\pi)$ and scaling they
are further transformed to $R_1(\pi)(X,-1/X,0)$ and 
$R_3(\pi)(0,0,1/n)$. Hence by introducing $\beta$ defined by
\Beq
\beta:=\alpha^{-1}\gamma=R_2(\pi)(X,1/X,1/n),
\Eeq
we obtain the following canonical form of the moduli matrix:
\Beq
\begin{matrix}{ccc}\bm{a}&\bm{b}&\bm{c}\end{matrix}
=\begin{matrix}{ccc}X& X& 0\\ -X^{-1} & X^{-1} & 0\\ 0 & 1/n
& 1/n \end{matrix}.
\Eeq
The deformation map coincides with that for $K^3(n)$. Therefore
the canonical 1-form is given by \Eq{Theta:T3(n):Isom(Nil)}, and 
$\Omega$ by $2/n$. 

\subsubsection{$T^3(n)/\ZR_k$($k=3,4,6$)}

By translations, scalings and rotations in 
$\HPDG^+(\Nil,\Isom(\Nil))$, we can completely eliminate the
continuous freedom of the moduli. Further by $R_1(\pi)$ 
we can transform $R_\gamma=R_3(-2\pi/k)$ to $R_3(2\pi/k)$.
Hence the moduli space is reduced to a single point
whose moduli matrix is given in Table 
\ref{tbl:CanonicalStr:Nil}.

The fundamental region is determined as follows.
First from the commutation relations in Table \ref{tbl:Nil:Pi1}
we obtain
\Beq
[\gamma^k,\alpha]=1,\quad
[\gamma^k,\beta]=1,\quad
[\alpha,\beta]=(\gamma^k)^n,
\Eeq
which leads to the exact sequence
\Beq
1\maps \pi_1(T^3(n))\maps \pi_1(M)\maps \{1,\gamma,\cdots,\gamma^{k-1}\}\maps 1.
\Eeq
Since $\{1,\gamma,\cdots,\gamma^{k-1}\}$ is isomorphic to $\ZR_k$
$\mod \gamma^k$, we obtain $M\approx T^3(n)/\ZR_k$.

In the case $k=3$, transformations in $K_0$ corresponding to 
the above moduli matrix are given by one of the following three:
\Beqr
&\alpha^p\beta^q\gamma^{3r}=
&\left(p-{q\over2},{\sqrt{3}\over2}q,{\sqrt{3}\over4}pq
+{\sqrt{3}\over 2n}r\right),\\
&\alpha^p\beta^q\gamma^{3r+1}=
&\left(p-{q\over2}-{1\over4},{\sqrt{3}\over2}q+{\sqrt{3}\over 4},
{\sqrt{3}\over4}pq+{\sqrt{3}\over 2n}(r+{1\over3})
+{\sqrt{3}\over8}p-{\sqrt{3}\over 48}\right)\nonumber\\
&&\quad\times R_3\left(2\pi\over3\right),\\
&\alpha^p\beta^q\gamma^{3r+2}=
&\left(p-{q\over2}-{1\over2},{\sqrt{3}\over2}q,
{\sqrt{3}\over4}pq+{\sqrt{3}\over 2n}(r+{2\over3})
-{\sqrt{3}\over8}q+{\sqrt{3}\over 48}\right)\nonumber\\
&&\quad \times R_3\left(-2\pi\over3\right).
\Eeqr
From this we find that the hexagon with the vertices
\Beq
\begin{matrix}{c}x_j\\ y_j\end{matrix}
=R\left({j\pi\over3}\right)\begin{matrix}{c}0\\ {1/\sqrt{3}}
\end{matrix}
+\begin{matrix}{c}-1/4\\ \sqrt{3}/4\end{matrix}
\quad(j=0,\cdots,5)
\Eeq
gives a fundamental region in the $x-y$ plane. Further
$\gamma^r$ leaves this region invariant. Hence noting
that $\gamma^3$ maps $z=0$ plane to $z=\sqrt{3}/2n$ plane,
we find that the volume of the fundamental region is
given by $\Omega=1/4n$.

Next in the case $k=4$, since the generators are simply given by
\Beq
\alpha=(1,0,0),\quad
\beta=(0,1,0),\quad
\gamma=R_3(\pi/2)(0,0,1/4n),
\Eeq
the fundamental region is given by $0\le x,y\le 1, 0\le z\le 1/4n$,
and its volume by $\Omega=1/4n$.

Finally in the case $k=6$, we find by an argument similar to the 
case $k=3$ that the hexagon with vertices
\Beq
\begin{matrix}{c}x_j\\ y_j\end{matrix}
=R\left({j\pi\over3}\right)\begin{matrix}{c} {1/\sqrt{3}}\\0
\end{matrix}
+\begin{matrix}{c}1\\ -\sqrt{3}\end{matrix}
\quad(j=0,\cdots,5)
\Eeq
gives a fundamental region in the $x-y$ plane, and $\gamma^6$
maps $z=0$ plane to $z=\sqrt{3}/2n$ plane. Hence 
the volume of the fundamental region is give by
$\Omega=1/8n$. 

In all the cases $\Theta$ and $H$ are give by the same
expressions as those for $T^3(n)$.

\begin{table}
\noindent
\begin{tabular}{llccccccccc}
\bf Symmetry &\bf Space & $Q$ & $P$ & $N_\r{m}$ & $N$ & $N_\r{c}$ &
$N_\r{d}$ & $N_\r{cc}$ & \small HPD & \small Modular\\
&&&&&&&&\\
$\Nil\sdp D_2$ 
& $T^3(n)$ 		& 3 & 3 & 2 & 8 & 8 & 0 & 1 & $\bigcirc$ & $\bigcirc$\\
& $K^3(n)$ 		& 3 & 3 & 0 & 6 & 6 & 0 & 1 & $\times$ &$\times$ \\
& $T^3(n)/\ZR_2$	& 3 & 3 & 2 & 8 & 8 & 0 & 1 & $\bigcirc$ & $\bigcirc$\\
& $T^3(2n)/\ZR_2\times\ZR_2$& 3 & 3 & 0 & 6 & 6 & 0 & 1& $\times$ &$\times$\\
&&&&&&&&\\
$\Isom(\Nil)$
& $T^3(n)$		& 2 & 2 & 2 & 6 & 4 & 2 & 1 & $\triangle$ & $\bigcirc$ \\
& $K^3(n)$ 		& 2 & 2 & 1 & 5 & 4 & 1 & 1 & $\bigcirc$ & $\times$\\
& $T^3(n)/\ZR_2$	& 2 & 2 & 2 & 6 & 4 & 2 & 1 & $\triangle$ & $\bigcirc$\\
& $T^3(n)/\ZR_2\times\ZR_2$& 2 & 2 & 1 & 5 & 4 & 1 & 1 &$\times$ &$\times$ \\
& $T^3/\ZR_k(k=3,4,6)$	& 2 & 2 & 0 & 4 & 4 & 0 & 1 &$\times$ &$\times$   \\
&&&&&&&&\\
\end{tabular}
\caption{\label{tbl:freedom:Nil}Canonical and degenerate degrees of
freedom for LHS of type $\Nil$}
\end{table}

\subsection{Summary}

As a summary of the results, the form of the reduced moduli matrix,
the volume $\Omega$ of the fundamental region for the base moduli 
point and the canonical 1-form $\Theta$ are listed for each
invariance group and topology in Table 
\ref{tbl:CanonicalStr:Nil}. 
Further the degrees of freedom of the covering data $Q$ and $P$, the
dimension $N_\r{m}$ of the reduced moduli space, the total dimension
$N$ of the invariant phase space, the non-degenerate and the
degenerate degrees of freedom of the canonical variables, $N_\r{c}$
and $N_\r{d}$, and the number of connected components of the invariant
phase space are listed in Table \ref{tbl:freedom:Nil}.

From these tables we see that the canonical structure of
$\Gamma^+_\r{LH,inv}(\Nil,\Nil\sdp D_2)$ is similar to that of
$\Gamma^+_\r{LH,inv}(E^3,\RF^3\sdp D_2)$, and no degeneracy occurs. On
the other hand, for $\Gamma^+_\r{LH,inv}(\Nil,\Isom(\Nil))$, the
canonical structure becomes completely degenerate in the moduli
sector. Though the dimensions of the phase space and the reduced
moduli are different, the degrees of degeneracy coincide with those
for $\Gamma^+_\r{LH,inv}(E^3,\RF^2\sdp\SO(2)\sdp D_2)$. This suggests
that in these cases the degeneracy of the canonical structure is
closely related with the structure of the isotropy groups of the
invariance groups, although the $\Gamma^+_\r{LH,inv}(\Nil,\Isom(\Nil))$
has no simple relation to $\Gamma^+_\r{LH,inv}(\Nil,\Nil\sdp D_2)$ as in
the case of $E^3$.

\section{LHS of type $\Sol$}

In this section we determine the canonical structure of locally
homogeneous pure gravity systems on compact closed orientable
manifolds of type $\Sol$.

\subsection{Basic properties}

\subsubsection{$\Gmax $ and $\HPDG^+(\Gmax )$}

Sol is a simply connected 3-dimensional group classified
as type VI$(0)$ in the Bianchi scheme, and homeomorphic to
$\RF^3$. We adopt the parameterization of the group such that
the product of two elements are written as
\Beq
(a,b,c)(x,y,z)=(a+e^{-c}x, b+e^c y, c+z).
\Eeq
In terms of the generators of its Lie algebra,
\Beq
\xi_1=\partial_x, \quad
\xi_2=\partial_y, \quad
\xi_3=\partial_z-x\partial_x+y\partial_y,
\Eeq
its structure is expressed as
\Beq
[\xi_1,\xi_2]=0,\quad
[\xi_3,\xi_1]=\xi_1,\quad
[\xi_3,\xi_2]=-\xi_2.
\Eeq
As the invariant basis we adopt
\Beqr
&& \chi^1=e^z dx+e^{-z}dy, \quad 
\chi^2=e^z dx - e^{-z}dy,\quad 
\chi^3=dz, 
\label{InvariantBasis:Sol}\\
&& d\chi^1=\chi^3\wedge\chi^2,\quad
d\chi^2=\chi^3\wedge\chi^1,\quad
d\chi^3=0,
\Eeqr
which is obtained by rotation of angle $\pi/4$ around 
the $z$-axis from the natural invariant basis.

The maximally symmetric metric on $\Sol$ is given by
\Beqr
&ds^2&=Q_1[(\chi^1)^2+(\chi^2)^2]+Q_3(\chi^3)^2 \nonumber\\
&& =2Q_1(e^{2z}dx^2+e^{-2z}dy^2) + Q_3 dz^2,
\Eeqr
where $Q_1$ and $Q_3$ are arbitrary positive constants. From
this we see that 
$\Isom(\Sol)$ has 8 connected components, and $\Sol$ is its
maximal connected subgroup. The other discrete components
are represented by the product of $\Sol$ and elements of
the discrete subgroup of rank 8 generated by $R_3(\pi)$,
$-R_1(\pi)$ and $J$ defined by
\Beq
J=\begin{matrix}{ccc}0&1&0\\1&0&0\\0&0&-1\end{matrix}
\Eeq
which is the rotation of angle $\pi$ around the line
$x=y,z=0$.  Among these 8 components, four of them consist
of orientation-reversing transformations, while the other
four give the maximal orientation-preserving symmetry 
group $\Isom^+(\Sol)=\Sol\sdp \{1,R_3(\pi),J,JR_3(\pi)\}
\cong \Sol \sdp D_2$. $f\in \Isom^+(\Sol)$ is written as
\Beq
f(\bm{x})=\{1,J\}\times\{1,R_3(\pi)\}\times\left[
\begin{matrix}{ccc}e^{-c^3} & 0 & 0\\0 &e^{c^3}&0\\0&0&1\end{matrix}
\bm{x} + \begin{matrix}{c}c^1\\c^2\\c^3\end{matrix}\right],
\Eeq
by which the invariant basis transforms as
\Beq
f^*\chi^I=A^I{}_J \chi^J;\quad
A=\{1,R_1(\pi)\}\times\{1,R_3(\pi)\}.
\Eeq

The automorphism group of $\Sol$ consists of elements which
transform the generators of the Lie algebra as
\Beq
\phi(\xi_1,\xi_2,\xi_3)=(\xi_1,\xi_2,\xi_3)\{1,J\}\times
\begin{matrix}{ccc} k_1&0&c^1\\ 0&k_2&-c^2\\0&0&1\end{matrix}.
\Eeq
They are induced from $f\in \HPDG^+(\Sol,\Sol)$ given by
\Beq
f(\bm{x})=\{1,J\}\times\left[
\begin{matrix}{ccc}k_1&0&0\\0&k_2&0\\0&0&1\end{matrix}\bm{x}
+\begin{matrix}{c}d^1-c^1e^{-z}\\ d^2+c^2e^z\\d^3\end{matrix}
\right].
\label{HPD:Sol}\Eeq
The corresponding transformation of the invariant basis is given by
\Beq
f^*\chi^I=A^I{}_J \chi^J;\quad
A=\{1,R_1(\pi)\}\times\begin{matrix}{ccc}k_+ & k_-& c_+\\
k_-& k_+ & c_-\\ 0&0&1\end{matrix},
\Eeq
where $k_\pm$ and $c_\pm$ are
\Beq
k_\pm:={1\over2}(k_1e^{d^3}\pm k_2 e^{-d^3}),\quad
c_\pm:=c^1 e^{d^3}\pm c^2 e^{-d^3}.
\Eeq

The subset $\HPDG^+(\Sol,\Isom^+(\Sol))$ is easily obtained 
from these HPDs. Its elements are expressed as
\Beq
f(\bm{x})=\{1,J\}\left[
\begin{matrix}{ccc}ke^{-d^3}&0&0\\0&ke^{d^3}&0\\0&0&1\end{matrix}\bm{x}
+\begin{matrix}{c}d^1\\ d^2\\d^3\end{matrix}
\right],
\label{HPD:Isom(Sol)}\Eeq
which transforms the invariant basis as
\Beq
f^*\chi^I=A^I{}_J \chi^J;\quad
A=\{1,R_1(\pi)\}\begin{matrix}{ccc}k & 0& 0\\
0& k & 0\\ 0&0&1\end{matrix}.
\Eeq

\subsubsection{Phase space of homogeneous covering data}

It is easy to see that by HPDs (\ref{HPD:Sol}) the components
$Q_{IJ}$ of the covering homogeneous metric $\tilde q$ with
respect to the invariant basis are put to the diagonal 
form $Q=[Q_1,1/Q_1,Q_3]$. For this metric the diffeomorphism
constraint (\ref{DC:H}) is written as
\Beq
P^{12}=P^{23}=P^{13}=0.
\Eeq
This implies that the $\Sol$-invariant covering data satisfying 
the diffeomorphism constraint is alway invariant by $\Isom^+(\Sol)$.
Hence the only non-empty phase space of homogeneous covering data
is
\Beq
\Gamma^+_\r{H,D}(\Sol,\Isom^+(\Sol))
=\left\{Q=[Q_1,Q_2,Q_3],P=[P^1,P^2,P^3]\right\}.
\Eeq

Thus, when expressed by the components with respect to the invariant
basis, the covering data have the same structure as those for
$E^3$ and $\Nil$. In particular the same parameterization of 
them (\ref{def:alphabeta:1})-(\ref{def:alphabeta:3}) and 
(\ref{def:palpha})-(\ref{def:p+}) can be used to
diagonalize $\Tr\dot QP$ in $\Theta$. Further, since the 
Ricci scalar curvature of the metric $\tilde q$ is given by
\Beq
R=-{(Q_1+Q_2)^2\over 2Q_1Q_2Q_3}=-2e^{-2\alpha+4\beta_+}
\cosh^2(2\sqrt{3}\beta_-),
\Eeq
the Hamiltonian is given by
\Beq
H={\kappa^2\over 12\Omega}Ne^{-3\alpha}
\left[-p_\alpha^2+p_-^2+p_+^2+{12\Omega\over \kappa^2}
e^{4(\alpha+\beta_+)}\cosh^2(2\sqrt{3}\beta_-)\right].
\label{H:Sol}\Eeq

\begin{table}
\begin{tabular}{ll}
\bf Space & \bf Fundamental group and  representation \\
\\
Sol($n;\omega_1,\omega_2$) 
& $[\alpha,\beta]=1,\gamma\alpha\gamma^{-1}=\alpha^p\beta^q,
\gamma\beta\gamma^{-1}=\alpha^r\beta^s;\quad
\begin{matrix}{cc}p&q\\r&s\end{matrix}\in\SL(2,\ZR)$\\
& \\
& $n=p+s, \omega_1={p-s+\sqrt{n^2-4}\over 2r},
\omega_2={p-s-\sqrt{n^2-4}\over 2r}$
\\
\\
$n>2:$ 
& $\alpha=(b^1\omega_1,b^2\omega_2,0)$\\
& $\beta=(b^1,b^2,0)\quad(b^1b^2\not=0)$\\
& $\gamma=(c^1,c^2,c^3); \quad e^{c^3}={n+\sqrt{n^2-4}\over 2}$\\
\\
$n<-2:$ 
& $\alpha=(b^1\omega_1,b^2\omega_2,0)$\\
& $\beta=(b^1,b^2,0)\quad(b^1b^2\not=0)$\\
& $\gamma=R_3(\pi)(c^1,c^2,c^3); \quad e^{c^3}={|n|+\sqrt{n^2-4}\over 2}$\\
\\
\end{tabular}
\caption{\label{tbl:Sol:Pi1}Fundamental groups and their
representation in Isom$^+$(Sol) of compact closed orientable
3-manifolds of type Sol}
\end{table}

\begin{table}
\begin{tabular}{llllc}
\bf Symmetry & \bf Topology & \bf Moduli & $\Omega$ & $\Theta$ \\
\\
Isom$^+$(Sol) & Sol($n;\omega_1,\omega_2$)
& $\begin{matrix}{ccc} \omega_1 & 1 & 0 \\ \pm\omega_2 & \pm 1 & 0\\
0 & 0 & c^3\end{matrix}$
& $|\omega_1-\omega_2|e^{c^3}$
& $\dot\alpha p_\alpha + \dot\beta_+p_+ +\dot\beta_-p_-$\\
\\
\end{tabular}
\caption{\label{tbl:CanonicalStr:Sol}Canonical structure of compact
orientable closed 3-manifold of type Sol}
\end{table}

\subsection{Topology and Moduli space}

All the orientable compact closed manifolds of type $\Sol$ are torus
bundles over $S^1$ with hyperbolic gluing maps. Conversely
any torus bundle over $S^1$ defined by a hyperbolic gluing map admit
a locally homogeneous structure modeled on $\Sol$.
Hence in terms of a matrix $Z=\begin{matrix}{cc}p&q\\r&s\end{matrix}
\in \SL(2,\ZR)$, the fundamental group $\pi_1(M)$ is described by
the relations among the generators of the fiber torus, $\alpha$ and
$\beta$, and the generator $\gamma$ for the base space $S^1$ as in
Table \ref{tbl:Sol:Pi1}. The condition that the gluing map is
hyperbolic is expressed in terms of $n=p+s$ as $|n|>2$. 

Abstract groups defined by $Z$ and $Z'$ in $\SL(2,\ZR)$ are isomorphic
if and only if there exist an integer matrix $V\in\GL(2,\ZR)$ such
that
\Beq
Z'=VZV^{-1}.
\label{ModularTrf:Sol}\Eeq
Since this modular transformation preserves the trace of $Z$,
fundamental groups with different values of $n$ are obviously 
non-isomorphic. However, the isomorphism class is not classified only
by $n$\footnote{In Ref.\cite{Koike.T&Tanimoto&Hosoya1994} it is stated 
that the isomorphism class is in one-to-one correspondence with the 
value of $n$, but it is not correct}. 
In fact, since the two roots of the quadratic
equation
\Beq
x={px +q \over r x + s}
\label{ModularRoots}
\Eeq
transform  as
\Beq
x'={a x + b\over c x + d}; \quad
V=\begin{matrix}{cc}a&b\\c&d\end{matrix},
\label{ModularTrf:roots}\Eeq
it can be easily shown that the isomorphism classes of the fundamental 
group are in one-to-one correspondence with the equivalence classes
of the roots of \Eq{ModularRoots} under the modular transformation 
(\ref{ModularTrf:roots}) for given $n$. From the theorem of the
continued fraction, the latter are determined by the divisor of
$n^2-4$ with the form $k^2$ and the equivalence classes of the reduced 
quadratic irrationals associated with the determinant $D=(n^2-4)/k^2$, 
whose count is finite. Thus for each $n$ there exist a finite number
of different isomorphism classes. For example, for $n=8$, there exist
two isomorphism classes represented by matrices 
$\begin{matrix}{cc}7&6\\1&1\end{matrix}$ and $\begin{matrix}{cc}7&3\\
2&1\end{matrix}$.

The nature of the embedding of the fundamental group differs for $n>2$ 
and $n<-2$; in the former case $\pi_1(M)$ is embedded into $\Sol$,
while in the latter case $\gamma$ is contained in the component
$R_3(\pi)\times\Sol$, as shown in Table \ref{tbl:Sol:Pi1}. 
Now we determine their canonical structure.

\subsection{Canonical Structure}

First we consider the case $n>2$. In this case we can put $c^1=c^2=0$
by a HPD of the form $(d^1,d^2,0)\in\Sol$. Then the HPD with
$d^1=d^2=0$ keeps the form of $\gamma$ and transforms the $(x,y)$
components of $\alpha$ and $\beta$ as
\Beq
f_*\begin{matrix}{cc}b^1\omega_1&b^1\\b^2\omega_2&b^2\end{matrix}
=\begin{matrix}{cc}ke^{-d^3}b^1&0\\0&ke^{d^3}b^2\end{matrix}
\begin{matrix}{cc}\omega_1&1\\ \omega_2&1\end{matrix}.
\Eeq
From the theorem of the continued fraction we can always find
$\omega_1$ from each equivalence class such that $\omega_1>1$ and
$-1<\omega_2<0$ corresponding to the reduced quadratic irrationals
with the helps of the combination of the modular transformation
$\gamma\maps\gamma^{-1}$ and the HPD $J$ if necessary. For this choice
$(b^1,b^2)$ can be transformed to $(1,1)$ for $b^1b^2>0$ and to $(1,-1)$ 
for $b^1b^2<0$, respectively. Hence the moduli matrix is reduced to
\Beq
\begin{matrix}{ccc}\bm{a}&\bm{b}&\bm{c}\end{matrix}=
\begin{matrix}{ccc} \omega_1 & 1 & 0 \\ \pm\omega_2 & \pm 1 & 0\\
0 & 0 & c^3\end{matrix},
\label{ModuliMatrix:Sol}\Eeq
where $e^{c^3}=(n+\sqrt{n^2-4})/2$. Thus the moduli space
$\M(M,\Isom^+(\Sol))$ reduces to two points. These two points are
connected only by orientation-reversing transformations. 
The canonical 1-form is simply given by 
\Beq
\Theta=\dot\alpha p_\alpha + \dot\beta_+ p_+ 
+\dot\beta_-p_-.
\label{Theta:Sol}\Eeq

For the moduli $K_0$ corresponding to \Eq{ModuliMatrix:Sol}, a generic
transformation belonging to it is simply expressed as
\Beq
\alpha^u\beta^v\gamma^w=(u\omega_1+v,\pm(u\omega_2+v),we^{c^3}).
\Eeq
Hence the fundamental region is given by a parallelepiped and its
volume by 
\Beq
\Omega=|\omega_1-\omega_2|e^{c^3}={n^2-4+n\sqrt{n^2-4}\over 2|r|}.
\Eeq
The Hamiltonian for the system is obtained just by inserting this into 
\Eq{H:Sol}.

The argument for the case $n<-2$ is quite similar. Though $\gamma$ now 
contains a factor $R_3(\pi)$, it does not affect the above derivation,
and we get the same result except that $n$ should be replaced by
$|n|$ in the expression for $\Omega$.

Thus for the locally homogeneous systems of type $\Sol$, there exists
no freedom of moduli and there occurs no degeneracy in the canonical
structure.

\section{Summary and Discussion}

In this paper we have developed a general algorithm to determine the
diffeomorphism-invariant phase space and its canonical structure of
locally homogeneous system on a compact closed 3-manifold, and have
applied it to locally homogeneous pure gravity systems of the Thurston
type $E^3$, $\Nil$ and $\Sol$. 

The main difference of our formulation from the conventional ones such
as that adopted in
Ref.\cite{Koike.T&Tanimoto&Hosoya1994,Tanimoto.M&Koike&Hosoya1997,%
Tanimoto.M&Koike&Hosoya1997a}
consists in the point that we have classified the diffeomorphism
classes of canonical data which contain both the configuration
variables and their conjugate momentums. Though this is a simple
extension of the conventional approach, it has enabled us to obtain
directly the canonical structure of the phase space of a locally
homogeneous system from that of a generic system on a compact
manifold. Further, together with a neat treatment of the invariance
group of the canonical data, it has given us a natural decomposition
of the diffeomorphism invariant phase space into the sector describing
the local structure of locally homogeneous data and the reduced moduli
sector describing the global structure of the data. This moduli sector
is in general smaller than that obtained by just considering the
structure of metric data. As we have shown, the dynamics of this
moduli sector is frozen. Though this point was already clearly stated
in Ref.\cite{Tanimoto.M&Koike&Hosoya1997} and was used in a crucial
way in their formulation, the statement and the proof in our paper is
more exact in the treatment of possible discrete components of the
invariance group of data, to which the HPD freedom and the
corresponding gauge-fixing in the moduli sector is very sensitive.

All the results of our analysis of the locally homogeneous pure
gravity systems are contained in tables \ref{tbl:CanonicalStr:E3},
\ref{tbl:freedom:E3}, \ref{tbl:CanonicalStr:Nil},
\ref{tbl:freedom:Nil} and \ref{tbl:CanonicalStr:Sol}. As these tables
show, the canonical structure and the dynamics of locally homogeneous
pure gravity systems are quite simple in our decomposition of the
phase space variables to the local sector and the moduli sector,
except for the degeneracy in the canonical structure. This degeneracy
implies that locally homogeneous
systems are not canonically closed in the full
diffeomorphism-invariant phase space of generic data. In particular
it is meaningless to discuss the quantum dynamics of a locally
VII(0)-homogeneous system. It will be an interesting problem to find a 
canonically closed minimal sector which contains the locally
VII(0)-system.

As we touched upon in the analysis of the locally homogeneous pure
gravity systems, the reduction of the moduli space by gauge-fixing
often leaves freedoms of discrete HPDs and modular
transformations. These residual discrete transformations
introduce non-trivial topological structures as well as conic
singularities into the phase space. The influence of these topological 
structures of the classical phase space on their quantum theory may be
an interesting problem.

Finally we comment on the other locally homogeneous pure gravity
systems. In general locally homogeneous systems are covered by Bianchi
models or the Kantowski-Sachs model. The former is further classified
into two subclasses, class A and class B. In this classification our
analysis only covers four of the class A Bianchi models, type I, II,
VI(0) and VII(0). However, these models exhaust all the interesting
cases except one as far as the canonical structure is concerned. First
locally homogeneous systems modeled on $S^3$ (type IX) have no moduli
freedom. Hence its canonical structure is essentially determined by
their homogeneous covers. Second, among the class B Bianchi models,
type V and type VII$(A\not=0)$ belong to the same Thurston type,
$H^3$, hence they do not have continuous moduli freedom either(see
Table \ref{tbl:ThurstonType}).  Among the remaining two Bianchi models
with compact quotients, VIII and III, III can be embedded into two
Thurston types, $H^2\times E^1$ and $\widetilde{SL_2\RF}$. In the
former case, the canonical structure is essentially determined by that
in the 2-dimensional systems on $H^2$. On the other hand, in the
latter case, the systems become locally VIII-homogeneous at the same
time. Thus they have both class A and class B symmetries. This is the
only important case we have not analyzed. Since our analysis shows
that the canonical structure becomes in general degenerate when the
invariance group has non-trivial isotropy groups, the analysis of
these locally homogeneous systems of type $\widetilde{SL_2\RF}$  will
give useful information on the origin of degeneracy of the canonical
structure as well as the dynamics of local homogeneous systems with
class B symmetry(cf. Ref.\cite{Tanimoto.M&Koike&Hosoya1997a}. 
This problem will be pursued in a future work.

\section*{Acknowledgment}

This work is supported by the Grant-In-Aid of for Scientific Research (C)
of the Ministry of Education, Science and Culture in Japan(05640340).

\end{document}